\documentclass[a4paper]{article} 
\usepackage{geometry}
	
\usepackage{tgpagella}
\usepackage{fontawesome5}
\usepackage[usenames,dvipsnames,svgnames,table]{xcolor}
\definecolor{mybg}{HTML}{282A36}
\definecolor{mycl}{HTML}{44475A}
\definecolor{myfg}{HTML}{F8F8F2}
\definecolor{mycomment}{HTML}{6272A4}
\definecolor{mycyan}{HTML}{8BE9FD}
\definecolor{mygreen}{HTML}{50FA7B}
\definecolor{myorange}{HTML}{FFB86C}
\definecolor{mypink}{HTML}{FF79C6}
\definecolor{mypurple}{HTML}{BD93F9}
\definecolor{myred}{HTML}{FF5555}
\definecolor{myyellow}{HTML}{F1FA8C}
\definecolor{VividPurple}{HTML}{3E0097}
\definecolor{SlateGrey}{HTML}{2E2E2E}
\definecolor{LightGrey}{HTML}{666666}
\definecolor{DarkPastelRed}{HTML}{450808}
\definecolor{PastelRed}{HTML}{8F0D0D}
\definecolor{GoldenEarth}{HTML}{E7D192}
\definecolor{awesome-emerald}{HTML}{00A388}
\definecolor{awesome-emerald-dark}{HTML}{00806A} 
\definecolor{awesome-skyblue}{HTML}{0395DE}
\definecolor{awesome-skyblue-dark}{HTML}{0376B0}
\definecolor{awesome-red}{HTML}{DC3522}
\definecolor{awesome-red-dark}{HTML}{B02A1C}
\definecolor{awesome-pink}{HTML}{EF4089}
\definecolor{awesome-pink-dark}{HTML}{EC136D}
\definecolor{awesome-orange}{HTML}{FF6138}
\definecolor{awesome-orange-dark}{HTML}{FF3300}
\definecolor{awesome-nephritis}{HTML}{27AE60}
\definecolor{awesome-nephritis-dark}{HTML}{219150}
\definecolor{awesome-concrete}{HTML}{95A5A6}
\definecolor{awesome-concrete-dark}{HTML}{74898B}
\definecolor{awesome-darknight}{HTML}{131A28}
\definecolor{awesome-darknight-dark}{HTML}{101623}
\definecolor{awesome-snowwhite}{HTML}{F9FBFD}
\definecolor{awesome-snowwhite-dark}{HTML}{F3F6FB}
\definecolor{awesome-blue-dark}{HTML}{0000FF}
\definecolor{awesome-golden}{HTML}{E1AD21}
\definecolor{awesome-silver}{HTML}{AAA9AD}
\definecolor{darktext}{HTML}{414141}
\definecolor{darktext-dark}{HTML}{262626}
\definecolor{text}{HTML}{333333}
\definecolor{graytext}{HTML}{5D5D5D}
\definecolor{lighttext}{HTML}{999999}
\definecolor{VividRed}{HTML}{7e2635}
\definecolor{DarkRed}{HTML}{a5402d}
\definecolor{SlateGrey}{HTML}{2E2E2E}
\definecolor{LightGrey}{HTML}{666666}
\UseRawInputEncoding
\usepackage[english]{babel}

\usepackage{lipsum}
\usepackage{amsmath, nccmath} 

\usepackage{gensymb}
\usepackage{amssymb}
\usepackage{graphics}

\usepackage{graphicx}
\usepackage{epsfig}

\usepackage{fancyhdr}
\usepackage{wasysym}
\usepackage{latexsym}
\usepackage{amsfonts}

\usepackage{float}
\usepackage{color,colortbl}
\usepackage{cite}
\usepackage{notoccite} 
\usepackage{soul}
\usepackage{multirow}
\usepackage{etoolbox}
\usepackage{authblk}
\usepackage{blindtext}
\usepackage{dcolumn}
\usepackage{bm}
\usepackage{qcircuit}
\usepackage{braket}
\usepackage{physics}
\usepackage{braket}
\usepackage{varwidth}
\usepackage{empheq}
\usepackage[]{algorithm2e}
						   
\usepackage{algorithmic}
\usepackage{framed}
\usepackage{enumitem}
\usepackage[utf8x]{inputenc} 
\usepackage[english]{babel}

\usepackage{url}
\usepackage{hyperref}
\usepackage[font={footnotesize}]{caption}
\usepackage{enumitem}
\usepackage{multirow}
\usepackage{textcomp}

\usepackage[T1]{fontenc}
\usepackage{listings}
\usepackage{subcaption}
\usepackage{textcomp}
\usepackage{mathrsfs}
\usepackage{hyperref}
\usepackage{caption}
\usepackage{cleveref}
\usepackage{xr}
\graphicspath{{./figures}}
\usepackage{amsthm}
\usepackage{enumerate}
\usepackage{array}
\usepackage[parfill]{parskip}
\usepackage{amscd}
\usepackage[]{units}
\usepackage{multicol}
\usepackage{tcolorbox}
\usepackage{media9}
\usepackage{pifont}
\usepackage{booktabs}
\usepackage[acronym]{glossaries}
\usepackage{alphalph}
\usepackage{mathtools}
\usepackage{blindtext}
\usepackage{subfiles}
\usepackage{fullpage}
\usepackage{makeidx}
\usepackage{fancyhdr}
\usepackage[calcwidth]{titlesec}
\usepackage[font={footnotesize},labelfont={bf}]{caption}
\usepackage{setspace}%
\providecommand{\U}[1]{\protect\rule{.1in}{.1in}}
\usepackage[symbol]{footmisc}
\usepackage{bookmark}
\usepackage[compat=1.1.0]{tikz-feynman}
\usepackage{tikz}
\usetikzlibrary{spy,calc}
\usetikzlibrary{matrix,backgrounds}
\usetikzlibrary{shapes.geometric, arrows}
\usetikzlibrary{decorations.pathmorphing}
\tikzstyle{startstop} = [rectangle, rounded corners, minimum width=3cm, minimum height=1cm,text centered, draw=black, fill=red!30]
\tikzstyle{io} = [trapezium, trapezium left angle=70, trapezium right angle=110, minimum width=3cm, minimum height=1cm, text centered, draw=black, fill=blue!30]
\tikzstyle{process} = [rectangle, minimum width=3cm, minimum height=1cm, text centered, text width=3cm, draw=black, fill=orange!30]
\tikzstyle{decision} = [diamond, minimum width=3cm, minimum height=1cm, text centered, draw=black, fill=green!30]
\tikzstyle{arrow} = [thick,->,>=stealth]

\tikzstyle{decision} = [diamond, draw, fill=blue!20, 
text width=4.5em, text badly centered, node distance=3cm, inner sep=0pt]
\tikzstyle{block} = [rectangle, draw, fill=blue!20, 
text width=5em, text centered, rounded corners, minimum height=4em]
\tikzstyle{line} = [draw, -latex']
\tikzstyle{cloud} = [draw, ellipse,fill=red!20, node distance=3cm,
minimum height=2em]		

\usetikzlibrary{shadows, patterns}
\tikzset{button/.style={
preaction={fill=blue,path fading=circle with fuzzy edge 20 percent,
opacity=.7,transform canvas={xshift=1mm,yshift=-1mm}},
preaction={pattern=#1,
path fading=circle with fuzzy edge 20 percent},
preaction={top color=white,
bottom color=red!50,
shading angle=180,
path fading=circle with fuzzy edge 20 percent,
opacity=0.4},
preaction={path fading=fuzzy ring 15 percent,
top color=black!5,
bottom color=black!80,
shading angle=180},
inner sep=2ex
},
button/.default=horizontal lines light blue,
circle}								 

\definecolor{anti-flashwhite}{rgb}{0.95, 0.95, 0.96}
\definecolor{codegreen}{rgb}{0,0.6,0}
\definecolor{codepurple}{rgb}{0.58,0,0.82}
\lstdefinelanguage{NeMO}{
keywords={},
ndkeywords={solver},
keywordstyle=\color{blue},
ndkeywordstyle=\color{codepurple},
commentstyle=\color{codegreen},
stringstyle=\color{cyan},
sensitive=true
}
\hypersetup{colorlinks=true,
linktocpage=true,
colorlinks=true,
linkcolor=awesome-blue-dark,
filecolor=awesome-red,
citecolor=awesome-pink-dark,      
urlcolor=awesome-orange-dark,
pdftitle={Electronic Transport in Electron-Phonon Gas of a Generalized Organic Molecular-crystal: Non-equilibrium Green's Function Formalism \& Boltzmann Transport Framework},
bookmarks=true,
pdfauthor={Bhupesh~Bishnoi},
bookmarks=true,				
bookmarksopen=true,
pdfpagemode=UseOutlines,
pdfstartview={FitH},	
unicode   =true,
pdftoolbar=true,
pdfmenubar=true,
}
\makeatletter
\def\@fnsymbol#1{\ensuremath{\ifcase#1\or \pmb\ddagger \else\@ctrerr\fi}}
\makeatother

\title {\Huge {Electronic Transport in Electron-Phonon Gas of Two-Dimensional Holstein's Organic Molecular-Crystal: Non-equilibrium Green's Function Formalism \& Boltzmann Transport Framework} \hspace{2cm} \\ \LARGE {(A Generalized Mathematical Solution)}}
\author{\bf{Bhupesh~Bishnoi\thanks{\href{mailto:bishnoi@computer.org}{\textcolor{PastelRed}{\texttt{\bf{bishnoi[At]computer[Dot]org}}}}~~~\href{mailto:bishnoi@ieee.org}{\textcolor{PastelRed}{\texttt{\bf{bishnoi[At]ieee[Dot]org}}}} }}}
\date{}

\begin{document}
\maketitle

\begin{abstract} 
We have presented a consistent electronic transport framework for the two-dimensional extended Holstein's organic molecular-crystal based upon complete quantum-mechanical treatment through the non-equilibrium Green's function (NEGF) formalism and corresponding one-to-one semi-classical framework based upon Boltzmann transport theory for the narrow-bandwidth electronic energy organic semiconductor crystal material and device. In this process, we have formulated electronic self-energy interaction with acoustic and polar optical phonon mode with one phonon and two phonon interactions through Feynman's diagrammatic techniques to investigate combined electronic transport. Furthermore, the aforementioned can readily expand for the three and four phonon interactions on choice based upon the particular class of organic semiconductor crystal and organic polymers for the modern organic device industry.
\end{abstract}

\section*{Introduction} 
\begin{figure}[H]
\centering
\includegraphics[scale=0.6]{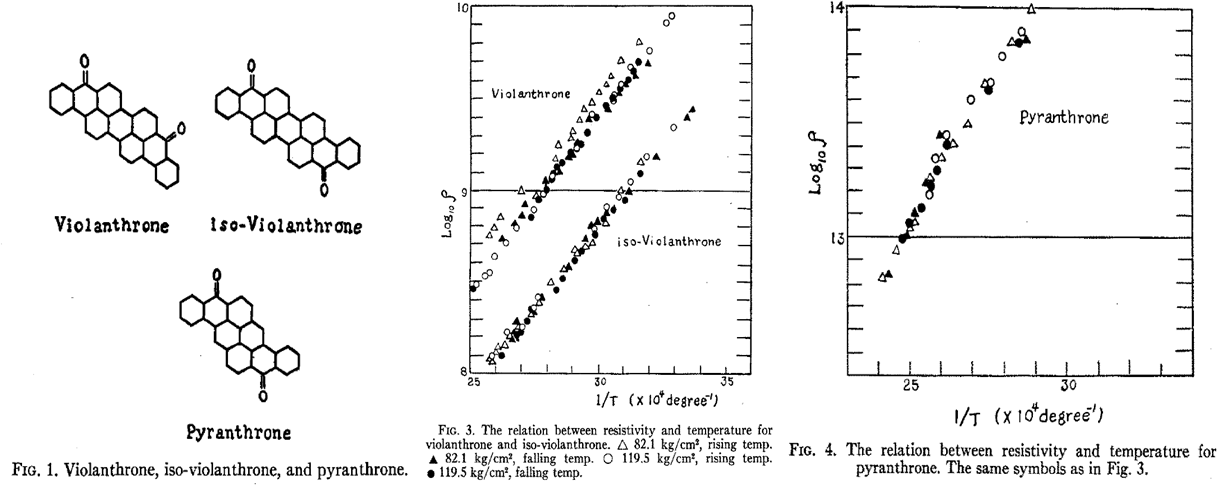}
\caption{Violanthrone, iso-violanthrone, and pyranthrone and relation between resistivity and temperature.\cite{akamatu_new_1947, akamatu_electrical_1950}}
\label{fig-0}
\end{figure}

Historically, during the 1940s, Akamatu, Nagamatsu, and Inokuchi, \textit{et al.} in japan, first investigated violanthrone, iso-violanthrone, and pyranthrone macromolecular lattice of organic solid-state hexagonal compounds of black carbon by X-ray diffraction method. Furthermore, for the first time in \cref{fig-0} measure these organic compounds' electrical resistivity at room temperature, resistivity variation with temperature, and calculated activation energy. Based on these investigations, it concludes that organic compounds are intrinsic semiconductors. \cite{akamatu_new_1947, akamatu_electrical_1950}

Later, in the 1950s, Holstein proposed the one-dimensional molecular-crystal model and polaron theory of transport in the molecular crystal in seminal articles. \cite{holstein_studies_1959-1, holstein_studies_1959, holstein_theory_1964} The original first proposal used classical phonon and Schr\"{o}dinger's wave equation in the organic polar crystal based on the relevant work of Yamashita \textit{et al.} on the polar inorganic crystal of $\text{NiO}$.  \cite{yamashita_electronic_1958} Meanwhile, Kepler, LeBlanc, \textit{et al.} in the 1960s, calculated the one-electron band structure and transport properties in the anthracene. \cite{leblanc_band_1961} Later on, Holstein and Friedman \textit{et al.} expand their work and investigate the Hall effect in the polaron band regime. \cite{friedman_hall_1963, friedman_transport_1964,friedman_density_1964,friedman_electron-phonon_1965, holstein_hall_1968,friedman_hall_1971} At the same time, Gosar and Sang Choi \textit{et al.} used Kubo's linear response theory\cite{kubo_statistical-mechanical_1957} and the Wannier representation \cite{wannier_structure_1937} in the aromatic crystal to estimate electron-phonon interaction integrals and estimate electronic mobility in the anthracene crystal. \cite{gosar_linear-response_1966 } Schnakenberg \textit{et al.} presented similar efforts for modeling the hopping and band-conductivity in the narrow-band semiconductor and reported the complete Boltzmann transport equation for the molecular-crystal. \cite{schnakenberg_derivation_1965, schnakenberg_quasiteilchen-spektrum_1966, schnakenberg_polaronic_1968,schnakenberg_electron-phonon_1969} 
Later, in the 1980s, Silbey and Munn \textit{et al.} tried to combine the polaronic theory and band theory in the molecular-crystal. \cite{ silbey_exchange_1965, silbey_general_1980, munn_theory_1985, munn_theory_1985-1, coropceanu_charge_2007, cheng_unified_2008 } At the same time, Kenkre and Dunlap \textit{et al.} modeled the dynamic and static disorder in organic molecular-crystal. \cite{kenkre_unified_1989, kenkre_charge_1992, kenkre_finite-bandwidth_2002,giuggioli_mobility_2003, shen_charge_2003 } Later sekirin \textit{et al.} modeled exciton phonon interaction couplings between excited states and lattice vibrations in the molecular crystals. \cite{tovstenko_excitonphonon_2002}

We have proposed here an electronic transport formalism in the organic crystal. The electronic transport and current calculation in the organic crystal are humongous tasks. The organic crystals constitute organic molecules and lack a rigid lattice structure. Historically, the critical transport time of flight is deduced based on two approaches. The electronic transport parameters calculate from the hopping jump probability method and the Boltzmann transport equation in the organic crystal. The non-equilibrium Green's function based electronic transport calculation formalism has seen tremendous growth in the past few decades and reported in the solid-state in-organic devices. However, the non-equilibrium Green's function formalism did not propose in the literature for the organic crystal devices to the best of the author's knowledge.
Furthermore, during the last two decades, there has been tremendous growth in the low-temperature fabrication process of organic molecules based on electronic and display devices and polymer-based electronics devices. Therefore it is natural to investigate the transport parameter from the bottom-up quantum mechanical approach, to deduce the mesoscopic charge density and electronic current from the microscopic propagative wavefunctions in the organic crystal. In this regard, we have formulated the current transport formalism based on the non-equilibrium Green's function formalism. There are some efforts to calculate the transport parameter in the organic devices based on the first principle, ab-initio density functional theory. And then extract the electron-phonon interaction coefficients to apply them in the second stage of Boltzmann transport theory to calculate the current parameters. The calculation of the Boltzmann transport equation is mainly in the relaxation time approximation. However, the calculated value from these methods is six to seven orders of difference from the measured values of current mobility. The underline bottleneck in these methods is that the density functional theory is a ground-state theory. The transport mechanism in the organic devices is a non-equilibrium system where the conduction path establishes and the continuous exchange of electronic wavefunction with the outside reservoir battery-cell with the system's heat conduction flow.
Moreover, there are other exotic effects, such as exciton transport in organic crystal transport. There is also the interaction with the light phonon in the organic display device. Furthermore, most of the organic molecular crystals are narrowband semiconductors. These methods are applicable in the band transport regimes where relaxation time approximation solves the Boltzmann transport equation held in a wideband transport regime. In the region of most modern molecular crystal devices, the narrow bandwidth is larger than the 0.2eV or at least ten times larger than the thermal kT value; the transport consider to be band regime, and polaronic band transport is minuscule. However, this generalized non-equilibrium Green's function method can easily incorporate the polaronic band hamiltonian. Therefore, the extension to the polaronic regime state forward for the low-temperature polaronic band. However, organic molecular crystal-based electronic devices use pentacene, anthracene, rubrene molecules with three or longer carbon chains and exhibit relatively high mobility in the tens of values while operating at room temperature. Therefore, they are far from the small polaron band regime. Transport governs by the strong coupling between the adjacent molecules site in the organic crystal, which gives larger mobility values. The molecular crystal with a stronger overlap coupling between transfer function in the atomic orbital in the nearest neighbor lattice site in the organic crystal observes a robust electrical conductivity. Compared to pentacene and anthracene, rubrene shows more strong coupling due to its distinct molecule structures and the electric current. These transfer overlap integral, or coupling coefficients in detail, are calculated by Mulliken in his seminar paper. \cite{mulliken_formulas_1949} Therefore, the organic crystal's molecular hamiltonian construction incorporates the overlapping atomic orbital of the adjacent nearest neighbor molecular crystal site and closely follows the tight-binding description of the linear combination of atomic orbitals (LCAO) in solid-state crystal.

Also, in most organic crystals, the conduction band is very narrow in the range of a few kT, and the valence bandwidth is at least twenty times or greater than the thermal voltage at room temperature. Most practical organic crystal devices are P-type, and the valence bandwidth governs the transport properties. Therefore, the possibility of a small polaronic band transport or jump probability-based hopping transport with discrete energy levels is minuscule. In the non-equilibrium Green's function formalism, we will calculate the greater Green's function for hole transport and the lesser Green's function for electronic transport. Also, the Fermi-function is $ (1-\mathcal{F}) $ form. Our method has not incorporated a more complex mechanism such as the phonon drag effect, where the electron traps in a potential well created by the phonon vibration. The electronic state is bound around the phonon and carried away by the phonon transport mechanism at room temperature. The internal vibration of molecules in the organic crystal will give intramolecular vibronic modes, giving rise to band-edged electronic band structures. However, such vibronic couplings are ignored in the present calculation.

The outline of this theoretical work is as follows. First, we briefly discuss the historical prospect of experimental work on organic molecular crystals and theoretical development in this field so far. Later we start our framework with Holstein's molecular-crystal model and division of hamiltonian in electronic, interaction, and Bosonic parts. However, as first derived and later on admitted in subsequent work by Holstein, the Schr\"{o}dinger's representation will not further progress as the small perturbation vector. Lattice translation vector will mix up, and for more than one-dimensional crystal, it is challenging to add up all the effects and propagate an Schr\"{o}dinger wave in such a crystal as well treating phonon in the quantum domain. To mitigate these difficulties, we will follow the second quantization language and Heisenberg representation and introduce the perturbation field in this molecular crystal, expand its effect on the electronic hamiltonian up to second-order and divide hamiltonian into the static part and dynamic part. Afterward, we will introduce electronic Green's function propagator in such a crystal and treat the dynamic part as a perturbation to the static part in the Green's function perturbation formulation. Finally, we expand Green's function perturbation to second order with one phonon and two phonon interactions in the narrow energy bandwidth, organic molecular crystal. Afterward, we discuss the possibility of all such Feynman diagrams of self-energy in linked, unlinked, reducible, and irreducible configuration in the graphical expansion scheme and calculate the interaction self-energy. After achieving the self-energy and related spectral function and broadening matrix from perturbation expansion, we extended the work using these electronic and Bosonic correlation functions into a Green's function propagator equation of motion in the layered organic crystal. Then, we wrote the coupled equation of motion in the Kadanoff-Keldysh non-equilibrium Green's function formalism to solve the direct calculation of carrier density and current continuity equation in the system, which is equivalent to the Landauer formula of current in a system. After completing the entire quantum domain framework, which is hugely computationally demanding, we switched to semi-classical Boltzmann transport theory to apply the non-equilibrium Green's function scattering self-energy formulation to re-formulate the Boltzmann transport equation conjoined the framework. Total net scattering rates derive for the one phonon and two phonon interactions with the acoustic and polar optical phonon mode. Furthermore, we proposed a Monte-Carlo-based stochastic solution to evaluate organic molecular crystals' conductivity and mobility at the end of this work and hope this mathematical framework is more physically insightful and microscopically detailed than the classical jump probability-based Marcus theory.

\section*{THEORY}
\subsection*{Model Hamiltonian}

Historically, a single electron interacting with the boson field in the ionic lattice crystal was modeled by Fr\"{o}hlich Hamiltonian. \cite{frohlich_electrons_1954,frohlich_xx_1950} Furthermore, a very similar hamiltonian for the molecular crystal is proposed by Holstein in molecular-crystal model (MCM). \cite{holstein_studies_1959-1} The interaction of Fermions with the Bosonic field in the ionic lattice, polar semiconductor, and molecules crystal was investigated to deduce the material's electronic properties in the weak and strong coupling region. The concept of large polaron quasi-particle was discussed and debated since the early 1940s with the advancement of the quantum theory of material and transport. \cite{landau_effective_1948,landau_collected_1965,lee_motion_1953,osaka_polaron_1959, pekar_theory_1969} Various mathematical treatments of quantum theory first applied to the polaron field, such as path-integral framework, \cite{feynman_slow_1955,feynman_mobility_1962} Wannier function, \cite{luttinger_effect_1951} Green's function, \cite{langreth_perturbation_1964} and Boltzmann transport equation, \cite{kadanoff_boltzmann_1963} for this interacting electron-phonon gas of one dimensional to the three-dimensional system. For completeness, we will write both the hamiltonian first and later develop the system from the starting point of Bloch wavefunction of the excess electron in the host crystal. Electrons in the interacting polar/non-polar crystal can also describe by the Wannier function. However, we will follow the Bloch representation and, later on, the tight-binding approximation to describe the crystal lattice. Our focus is to describe the non-equilibrium system electrical dynamics in one-electron, N-phonon Green's function representation for evaluating the organic molecular crystal device's conductivity at finite temperature.

The physical situation in the organic molecular crystal semiconductor divides into three principal Hamiltonian features. Furthermore, the related Hamiltonian described each physical state describing the governing dynamics and interaction with the others.
The system's total Hamiltonian is the sum of three terms, electronic Hamiltonian $ H_{\mathrm{E}} $, lattice Hamiltonian $  H_{\mathrm{P}} $, and interaction Hamiltonian $ H_{\mathrm{INT}} $, respectively, representing: The $ H_{\mathrm{E}} $ electronic component, which consists of the electron's kinetic energy effective one-electron periodic potential. $ H_{\mathrm{P}} $ is a lattice component and sum of lattice phonon kinetic energy and lattice potential energy in the function of phonon displacements from their equilibrium positions. And $ H_{\mathrm{INT}} $, electron-phonon interaction, and function of electron coordinate and phonon displacement. We will treat $ H_{\mathrm{INT}} $ essentially in the standard theory of slow-moving electrons in semiconducting crystals as a small perturbation where phonon-vibration quanta are simultaneously absorbed or emitted with electron and give rise to scattering transitions to the electron. We have formulated this generalized technique based upon the two-dimensional Holstein's molecular-crystal model for the narrow-band semiconductor. The molecular crystal lattice model consists of an N-identical polyatomic molecule where the center of gravity and orientation of the molecules is fixed in the lattice. However, internuclear separation varies due to the lattice vibration of the individual molecules. As mentioned above, the generalization consists of extending the  molecular-crystal model to a two-dimensional crystal lattice and apply a dc electric field for the transport calculation.

\begin{figure}[H]
\centering
\includegraphics[scale=0.6]{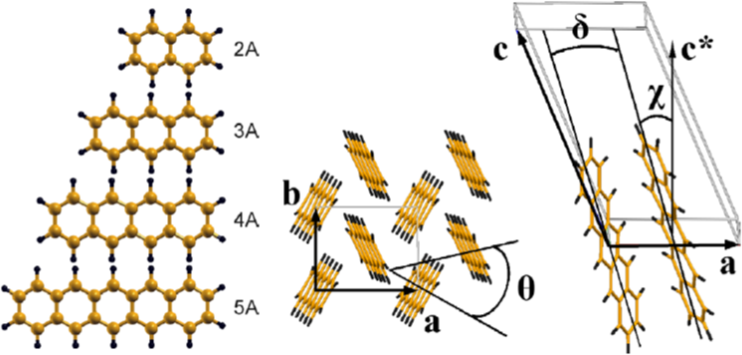}
\caption{Oligomers of the acene series and their herringbone packing. Left, from top to bottom: naphthalene (2A), anthracene (3A), tetracene (4A), and pentacene (5A). Right: the herringbone arrangement in the ab plane and the layered stacking perpendicular to it.\cite{ambrosch-draxl_role_2009}}
\label{fig-14}
\end{figure}

\begin{figure}[H]
\centering
\includegraphics[scale=0.6]{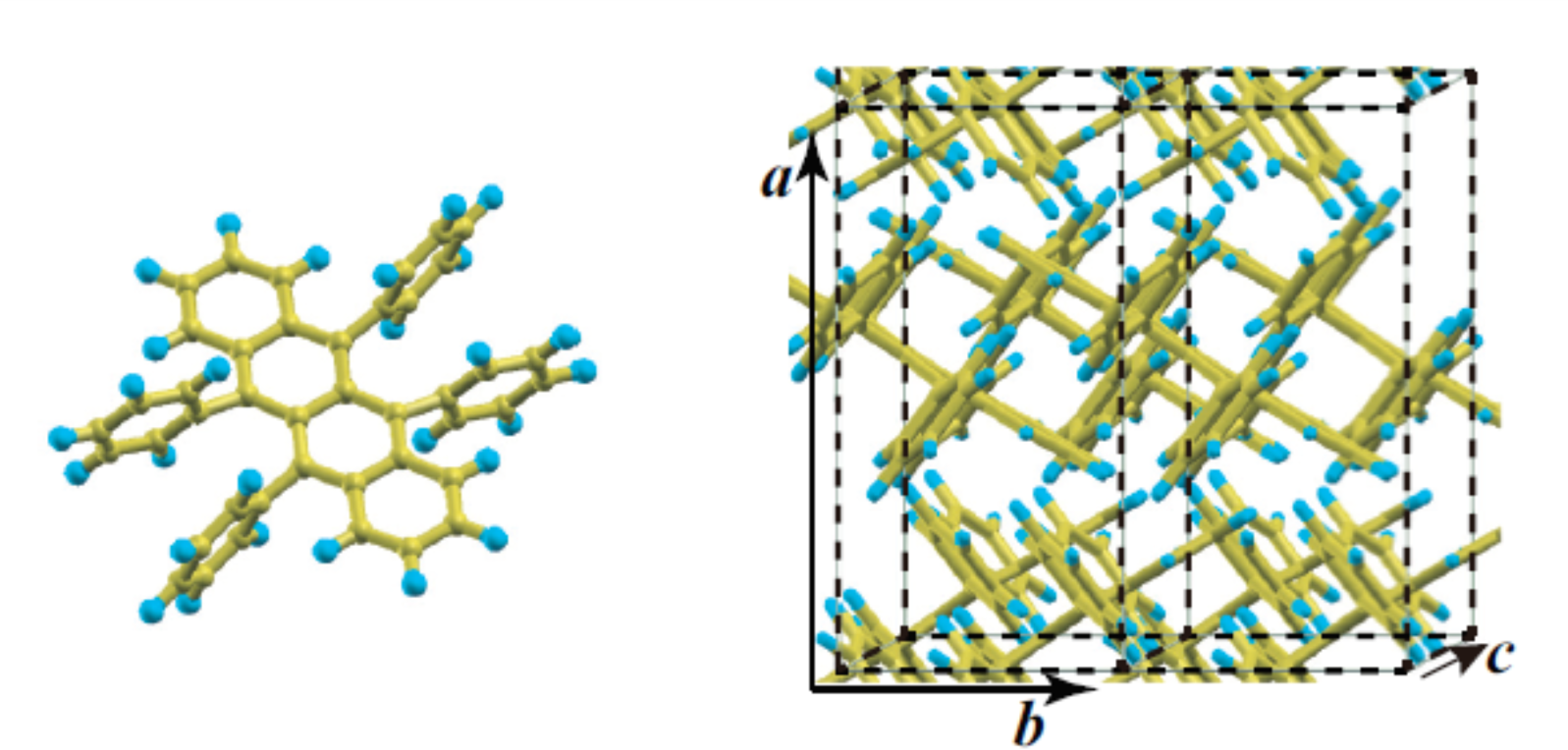}
\caption{Atomic structure of isolated rubrene molecules and rubrene single crystals. The C and H atoms are indicated by yellow and light blue spheres, respectively. In the single crystal, the tilted long axes of the molecules are stacked along the b direction. \cite{yanagisawa_homo_2013}}
\label{fig-13}
\end{figure}

\begin{figure}[H]
\centering
\includegraphics[scale=0.6]{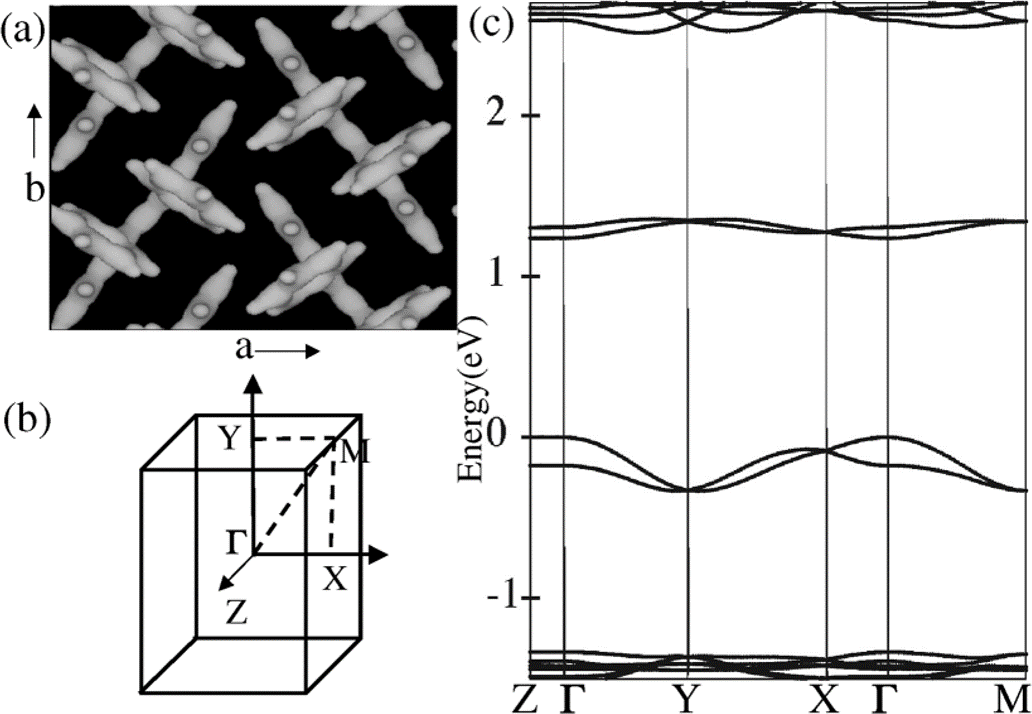}
\caption{The charge density isosurface enclosing 40\% of the total charges obtained from DFT, which illustrates the crystal structure in the ab plane with 2 inequivalent rubrene molecules arranged in a herringbone structure. (b) The reciprocal lattice; X, Y, and Z correspond to the a, b, and c crystalline direction. (c) The band structure of rubrene using DFT-GGA. \cite{li_light_2007}}
\label{fig-15}
\end{figure}

\begin{figure}[H]
\centering
\includegraphics[scale=0.6]{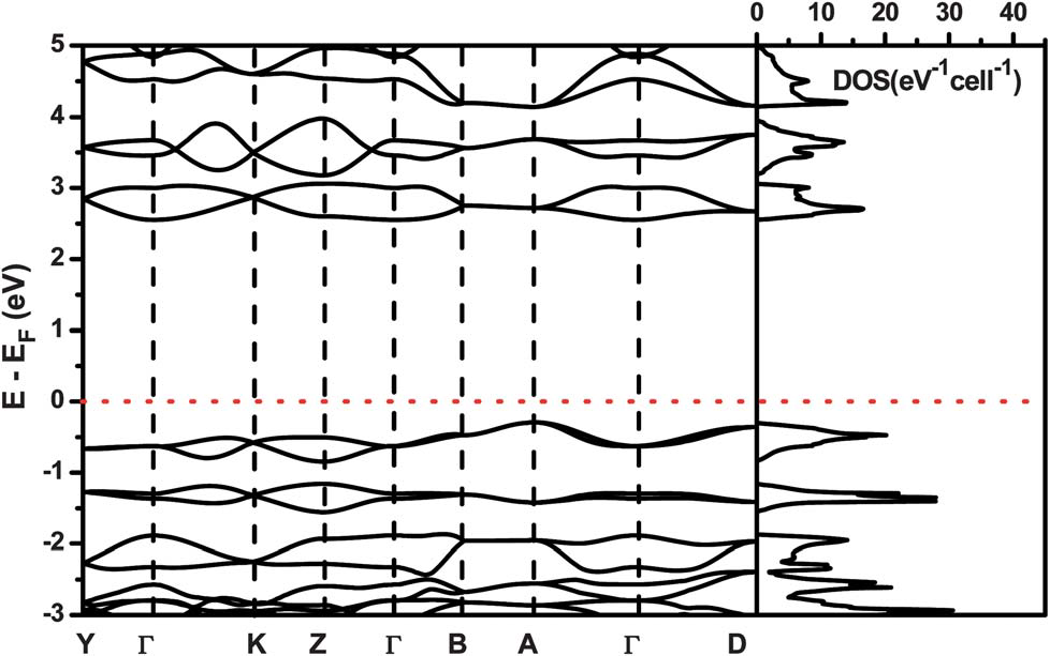}
\caption{LDA band structure and DOS of naphthalene. The reciprocal coordinates of high-symmetry points are G(0,0,0), Y(0.5,0,0), K(0.5,0,0.5), Z(0,0,0.5), B(0,0.5,0), A(0.5,0.5,0), D(0.5,0.5,0.5) respectively. The red dashed line represents the position of the Fermi level.\cite{xi_first-principles_2012}}
\label{fig-16}
\end{figure}

\begin{figure}[H]
\centering
\includegraphics[scale=0.6]{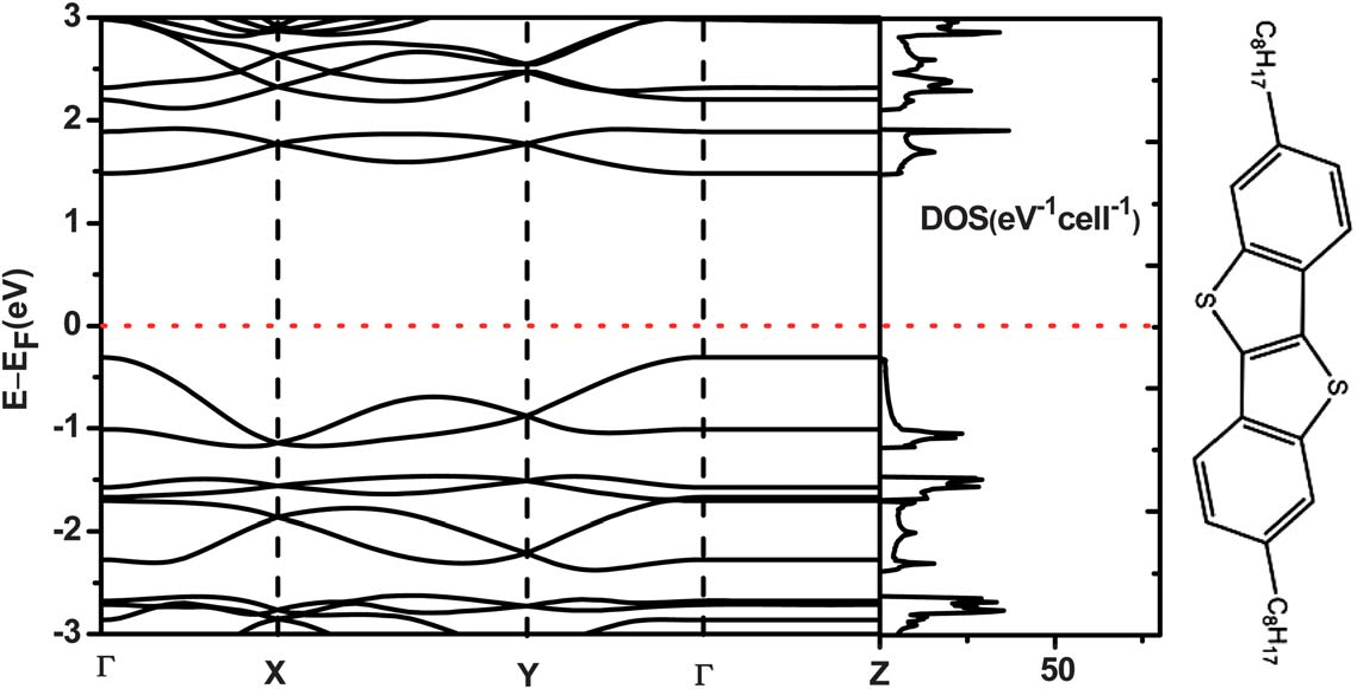}
\caption{The DFT-calculated band structure and DOS of C8-BTBT. The reciprocal coordinates of the high-symmetry points are G(0,0,0), X(0.5,0,0), Y(0,0.5,0), Z(0,0,0.5) respectively. The red dashed line is the position of the Fermi level.\cite{xi_first-principles_2012}}
\label{fig-17}
\end{figure}

\begin{figure}[H]
\centering
\includegraphics[scale=0.5]{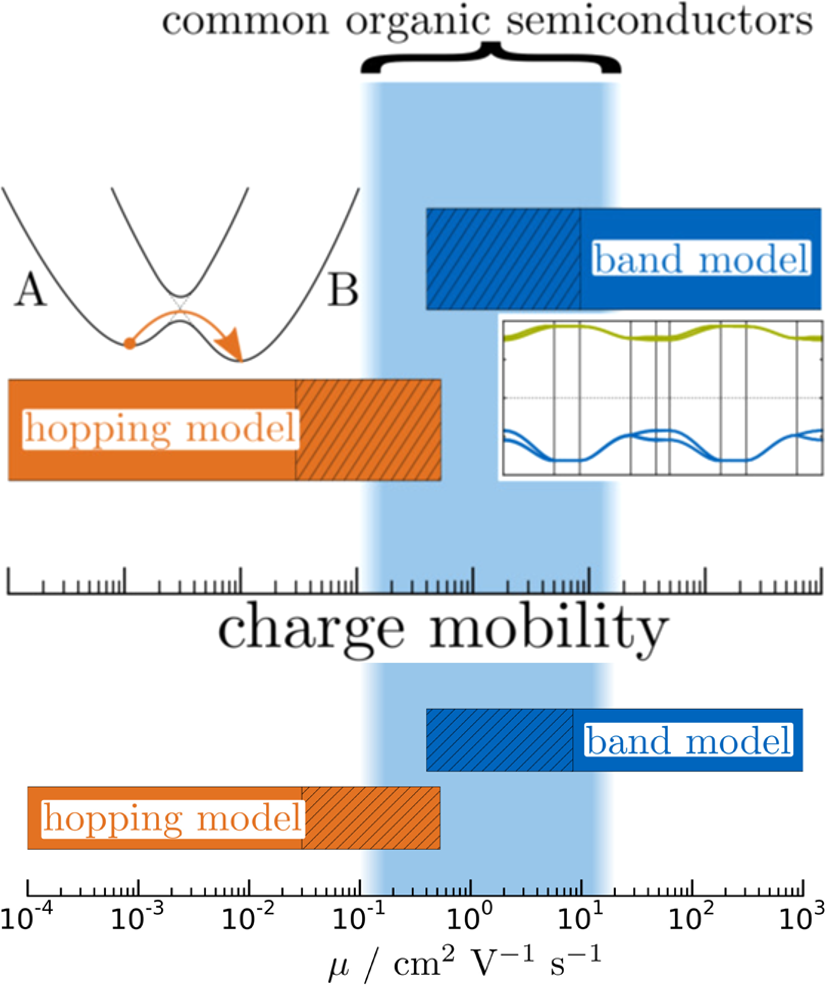}
\caption{Charge transport in molecular materials: an assessment of computational methods.\cite{oberhofer_charge_2017}}
\label{fig-18}
\end{figure}

\begin{figure}[H]
\centering
\includegraphics[scale=0.4]{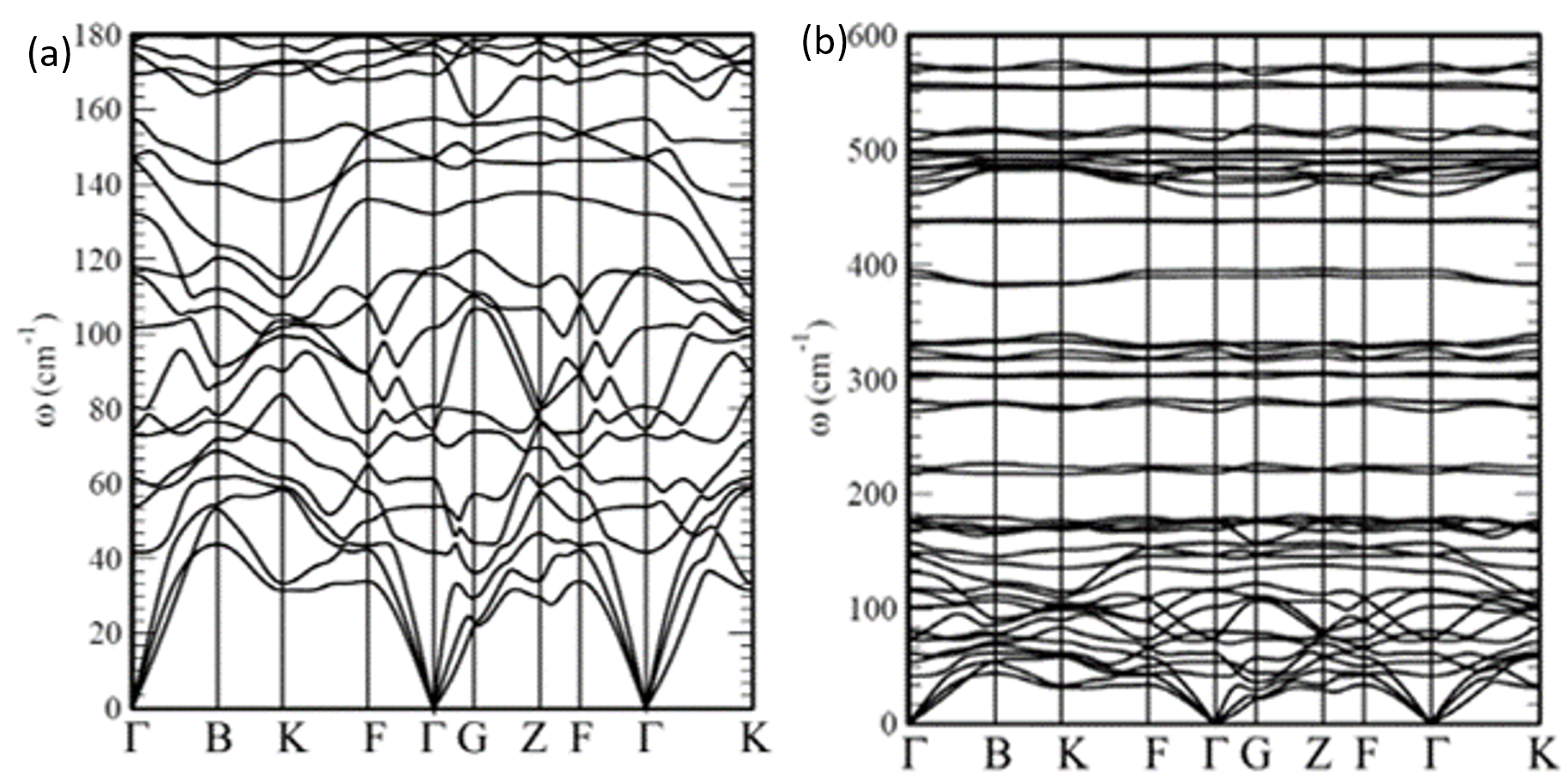}
\caption{Theoretically DFT-LDA calculated dispersion relations for LT tetracene. Intermolecular modes (low energy) in (a), Intermolecular and low-laying intramolecular (high energy) modes in (b)\cite{abdulla_a_2015}}
\label{fig-31}
\end{figure}

\begin{figure}[H]
\centering
\includegraphics[scale=0.5]{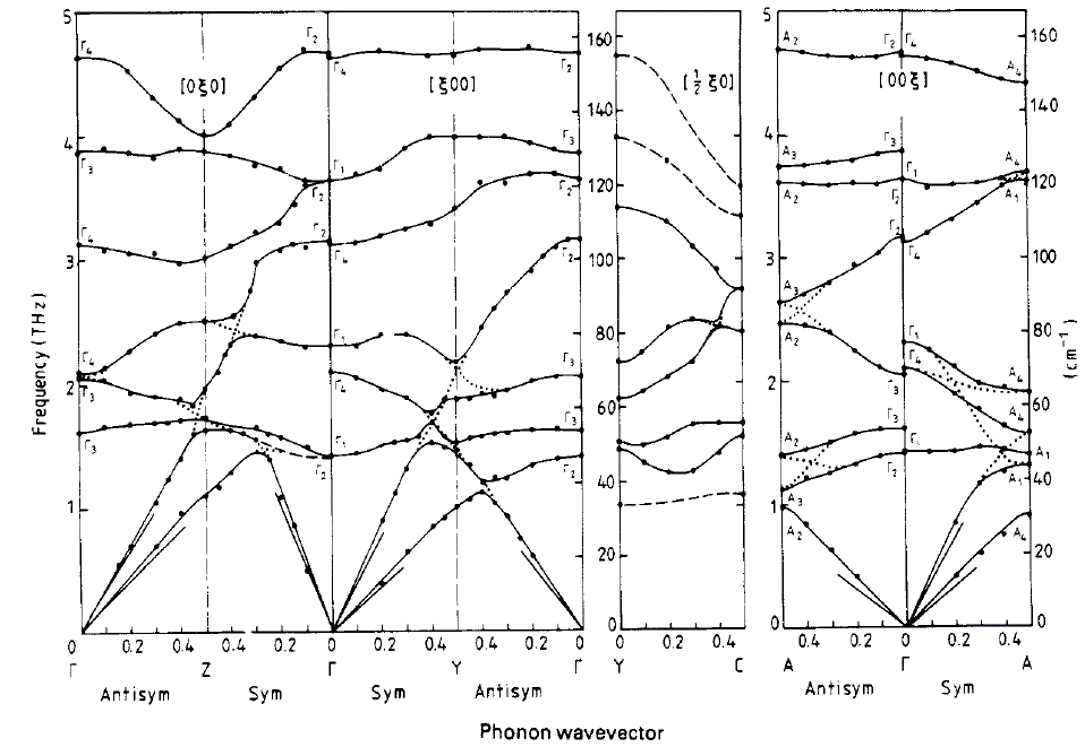}
\caption{Measured dispersion curves for the 12 external and the 4 lowest internal modes in anthracene at 12 K.\cite{dorner_the_1982}}
\label{fig-32}
\end{figure}

The characteristic feature of these organic molecular crystals is narrow band-width, flat-band, hole transport, low density of state, low mobility and band transport at room temperature. Bloch electron spread in crystal with slightest crystal defect, impurity, Polaron band transport only in extremely low temperature feasible, hopping transport in a classical picture in individual molecules with poor mobility. Some of these features shown in the figures \cref{fig-0},\cref{fig-14},\cref{fig-13},\cref{fig-15},\cref{fig-16},\cref{fig-17},\cref{fig-18},\cref{fig-31},\cref{fig-32} from the literature. Moreover, relevant elements are discussed throughout the mathematical development of this work.

Organic molecule crystal phonon band spectra from the theoretical and experimental work. The experiment is very few reported due to difficulty in crystal formulation.

Most theoretical studies on phonon spectra in the periodic molecular crystal are on an isolated single molecule and interpret it as a crystal phonon. This approach is valid for high-frequency intramolecular phonons as they are dispersion-less and consequently very localized. However, low-frequency intermolecular phonon gives inaccurate results by empirical force fields. Moreover, accurate DFT calculations are computationally expansive as in an organic crystal, and the standard unit cell consists of hundreds of atoms. For low-energy phonons spectra measurement, experimentally, terahertz time-domain spectroscopy is widely used. However, it will give gamma phonon energy and not validate the dispersion curve of acoustic phonon. High-resolution inelastic neutron scattering is employed to mitigate these shortcomings, which provide information on low-energy phonons. However, this scheme required single organic crystals, which are challenging to grow. Therefore, phonon band spectra measurement data is available only for a few small organic molecules crystals such as anthracene and naphthalene due to these experimental difficulties.\cite{dorner_the_1982,silinsh_molecular_1995,nematiaram_modeling_2020}

We have investigated a non-equilibrium Green's function based electronic transport framework on the n-dimensional generalized Holstein's molecular-crystal model for the narrow-band semiconductor. We have formulated electronic interaction with acoustic and polar phonons in the narrow energy bandwidth organic molecular crystal to investigate the electronic transport through non-equilibrium Green's function formalism. The self-energy interaction term in the Green's function is formulated based on the Feynman diagram graphical expansion technique. \cite{feynman_space-time_1948,schwinger_on_1951,konstantinov_diagram_1961,mattuck_guide_1976,fanchenko_generalized_1983} In the narrow energy bandwidth, organic molecular crystal interactions incorporated two or more phonon scattering processes; therefore, we have included both one phonon and two phonons in our formalism.

\begin{figure}[H]
\centering
\includegraphics[scale=0.4]{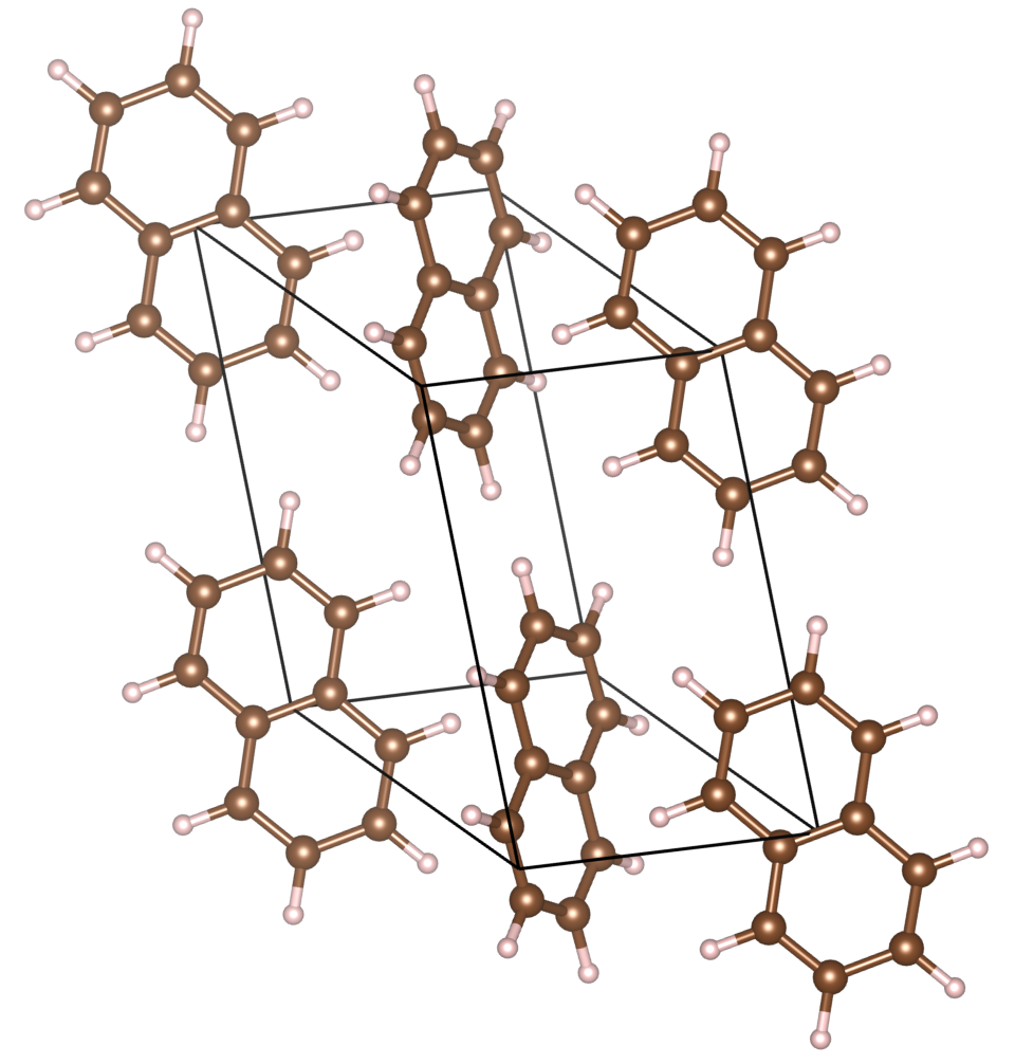}
\caption{Three-dimensional Naphthalene crystal; The monoclinic unit cell has two molecules with an angle of 51.9 $\degree $. Thus, unit cells have a total of 36 atoms. Transport calculation mainly through classical Marcus theory through jump probability.}
\label{fig-0-0-1}
\end{figure}

\begin{figure}[H]
\centering
\includegraphics[scale=0.3]{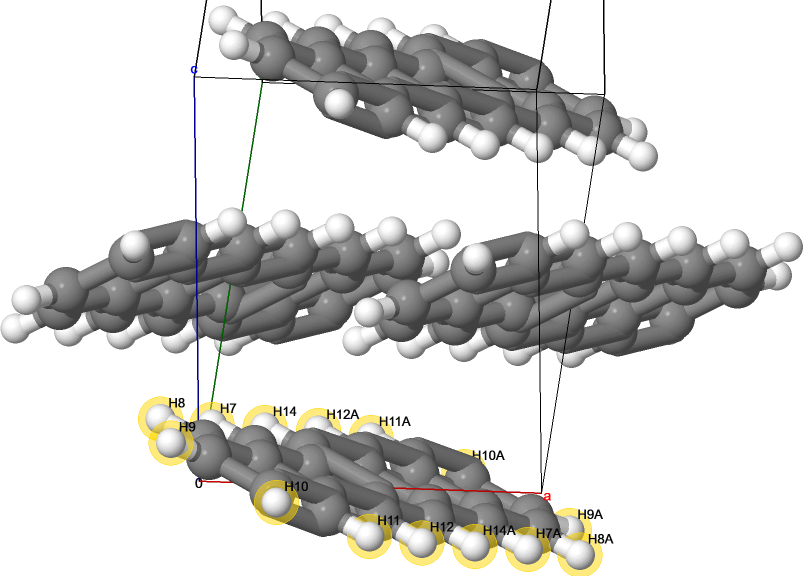}
\caption{Pentacene Unit cell,\cite{holmes_on_1999} }
\label{fig-20}
\end{figure}

\begin{figure}[H]
\centering
\includegraphics[scale=0.5]{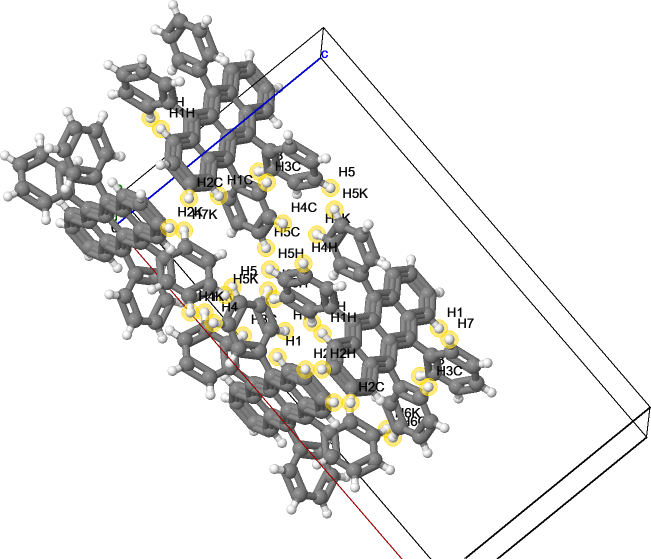}
\caption{Rubrene Unit cell,\cite{jurchescu_low_2006}}
\label{fig-21}
\end{figure}

\begin{figure}[H]
\centering
\includegraphics[scale=0.6]{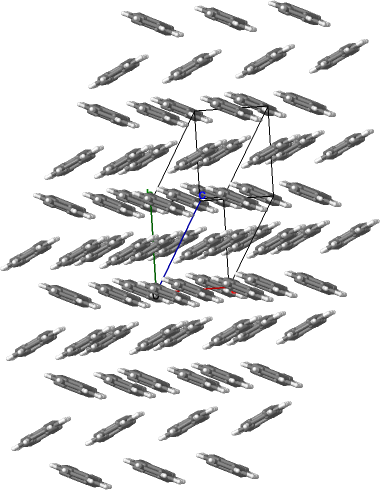}
\caption{Pentacene 3x3x3 cell}
\label{fig-22}
\end{figure}

\begin{figure}[H]
\centering
\includegraphics[scale=0.6]{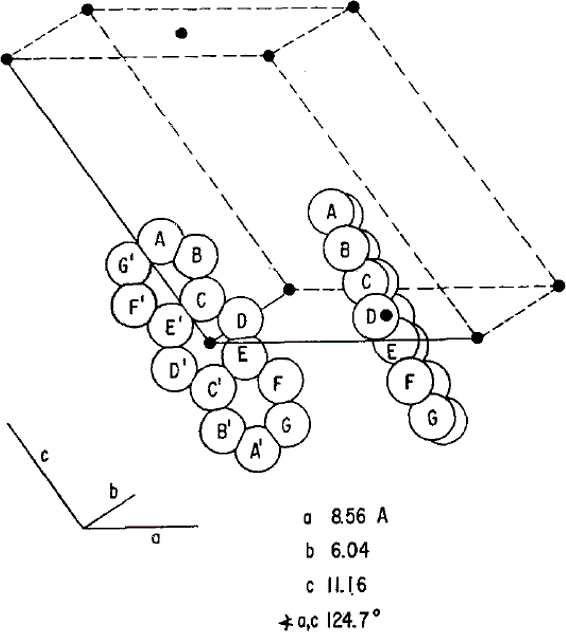}
\caption{Anthracene crystal. Le-blanc first calculation of resonance integral in 1961.\cite{leblanc_band_1961}}
\label{fig-23}
\end{figure}

We will consider the generalized Holstein's molecular-crystal model for the narrow energy bandwidth semiconductor, where at equilibrium lattice positions $\boldsymbol l$ have an atomic or molecular potentials $U(\boldsymbol r)$ and have the lattice symmetry property along it's inversion center. Throughout the work, the bold mathematical symbol used to represent vector position unless otherwise stated. Each such lattice positions $\boldsymbol l$ is describe by site index $ \boldsymbol n $, $ \boldsymbol m $ in the complete crystal and connected through $ \boldsymbol m= \boldsymbol l \pm \boldsymbol n $ where $ \boldsymbol n $ is site index for next nearest neighbor. Arbitrary site index  $ \boldsymbol n $ or $ \boldsymbol m $  is connected through a generalized vector index $ \boldsymbol r = \sum_\text{i,j,k} r_{\text{i,j,k}}\boldsymbol a_{\text{i,j,k}} $, which is sum of basic set of lattice displacement vectors $ \boldsymbol a_{ijk} $ and integers $ r_{ijk} $ see  \cref{fig-0-0},\cref{fig-0-0-1},\cref{fig-20},\cref{fig-21},\cref{fig-22},\cref{fig-23} for geometrical description of model. In the nearest neighbors configuration for two-dimensional case one such arbitrary site index  $ \boldsymbol n $ is connected to  $ \boldsymbol m $ thorough $ \boldsymbol r=r_{1}\boldsymbol {a}_{1}+r_{2}\boldsymbol {a}_{2} $ and denote the position of next neighbor atomic or molecular potentials. In such a molecular-crystal isolated potential $U(\boldsymbol r)$ construct a Bloch-type wave function $\psi(\boldsymbol k, \boldsymbol r)$ with a wave vector $ \boldsymbol k $ spread across the crystal as a linear combination of it's normalized atomic or molecular Wannier eigenfunctions  $\phi(\boldsymbol r)$ is, \cite{kohn_analytic_1959,giustino_electron_2007,marzari_maximally_2012}

\begin{equation}\label{eq-1.1}
\psi(\boldsymbol k, \boldsymbol r)=\sum_{\boldsymbol n'}e^{i\boldsymbol k\boldsymbol n^{\prime}}\phi(\boldsymbol r-\boldsymbol n^{\prime}) = \sum_{\boldsymbol n'}\phi(\boldsymbol r-\boldsymbol n^{\prime})a_{\boldsymbol n^{\prime}}   
\end{equation}

Where $ a_{\boldsymbol n^{\prime}} $ denotes the annihilation operators at lattice position vector $ \boldsymbol n^{\prime} $.

\begin{figure}[H]
\centering  
\begin{tikzpicture}
\tikzstyle{B_circle} = [font={\bfseries}, shape=circle, minimum size=1cm, text=green, very thick, draw=black, inner color=blue,outer color=blue, text width=1.1cm, align=center]
\tikzstyle{B1_circle} = [font={\bfseries}, shape=circle, minimum size=1cm, text=green, very thick, draw=black, inner color=blue,outer color=blue!50, text width=1.1cm, align=center]

\tikzstyle{R_circle} = [font={\bfseries}, shape=circle, minimum size=1cm, circular drop shadow, text=white, very thick, draw=black, inner color=red,outer color=red, text width=1.1cm, align=center]

\tikzstyle{RL_circle} = [font={\bfseries}, shape=circle, minimum size=1cm, circular drop shadow, text=white, very thick, draw=black, left color=red,right color=blue, text width=1.1cm, align=center]
\tikzstyle{RR_circle} = [font={\bfseries}, shape=circle, minimum size=1cm, circular drop shadow, text=white, very thick, draw=black, left color=blue,right color=red, text width=1.1cm, align=center]
\tikzstyle{RT_circle} = [font={\bfseries}, shape=circle, minimum size=1cm, circular drop shadow, text=white, very thick, draw=black, top color=red,bottom color=blue, text width=1.1cm, align=center]
\tikzstyle{RB_circle} = [font={\bfseries}, shape=circle, minimum size=1cm, circular drop shadow, text=white, very thick, draw=black, top color=blue,bottom color=red, text width=1.1cm, align=center]

\tikzstyle{RL5_circle} = [font={\bfseries}, shape=circle, minimum size=1cm, circular drop shadow, text=white, very thick, draw=black, left color=red!40,right color=blue, text width=1.1cm, align=center]
\tikzstyle{RR5_circle} = [font={\bfseries}, shape=circle, minimum size=1cm, circular drop shadow, text=white, very thick, draw=black, left color=blue,right color=red!40, text width=1.1cm, align=center]
\tikzstyle{RT5_circle} = [font={\bfseries}, shape=circle, minimum size=1cm, circular drop shadow, text=white, very thick, draw=black, top color=red!40,bottom color=blue, text width=1.1cm, align=center]
\tikzstyle{RB5_circle} = [font={\bfseries}, shape=circle, minimum size=1cm, circular drop shadow, text=white, very thick, draw=black, top color=blue,bottom color=red!40, text width=1.1cm, align=center]

\tikzstyle{R_edge} = [draw=red, line width=2]
\tikzstyle{OR_edge} = [draw=magenta, line width=2]
\tikzstyle{YE_edge} = [draw=orange, line width=2]
\tikzstyle{GR_edge} = [draw=green, line width=2]

\node (v1) at (0,0) [B_circle] {\tiny$M_{r_{i-n,j-n}}$};
\node (v2) at (2,0) [B1_circle] {\tiny$M_{r_{i-1,j-n}}$};
\node (v3) at (4,0) [B1_circle] {\tiny$M_{r_{i,j-n}}$};
\node (v4) at (6,0) [B1_circle] {\tiny$M_{r_{i+1,j-n}}$};
\node (v5) at (8,0) [B_circle] {\tiny$M_{r_{i+n,j-n}}$};

\node (v6) at (0,2) [B1_circle] {\tiny $M_{r_{i-n,j-1}}$};
\node (v7) at (2,2) [RR5_circle] {\tiny$M_{r_{i-1,j-1}}$};
\node (v8) at (4,2) [RT_circle] {\tiny$M_{r_{i,j-1}}$};
\node (v9) at (6,2) [RL5_circle] {\tiny$M_{r_{i+1,j-1}}$};
\node (v10) at (8,2) [B1_circle] {\tiny$M_{r_{i+n,j-1}}$};

\node (v11) at (0,4) [B1_circle] {\tiny$M_{r_{i-n,j}}$};
\node (v12) at (2,4) [RR_circle] {\tiny$M_{r_{i-1,j}}$};
\node (v13) at (4.2,3.8) [R_circle] {\tiny$M_{r_{i,j}}$};
\node (v14) at (6,4) [RL_circle] {\tiny$M_{r_{i+1,j}}$};
\node (v15) at (8,4) [B1_circle] {\tiny$M_{r_{i+n,j}}$};

\node (v16) at (0,6) [B1_circle] {\tiny$M_{r_{i-n,j+1}}$};
\node (v17) at (2,6) [RR5_circle] {\tiny$M_{r_{i-1,j+1}}$};
\node (v18) at (4,6) [RB_circle] {\tiny$M_{r_{i,j+1}}$};
\node (v19) at (6,6) [RL5_circle] {\tiny$M_{r_{i+1,j+1}}$};
\node (v20) at (8,6) [B1_circle] {\tiny$M_{r_{i+n,j+1}}$};

\node (v21) at (0,8) [B_circle] {\tiny$M_{r_{i-n,j+n}}$};
\node (v22) at (2,8) [B1_circle] {\tiny$M_{r_{i-1,j+n}}$};
\node (v23) at (4,8) [B1_circle] {\tiny$M_{r_{i,j+n}}$};
\node (v24) at (6,8) [B1_circle] {\tiny$M_{r_{i+1,j+n}}$};
\node (v25) at (8,8) [B_circle] {\tiny$M_{r_{i+n,j+n}}$};

\draw[GR_edge] (v1) edge (v2);
\draw[GR_edge] (v2) edge (v3);
\draw[GR_edge] (v3) edge (v4);
\draw[GR_edge] (v4) edge (v5);

\draw[YE_edge] (v6) edge (v7);
\draw[OR_edge] (v7) edge (v8);
\draw[OR_edge] (v8) edge (v9);
\draw[YE_edge] (v9) edge (v10);

\draw[YE_edge] (v11) edge (v12);
\draw[R_edge] (v12) edge (v13);
\draw[R_edge] (v13) edge (v14);
\draw[YE_edge] (v14) edge (v15);

\draw[YE_edge] (v16) edge (v17);
\draw[OR_edge] (v17) edge (v18);
\draw[OR_edge] (v18) edge (v19);
\draw[YE_edge] (v19) edge (v20);

\draw[GR_edge] (v1) edge (v6);
\draw[YE_edge] (v2) edge (v7);
\draw[YE_edge] (v3) edge (v8);
\draw[YE_edge] (v4) edge (v9);
\draw[GR_edge] (v5) edge (v10);

\draw[GR_edge] (v6) edge (v11);
\draw[OR_edge] (v7) edge (v12);
\draw[R_edge] (v8) edge (v13);
\draw[OR_edge] (v9) edge (v14);
\draw[GR_edge] (v10) edge (v15);

\draw[GR_edge] (v16) edge (v21);
\draw[YE_edge] (v17) edge (v22);
\draw[YE_edge] (v18) edge (v23);
\draw[YE_edge] (v19) edge (v24);
\draw[GR_edge] (v20) edge (v25);

\draw[GR_edge] (v16) edge (v11);
\draw[OR_edge] (v17) edge (v12);
\draw[R_edge] (v18) edge (v13);
\draw[OR_edge] (v19) edge (v14);
\draw[GR_edge] (v20) edge (v15);

\draw[GR_edge] (v21) edge (v22);
\draw[GR_edge] (v22) edge (v23);
\draw[GR_edge] (v23) edge (v24);
\draw[GR_edge] (v24) edge (v25);

\foreach \border in {0.09}
\useasboundingbox (current bounding box.south west)+(-\border,-\border) rectangle (current bounding box.north east)+(\border,\border);

\shadedraw [inner color=blue,outer color=blue!80,draw=black, thick] plot[smooth, tension=.7] coordinates {(-0.5,9) (0,10) (0.5,9)};
\shadedraw [inner color=blue,outer color=blue!30,draw=black, thick] plot[smooth, tension=.7] coordinates {(1.5,9) (2,10) (2.5,9)};
\shadedraw [inner color=blue,outer color=blue!30,draw=black, thick] plot[smooth, tension=.7] coordinates {(3.5,9) (4,10) (4.5,9)};
\shadedraw [inner color=blue,outer color=blue!30,draw=black, thick] plot[smooth, tension=.7] coordinates {(5.5,9) (6,10) (6.5,9)};
\shadedraw [inner color=blue,outer color=blue!80,draw=black, thick] plot[smooth, tension=.7] coordinates {(7.5,9) (8,10) (8.5,9)};
\shadedraw [inner color=blue,outer color=blue!30,draw=black, thick] plot[smooth, tension=.7] coordinates {(9,5.5) (9.5,6.5) (10,5.5)};
\shadedraw [inner color=blue,outer color=blue!30,draw=black, thick] plot[smooth, tension=.7] coordinates {(9,3.5) (9.5,4.5) (10,3.5)};
\shadedraw [inner color=blue,outer color=blue!30,draw=black, thick] plot[smooth, tension=.7] coordinates {(9,1.5) (9.5,2.5) (10,1.5)};
\shadedraw [inner color=blue,outer color=blue!30,draw=black, thick] plot[smooth, tension=.7] coordinates {(-2,5.5) (-1.5,6.5) (-1,5.5)};
\shadedraw [inner color=blue,outer color=blue!30,draw=black, thick] plot[smooth, tension=.7] coordinates {(-2,3.5) (-1.5,4.5) (-1,3.5)};
\shadedraw [inner color=blue,outer color=blue!30,draw=black, thick] plot[smooth, tension=.7] coordinates {(-2,1.5) (-1.5,2.5) (-1,1.5)};
\shadedraw [inner color=blue,outer color=blue!80,draw=black, thick] plot[smooth, tension=.7] coordinates {(-0.5,-2) (0,-1) (0.5,-2)};
\shadedraw [inner color=blue,outer color=blue!30,draw=black, thick] plot[smooth, tension=.7] coordinates {(1.5,-2) (2,-1) (2.5,-2)};
\shadedraw [inner color=blue,outer color=blue!30,draw=black, thick] plot[smooth, tension=.7] coordinates {(3.5,-2) (4,-1) (4.5,-2)};
\shadedraw [inner color=blue,outer color=blue!30,draw=black, thick] plot[smooth, tension=.7] coordinates {(5.5,-2) (6,-1) (6.5,-2)};
\shadedraw [inner color=blue,outer color=blue!80,draw=black, thick] plot[smooth, tension=.7] coordinates {(7.5,-2) (8,-1) (8.5,-2)};

\shadedraw [left color=blue,right color=red!40,draw=black, thick]  plot[smooth, tension=.7] coordinates {(0.5,6.75) (1,7.75) (1.5,6.75)};
\shadedraw [top color=blue,bottom color=red, draw=black, thick] plot[smooth, tension=.7] coordinates {(2.5,6.75) (3,7.75) (3.5,6.75)};
\shadedraw [left color=red!40,right color=blue,draw=black, thick]  plot[smooth, tension=.7] coordinates {(6.5,6.75) (7,7.75) (7.5,6.75)};
\shadedraw [left color=blue,right color=red!40, draw=black, thick] plot[smooth, tension=.7] coordinates {(0.5,2.75) (1,3.75) (1.5,2.75)};
\shadedraw [left color=red!40,right color=blue, draw=black, thick] plot[smooth, tension=.7] coordinates {(6.5,2.75) (7,3.75) (7.5,2.75)};
\shadedraw [top color=red,bottom color=blue, draw=black, thick] plot[smooth, tension=.7] coordinates {(2.5,2.75) (3,3.75) (3.5,2.75)};
\shadedraw [left color=red,right color=blue,draw=black, thick] plot[smooth, tension=.7] coordinates {(6.5,4.75) (7,5.75) (7.5,4.75)};

\shadedraw [inner color=red,outer color=red,draw=black, thick]plot[smooth, tension=.7] coordinates {(3.75,4.5) (3.25,5.5) (2.75,4.5)};
\shadedraw [left color=blue,right color=red,draw=black, thick] plot[smooth, tension=.7] coordinates {(1.5,4.75) (1,5.75) (0.5,4.75)};

\draw[-latex,ultra thick,black] (4.5,5) edge node[below] {$\hat l$}(6.5,5);
\node at (6.4,5.2) {$ R_i$};
\draw[-latex,ultra thick,black,] (5,4.5) edge node[near end, right] {$\hat l$}(5,6.5);
\node at (5,6.75) {$ R_j$};
\draw[-latex,ultra thick,orange] (4.8,3.4)--(5.5,2.5);
\node at (5,2.5) {\Large $ y_{\boldsymbol n} $};

\draw (0,8.85) node[text=black] { $ |\phi(\boldsymbol r-\boldsymbol n^{\prime}\rangle $};
\draw (2,8.85) node[text=black] { $ |\phi(\boldsymbol r-\boldsymbol n^{\prime}\rangle $};
\draw (4,8.85) node[text=black] { $ |\phi(\boldsymbol r-\boldsymbol n^{\prime}\rangle $};
\draw (6,8.85) node[text=black] { $ |\phi(\boldsymbol r-\boldsymbol n^{\prime}\rangle $};
\draw (8,8.85) node[text=black] { $ |\phi(\boldsymbol r-\boldsymbol n^{\prime}\rangle $};
\draw (-1.5,5.25) node[text=black] { $ |\phi(\boldsymbol r-\boldsymbol n^{\prime}\rangle $};
\draw (9.5,5) node[text=black] { $ |\phi(\boldsymbol r-\boldsymbol n^{\prime}\rangle $};
\draw (-1.5,3.25) node[text=black] { $ |\phi(\boldsymbol r-\boldsymbol n^{\prime}\rangle $};
\draw (9.5,3.25) node[text=black] { $ |\phi(\boldsymbol r-\boldsymbol n^{\prime}\rangle $};
\draw (-1.5,1.25) node[text=black] { $ |\phi(\boldsymbol r-\boldsymbol n^{\prime}\rangle $};
\draw (9.5,1.25) node[text=black] { $ |\phi(\boldsymbol r-\boldsymbol n^{\prime}\rangle $};
\draw (0,-2.25) node[text=black] { $ |\phi(\boldsymbol r-\boldsymbol n^{\prime}\rangle $};
\draw (2,-2.25) node[text=black] { $ |\phi(\boldsymbol r-\boldsymbol n^{\prime}\rangle $};
\draw (4,-2.25) node[text=black] { $ |\phi(\boldsymbol r-\boldsymbol n^{\prime}\rangle $};
\draw (6,-2.25) node[text=black] { $ |\phi(\boldsymbol r-\boldsymbol n^{\prime}\rangle $};
\draw (8,-2.25) node[text=black] { $ |\phi(\boldsymbol r-\boldsymbol n^{\prime}\rangle $};

\draw  (1,7.25) node[text=white, rotate=90] {\tiny $ |\phi(\boldsymbol r-\boldsymbol n^{\prime} \rangle$};
\draw  (3,7.25) node[text=white, rotate=90] {\tiny  $ |\phi(\boldsymbol r-\boldsymbol n^{\prime}\rangle $};
\draw  (7,7.25) node[text=white, rotate=90] {\tiny  $ |\phi(\boldsymbol r-\boldsymbol n^{\prime}\rangle $};
\draw  (1,5.25) node[text=white, rotate=90] {\tiny  $ |\phi(\boldsymbol r-\boldsymbol n^{\prime}\rangle $};
\draw  (3.25,5) node[text=white, rotate=90] {\tiny  $ \boldsymbol {|\phi(r) \rangle}$};
\draw  (7,5.25) node[text=white, rotate=90] {\tiny  $ |\phi(\boldsymbol r-\boldsymbol n^{\prime}\rangle $};
\draw  (1,3.25) node[text=white, rotate=90] {\tiny $ |\phi(\boldsymbol r-\boldsymbol n^{\prime}\rangle $};
\draw  (3,3.25) node[text=white, rotate=90] {\tiny $ |\phi(\boldsymbol r-\boldsymbol n^{\prime}\rangle $};
\draw  (7,3.25) node[text=white, rotate=90] {\tiny  $ |\phi(\boldsymbol r-\boldsymbol n^{\prime}\rangle $};
\draw  (3.5,-2.75) node[text=black] {\large Lattice grid point two-dimensional extended Holstein's molecular-crystal model};
\end{tikzpicture}
\caption{Real space molecular orbitals in two-dimensional Holstein's molecular-crystal model}
\label{fig-0-0}
\end{figure}

The Schr\"{o}dinger's equation of the Bloch wave function of molecular-crystal with electronic energy $\varepsilon(\boldsymbol k)$ is,

\begin{equation}\label{eq-1.2}
H_{\mathrm{E}}\psi(\boldsymbol k, \boldsymbol r)=\varepsilon(\boldsymbol k)\psi(\boldsymbol k, \boldsymbol r)   
\end{equation}

Therefore, in second quantization language, the Hamiltonian operator of an electron with electronic energy part with an atom-like periodic potential having a center of symmetry and real $ U(\boldsymbol r-\boldsymbol l) $  with respect to  center $ (\boldsymbol r-\boldsymbol l) $ in the externally applied electric field vector $ \boldsymbol E_{F}  $ is, \cite{stefanucci_vanleeuwen_2013,shavitt_bartlett_2009}

\begin{equation}\label{eq-1.3}
H_{\mathrm{E}}=\int \mathrm{d}^{3}(\boldsymbol r-\boldsymbol l)\psi^{\dagger}(\boldsymbol k, \boldsymbol r-\boldsymbol l)\Bigg\{-\frac{\hbar^{2}}{2m}\Big[\frac{\partial^{2} }{\partial (\boldsymbol r-\boldsymbol l)}+\mathrm{q}\boldsymbol E_{F} \cdot \boldsymbol r \Big]+\sum_{\boldsymbol l}U(\boldsymbol r-\boldsymbol l)-\mu\Bigg\} \psi(\boldsymbol k, \boldsymbol r-\boldsymbol l)  
\end{equation}

Where the term $(\mathrm{q}\boldsymbol E_{F} \cdot \boldsymbol r)  $ is externally applied electric field-induced energy at site $\boldsymbol r$ and $ \mu $ is the chemical potential. At the first-order treatment, externally applied electric field vector $ \boldsymbol E_{F}  $ may operate on the lattice molecule site and may change their mean position $ (\boldsymbol r-\boldsymbol l) $ and also applied electric field vector $ \boldsymbol E_{F}  $  may 
interact with molecule potential energy $ U(\boldsymbol r-\boldsymbol l) $. However, all such phonon drag effect and London effect interaction are omitted at first-order treatment.\cite{yamashita_heitler-london_1960,kurosawa_heitler-london_1960} Moreover, we will not mention specifically externally applied electric field-induced energy $(\mathrm{q}\boldsymbol E_{F} \cdot \boldsymbol r) $ in subsequent treatment and introduce again in the transport equation in terms of non-equilibrium Green's function.

The energy eigenvalues $\varepsilon_{0}$ is given by, 

\begin{equation}\label{eq-1.4}
\Bigg\{-\frac{\hbar^{2}}{2m}\frac{\partial^{2} }{\partial (\boldsymbol r)}+U(\boldsymbol r)-\varepsilon_{0}+\mu\Bigg\}\phi(\boldsymbol r)=0 
\end{equation}

Inserting \cref{eq-1.1}, \cref{eq-1.3} and \cref{eq-1.4} in Schr\"{o}dinger's equation \cref{eq-1.2} and multiply with $\phi^{*}(\boldsymbol r-\boldsymbol n)$ and performing integration with two-center integrals over $\boldsymbol r$, the electronic energy $\varepsilon(\boldsymbol k)$ following the schnakenberg \textit{et al.} \cite{schnakenberg_derivation_1965,schnakenberg_quasiteilchen-spektrum_1966,schnakenberg_electron-phonon_1969} treatment of Holstein's molecular-crystal model is, 

\begin{equation}\label{eq-1.5}
\begin{aligned}
\varepsilon(\boldsymbol k)&=\varepsilon_{0}-\mu+\frac{\sum\limits_{\boldsymbol n,\boldsymbol n^{\prime}}{e}^{i\boldsymbol k(\boldsymbol n^{\prime}-\boldsymbol n)}  \int {d}^{3}r\phi^{*}(\boldsymbol r-\boldsymbol n)\sum\limits_{\boldsymbol l\neq {\boldsymbol n}^{\prime}}U(\boldsymbol r-\boldsymbol l)\phi(\boldsymbol r-\boldsymbol n^{\prime})}{\sum\limits_{\boldsymbol n-\boldsymbol n^{\prime}}{e}^{i\boldsymbol k(\boldsymbol n^{\prime}-\boldsymbol n)}\underbrace{\int {d}^{3}r\phi^{*}(\boldsymbol r-\boldsymbol n)\phi(\boldsymbol r-\boldsymbol n^{\prime})}_{S_{\boldsymbol n \boldsymbol n'}}}\\
&=\varepsilon_{0}-\mu+\sum_{\boldsymbol n,\boldsymbol n^{\prime}} K_{\boldsymbol n,\boldsymbol n^{\prime}}{e}^{i\boldsymbol k(\boldsymbol n^{\prime}-\boldsymbol n)}
\end{aligned}
\end{equation}

$ K_{\boldsymbol n,\boldsymbol n^{\prime}} $ is the Fourier coefficients of the narrow band overlap integrals, $ \mu $ is chemical potential of molecules crystal, $\phi(\boldsymbol r)$ is normalized and in the denominator of \cref{eq-1.5} with the lowest order integral include functions centered at distinct molecular site index $\boldsymbol n, \boldsymbol n^{\prime}$ and  approximate as Kronecker delta $ \delta $, \cite{bateman_tables_1954,feynman_feynman_1965,gradshteyn_table_1980,galitski_exploring_nodate,watanabe_definition_2015,grosche_handbook_1998,berestetskii_preface_1982,bonch-bruevich_green_1962}

\begin{equation}\label{eq-1.6}	
\int {d}^{3}r\phi^{*}(\boldsymbol r-\boldsymbol n)\phi(\boldsymbol r-\boldsymbol n^{\prime}) = S_{\boldsymbol n \boldsymbol n'}  \approx\delta_{\boldsymbol n \boldsymbol n^{\prime}}   
\end{equation}

Furthermore, in the numerator, by identical reasoning, we confine to two-center integrals, and overlap integrals $ K_{\boldsymbol n,\boldsymbol n^{\prime}} $ is expanded as resonance integral $U_J(\boldsymbol n^{\prime}-\boldsymbol n)$ and Coulomb integral $U_C$ as follow,

\begin{equation}\label{eq-1.7}
\varepsilon(\boldsymbol k)=\varepsilon_{0}-\mu+\underbrace{\int {d}^{3}r|\phi(\boldsymbol r)|^{2}\sum_{\boldsymbol l\neq \boldsymbol n}U(\boldsymbol r-\boldsymbol l+\boldsymbol n)}_{U_C}+\sum_{\boldsymbol n-\boldsymbol n'}{e}^{i\boldsymbol k(\boldsymbol n^{\prime}-\boldsymbol n)}\underbrace{\int {d}^{3}r\phi^{*}(\boldsymbol r)U(\boldsymbol r)\phi(\boldsymbol r-\boldsymbol n^{\prime}+\boldsymbol n)}_{U_J(\boldsymbol n^{\prime}-\boldsymbol n)}  
\end{equation}

The second term in \cref{eq-1.7} is defined as Coulomb integral $U_C$ as,

\begin{equation}\label{eq-1.8}
U_C=\int {d}^{3}r|\phi(\boldsymbol r)|^{2}\sum_{\boldsymbol l\neq \boldsymbol n}U(\boldsymbol r-\boldsymbol l+\boldsymbol n) =\int {d}^{3}r|\phi(\boldsymbol r)|^{2} U(\boldsymbol r-\boldsymbol m) 
\end{equation}

The third term in \cref{eq-1.7} under integral without the exponential prefactor is defined as resonance integral $U_J(\boldsymbol n^{\prime}-\boldsymbol n)$ as,

\begin{equation}\label{eq-1.9}	
\begin{aligned}
U_J(\boldsymbol n^{\prime}-\boldsymbol n) & =\int {d}^{3}r\phi^{*}(\boldsymbol r)U(\boldsymbol r)\phi(\boldsymbol r-\boldsymbol n^{\prime}+\boldsymbol n)\ ,\ \boldsymbol n\neq \boldsymbol n',\\
U_J(\boldsymbol n^{\prime}-\boldsymbol n) & = 0, \ \boldsymbol n = \boldsymbol n' \\
U_J(\boldsymbol n^{\prime}-\boldsymbol n) & = U_J(\boldsymbol n - \boldsymbol n^{\prime}), \ \mathrm{from \ the \ inversion \ symmetric \ of \ lattice} 
\end{aligned}
\end{equation}

The Number of molecular orbitals in Pentacene C22H14 is $P_{z}$ orbital per carbon atom and 6 $P_{z}$ molecular orbital per ring. Therefore 22 molecular orbitals of carbon and a unit cell have four molecules, consequently 22X4=88 molecular orbitals per unit cell. Rubrene C42H28 has 42 carbon $P_{z}$ orbital, and the unit cell has four molecules. 42X4=168 molecular orbital per unit cell. Mulliken in 1949 tabulated the formula, \cite{mulliken_formulas_1949}, and LeBlanc in 1961 calculated the resonance integrals of molecular orbital for anthracene crystal. \cite{leblanc_band_1961}  For calculating the molecular orbitals out of one unit cell, more advanced prediction can be made through deep learning and machine learning-based approach. The learning can efficiently estimate any linear combination of orbital in principal and populate the hamiltonian for the entire device under investigation for transport calculation.  \cite{tsubaki_quantum_2020,burkle_deep-learning_2021}

Now writing the complete electronic Hamiltonian in the second quantization language, 

\begin{equation}\label{eq-1.10}
H_{\mathrm{E}}=\sum_{\boldsymbol k}\varepsilon(\boldsymbol k)a^{\dagger}_{\boldsymbol k}a_{\boldsymbol k}=(\varepsilon_{0} -\mu +U_C)\sum_{\boldsymbol n}a_{\boldsymbol n}^{\dagger}a_{\boldsymbol n}+\sum_{\boldsymbol n,{\boldsymbol n'}}U_J(\boldsymbol n^{\prime}-\boldsymbol n)a_{\boldsymbol n}^{\dagger}a_{\boldsymbol n}   
\end{equation}

Where $a_{\boldsymbol k}$, $a_{\boldsymbol k}^{\dagger}$ are annihilation and creation operators of Bloch type propagating band wave $\psi(\boldsymbol k, \boldsymbol r)$ with wavevestor $\boldsymbol k$.
Annihilation $a_{\boldsymbol n}$ and creation $a_{\boldsymbol n}^{\dagger}$ are relates to lattice Wannier functions. However, the molecular eigenfunctions of $U(\boldsymbol r)$ are identical to Wannier functions in the lowest order Bloch's estimation. \cite{wannier_structure_1937} Therefore, $a_{\boldsymbol n}, a_{\boldsymbol n}^{\dagger}$ are annihilation and creation operators of eigenfunctions $\phi(\boldsymbol r)$ with eigenstates of the isolated potentials $U(\boldsymbol r)$. The second term in \cref{eq-1.10} is equivalent to the $ H_{int}$ of the Holstein model.  

The effect of Bosonic field, i.e., acoustic and polar phonons, on the electronic Hamiltonian in the molecular crystal is incorporated by expanding the Coulomb integral $U_C$ and resonance integral $U_J(\boldsymbol n^{\prime}-\boldsymbol n)$ in terms of instantaneous positions of the molecules defined as $r_{\boldsymbol n}=\boldsymbol n+y_{\boldsymbol n}$ in the lattice, where deviations $y_{\boldsymbol n}$ is variations from the equilibrium positions of lattice index $\boldsymbol n$. Writing molecules positions in terms of instantaneous positions are an approximation in first order and this approximation's validity is later proclaimed.

We will introduce the phonons part into our electron-phonon gas system by the free Bosonic Hamiltonian. It should be noticed that the phonon annihilators and creators operator here is not destroying or creating the mass or molecules here but represent the creation and destruction of associated phonon energy mode as defined by,

\begin{equation}\label{eq-1.11}
H_{\mathrm{P}}=\sum_{i=1}^{2}\sum_{\boldsymbol q}\omega_{\boldsymbol q i}b_{\boldsymbol q i}^{\dagger}b_{\boldsymbol q i}
\end{equation}

Where $ i=1,2 $ referring to  acoustic and polar phonons respectively, $b_{\boldsymbol q }$ and $b_{\boldsymbol q }^{\dagger}$ are phonon annihilators and creators and $\omega_{\boldsymbol q i}$ is the frequency of phonons of type $\mathrm{i}$ includes both a polarization index $v$ and phonon wave vector $\boldsymbol q$ of the lattice. For $ i=1 $ referring to longitudinal acoustic phonon, frequency of phonon is $\omega_{\boldsymbol q} = \omega_{\mathrm{ac}}$ and for $ i=2 $ referring to longitudinal polar optical phonon, frequency of phonon is $\omega_{\boldsymbol q} =\omega_{lo}$.

In the second quantization language, the amplitude of the molecular vibration $y_{\boldsymbol n}$ which deviate from the equilibrium positions site $\boldsymbol n$ within the unit cell, after transforming the classical small molecular displacements $y_{\boldsymbol n}$ into the quantize phonons normal modes to treat with propagating electron in the Bloch basis for the perturbation, \cite{mahan_many-particle_1990, cohen_louie_2016}

\begin{equation}\label{eq-1.12}
y_{\boldsymbol n}=\sum_{\boldsymbol q,\mathrm {v}}\frac{\hbar^{1/2}}{(2MN\omega_ {\boldsymbol q,\mathrm {v}})^{1/2}}{\boldsymbol e}_{\boldsymbol q,\mathrm {v}}\big({e}^{i(\boldsymbol q,\mathrm {v})\boldsymbol n}b_{\boldsymbol q,\mathrm {v}}+{e}^{-i(\boldsymbol q,\mathrm {v})\boldsymbol n}b_{\boldsymbol q,\mathrm {v}}^{\dagger}\big)
\end{equation}

Where $M$ reduced mass of molecular grid cells, $N$ total number of molecule lattice sites per unit volume, $w_{\boldsymbol q,\mathrm {v}}$ phonon frequency for inter-molecular or intra-molecular vibration in the branch index $ \mathrm {v} $, ${\boldsymbol e}_{\boldsymbol q,\mathrm {v}}$ phonon polarization vector, $b_{\boldsymbol q,\mathrm {v}}, b_{\boldsymbol q,\mathrm {v}}^{\dagger}$ phonon annihilation and creation operators, with a phonon wave vector $\boldsymbol q$ and a phonon polarization branch index $ \mathrm {v}$.  For the non-polar organic molecular crystal, the value of polarization vector and index can be assumed unity. For the polar organic or inorganic crystal, ${\boldsymbol e}_{\boldsymbol q,\mathrm {v}}$ an appropriate polarization index should be treated. Here we start with a general description and narrow down the framework in the subsequent discussion without explicitly mentioning the different phonon branches and drop the branch index $ \mathrm {v}$. Later in the Boltzmann transport equation for counting the scattering rate between different electron and phonon branches, we will reintroduce the branch index $\mathrm {n}$, $\mathrm {m}$, $ \mathrm {v}$, respectively.

The Coulomb integral $U_C$ in terms of deviations $y_{\boldsymbol n}$ is,

\begin{equation}\label{eq-1.13}
U_C=\int {d}^{3}r|\phi(\boldsymbol r)|^{2}\sum_{\boldsymbol l\neq \boldsymbol n}U(\boldsymbol r-\boldsymbol r_{\boldsymbol l}+\boldsymbol r_{\boldsymbol n})    
\end{equation}

And resonance integral $U_J(\boldsymbol n^{\prime}-\boldsymbol n)$ in terms of deviations $y_{\boldsymbol n}$ is,

\begin{equation}\label{eq-1.14}	
U_J (\boldsymbol r_{\boldsymbol n}, \boldsymbol r_{\boldsymbol n'})=\int {d}^{3}\boldsymbol r\phi^{*}(\boldsymbol r)U(\boldsymbol r)\phi(\boldsymbol r-\boldsymbol r_{\boldsymbol n^{\prime}}+\boldsymbol r_{\boldsymbol n})  
\end{equation}

Where $ U(\boldsymbol r-\boldsymbol r_{\boldsymbol l}+\boldsymbol r_{\boldsymbol n}) $ and $ U(\boldsymbol r)  $ in Coulomb integral $U_C$ and resonance integral $U_J(\boldsymbol n^{\prime}-\boldsymbol n)$ is additional potential energy of electron due to interaction with vibrating molecular grid cell.
In the simplest approximation $ U(\boldsymbol r)  $ and $ U(\boldsymbol r-\boldsymbol r_{\boldsymbol l}+\boldsymbol r_{\boldsymbol n}) $ is
additive summation due to the potentials from all the nearest neighbor individual polar molecules and therefore proportional to the deflections $y_{\boldsymbol n}$ in the respective grid cell, and the shape of $ U(\boldsymbol r)  $ with distance $ \boldsymbol r  $ is $ 1 / \boldsymbol r^2  $ dependence in the simplest approximation,

\begin{equation}\label{eq-1.14b}
U(\boldsymbol r-\boldsymbol r_{\boldsymbol l}+\boldsymbol r_{\boldsymbol n}) ; U(\boldsymbol r)\approx\sum_{\mathrm{all}\,  \boldsymbol n}U(\boldsymbol r- \boldsymbol n)y_{\boldsymbol n} \approx \frac{1}{\boldsymbol r^2}  y_{\boldsymbol n}
\end{equation}

In the narrow bandwidth semiconductor, to incorporate the full effect of interaction processes between Fermionic and Bosonic field, we will expand the resonance integral $U_J(\boldsymbol n^{\prime}-\boldsymbol n)$ and Coulomb integral $U_C$ in powers of second-order deviations terms to include up to two phonon scattering interaction. The organic semiconductor has a narrow electronic bandwidth. Moreover, the acoustic phonon has high energy, and the polar optical phonon has even more high energy. Therefore extremely narrow electronic bandwidth semiconductors cannot interact with one phonon process alone and balance the energy conservation. Therefore two or more phonon processes are inevitable. Here we restrict our self to two phonon processes, but it can be, in principle, expanded to the third and fourth-order with a more mathematical cumbersome equation. In a practical organic device, additional imperfections such as lattice point defects, charged or uncharged impurity centers, adsorbent atoms, molecules in the nearby layer, and the surface may influence the charge transport. All of those interactions with electrons depend upon the crystal lattice's finite distance point. Therefore depend upon the phonon field due to the distance fluctuation between diverse lattice points. However, these interactions can be modeled as Coulomb integral expansion $U_C$ and resonance integral expansion $U_J(\boldsymbol n^{\prime}-\boldsymbol n)$ type in the relevant lattice scenario. We have expanded one phonon and two phonon processes. We can expand the framework in terms of three and four phonon simultaneous processes. Moreover, two phonons non-simultaneous interaction, a combination of consecutive two, one-phonon interaction at distinct times, can also be constructed. However,  their contribution in comparison to simultaneous processes is minuscule and can be neglected. However, in the Boltzmann transport equation framework, the treatment of two or more simultaneous and non-simultaneous scattering interactions is impossible to treat. 

Resonance integral $U_J$ and Coulomb integral $U_C$ is a function of the instantaneous positions of lattice sites in the molecular crystal lattice vibration. The lattice vibration effect through acoustic phonon interaction is incorporated through $U_C$  and $U_J$ are expanded to second order in powers of  $y_{\boldsymbol n}$ dilation in the equilibrium state. Expanding the Coulomb integral $U_C$ in the space components $(i, j)$ of the vector for up to second-order deviations terms by acoustic phonons interaction,

\begin{equation}\label{eq-1.15}
\begin{aligned}
U_C & =\sum_{\boldsymbol l}\underbrace{\int {d}^{3}r|\phi(\boldsymbol r)|^{2}U(\boldsymbol r-\boldsymbol l+\boldsymbol n)}_{U_{C_0}}\boldsymbol{-}\sum_{\boldsymbol l}\sum_{i}\underbrace{\frac{\partial }{\partial (\boldsymbol l_{i}-\boldsymbol n_{i})} \int {d}^{3}r|\phi(\boldsymbol r)|^{2}U(\boldsymbol r-\boldsymbol l+\boldsymbol n)}_{U_{C_{i}}(\boldsymbol l-\boldsymbol n)}\cdot(y_{\boldsymbol ni}-y_{\boldsymbol li}) \\
& \boldsymbol{+}\frac{1}{2}\sum_{\boldsymbol l}\sum_{i,j} \underbrace{\frac{\partial^{2} }{{\partial (\boldsymbol l_{i}-\boldsymbol n_{i})\partial (\boldsymbol l_{j}-\boldsymbol n_{j})}} \int {d}^{3}r|\phi(\boldsymbol r)|^{2}U(\boldsymbol r-\boldsymbol l+\boldsymbol n)}_{U_{C_{ij}}(\boldsymbol l-\boldsymbol n)}\cdot(y_{\boldsymbol ni}-y_{\boldsymbol li})\cdot(y_{\boldsymbol nj}-y_{\boldsymbol lj})
\end{aligned}
\end{equation}

In \cref{eq-1.8} the first term is the unperturbed Coulomb integral. The second term is first-order deviation terms. The third term is second-order deviation terms where $\boldsymbol  n $ is the site index of molecules and $\boldsymbol  l $ is the mean equilibrium position. From the second and third term respectively, we define $ U_{C_{i}}(\boldsymbol l-\boldsymbol n) $ and $ U_{C_{ij}}(\boldsymbol l-\boldsymbol n) $, 

\begin{equation}\label{eq-1.16}	
U_{C_{i}}(\boldsymbol l-\boldsymbol n)=\frac{\partial }{\partial (\boldsymbol l_{i}-\boldsymbol n_{i})} \int {d}^{3}r|\phi(\boldsymbol r)|^{2}U(\boldsymbol r-\boldsymbol l+\boldsymbol n)   
\end{equation}

\begin{equation}\label{eq-1.17}	
U_{C_{ij}}(\boldsymbol l-\boldsymbol n)=\frac{\partial^{2} }{{\partial (\boldsymbol l_{i}-\boldsymbol n_{i})\partial (\boldsymbol l_{j}-\boldsymbol n_{j})}} \int {d}^{3}r|\phi(\boldsymbol r)|^{2}U(\boldsymbol r-\boldsymbol l+\boldsymbol n)   
\end{equation}

Expanding the resonance integral $U_J$ in the space components $(i, j)$ of the vector for up to second-order deviations terms by acoustic phonons interaction,

\begin{equation}\label{eq-1.18}
\begin{aligned}
U_J(\boldsymbol r_{\boldsymbol n}, \boldsymbol r_{\boldsymbol n^{\prime}}) & =\underbrace{\int {d}^{3}r\phi^{*}(\boldsymbol r)U(\boldsymbol r)\phi(\boldsymbol r-\boldsymbol n^{\prime}+\boldsymbol n)}_{U_{J}(\boldsymbol n^{\prime}-\boldsymbol n)}\\ &\boldsymbol{-}\sum_{i}\underbrace{\frac{\partial} {\partial (\boldsymbol n^{\prime}_{i}-\boldsymbol n_{i})}\int {d}^{3}r\phi^{*}(\boldsymbol r)U(\boldsymbol r)\phi(\boldsymbol r-\boldsymbol n^{\prime}+\boldsymbol n)}_{U_{J_{i}}(\boldsymbol n^{\prime}-\boldsymbol n)}\cdot(y_{\boldsymbol ni}-y_{\boldsymbol n^{\prime}i}) \\
& \boldsymbol{+}\frac{1}{2}\sum_{i,j}\underbrace{\frac{\partial^{2}} {\partial (\boldsymbol n^{\prime}_{i}-\boldsymbol n_{i})\partial (\boldsymbol n^{\prime}_{j}-\boldsymbol n_{j})}\int {d}^{3}r\phi^{*}(\boldsymbol r)U(\boldsymbol r)\phi(\boldsymbol r-\boldsymbol n^{\prime}+\boldsymbol n)}_{U_{J_{ij}}(\boldsymbol n^{\prime}-\boldsymbol n)}\cdot(y_{\boldsymbol ni}-y_{\boldsymbol n^{\prime}i})\cdot(y_{\boldsymbol nj}-y_{\boldsymbol n^{\prime}j}) 
\end{aligned}
\end{equation}

In \cref{eq-1.18} first term is the unperturbed resonance integral $U_J(\boldsymbol n^{\prime}-\boldsymbol n)$. The second term is first-order deviation terms, and the third term is second-order deviation terms where $\boldsymbol  n $ is the site index of molecules and $ \boldsymbol  n^{\prime} $ is the next neighbor site. From the second and third term respectively, we define $ U_{J_{i}}(\boldsymbol n^{\prime}-\boldsymbol n) $ and $ U_{J_{ij}}(\boldsymbol n^{\prime}-\boldsymbol n) $,

\begin{equation}\label{eq-1.19}	
U_{J_{i}}(\boldsymbol n^{\prime}-\boldsymbol n)=\frac{\partial} {\partial (\boldsymbol n^{\prime}_{i}-\boldsymbol n_{i})}\int {d}^{3}r\phi^{*}(\boldsymbol r)U(\boldsymbol r)\phi(\boldsymbol r-\boldsymbol n^{\prime}+\boldsymbol n)
\end{equation}

\begin{equation}\label{eq-1.20}
U_{J_{ij}}(\boldsymbol n^{\prime}-\boldsymbol n)= \frac{\partial^{2}} {\partial (\boldsymbol n^{\prime}_{i}-\boldsymbol n_{i})\partial (\boldsymbol n^{\prime}_{j}-\boldsymbol n_{j})}\int {d}^{3}r\phi^{*}(\boldsymbol r)U(\boldsymbol r)\phi(\boldsymbol r-\boldsymbol n^{\prime}+\boldsymbol n)
\end{equation}

Now, The complete electronic Hamiltonian in the expansion of one and two-phonon electron-phonon scattering interaction by inserting \cref{eq-1.15} and \cref{eq-1.18} into \cref{eq-1.10}. The complete electronic Hamiltonian $ H_{\mathrm{E}} $ is sum of diagonal Coulomb Hamiltonian $H_{\mathrm{C}}$ and non-diagonal resonance Hamiltonian $H_{\mathrm{J}}$ as,

\begin{equation}\label{eq-1.21}
\begin{aligned}
H_{\mathrm{E}}= H_{\mathrm{C}} + H_{\mathrm{J}}	
\end{aligned}
\end{equation}

The Coulomb Hamiltonian $H_{\mathrm{C}}$ part of electronic Hamiltonian $H_{\mathrm{E}}$ is where electron's interaction with phonon is diagonal in the space coordinates. Furthermore, it describes the fluctuation in the entire electronic band due to the Bosonic field and gives rise to a shift in the electronic band,

\begin{equation}\label{eq-1.22}
\begin{split}
H_{\mathrm{C}} =\sum_{\boldsymbol n}\sum_{\boldsymbol l}\Bigg\{&\underbrace{\int {d}^{3}r|\phi(\boldsymbol r)|^{2}U(\boldsymbol r-\boldsymbol l+\boldsymbol n)}_{U_{C_0}}\\ \boldsymbol{+}\sum_{\boldsymbol q}&\underbrace{
\frac{-\hbar^{1/2}}{(2MN\omega_ {\boldsymbol q})^{1/2}}\sum_{i}e_{\boldsymbol q i}\frac{\partial }{\partial (\boldsymbol l_{i}-\boldsymbol n_{i})} \int  {d}^{3}r|\phi(\boldsymbol r)|^{2}U(\boldsymbol r-\boldsymbol l+\boldsymbol n)}_{U^{C}_{\boldsymbol q;\mathrm{ac}}(\boldsymbol l-\boldsymbol n)}\cdot \underbrace{\big[({e}^{i\boldsymbol {qn}}-{e}^{i\boldsymbol q \boldsymbol l})b_{\boldsymbol q}+({e}^{-i\boldsymbol {qn}}-{e}^{-i\boldsymbol q \boldsymbol l})b_{\boldsymbol q}^{\dagger} \big]}_{B_{\boldsymbol n \boldsymbol l,\boldsymbol q;ac}} \\
&\boldsymbol{+}\frac{1}{2}\sum_{\boldsymbol q,\boldsymbol q'}\underbrace{ \frac{\hbar}{(4M^{2}N^{2}\omega_{\boldsymbol q}\omega_{\boldsymbol q^{\prime}})^{1/2}}\sum_{i,j}e_{\boldsymbol q i}e_{\boldsymbol q'j}\frac{\partial^{2} }{{\partial (\boldsymbol l_{i}-\boldsymbol n_{i})\partial (\boldsymbol l_{j}-\boldsymbol n_{j})}}\cdot \int {d}^{3}r|\phi(\boldsymbol r)|^{2}U(\boldsymbol r-\boldsymbol l+\boldsymbol n)}_{U^{C}_{\boldsymbol q\boldsymbol q^{\prime};\mathrm{ac}}(\boldsymbol l-\boldsymbol n)}\\& \cdot\underbrace{\big[({e}^{i\boldsymbol {qn}}-{e}^{i\boldsymbol q \boldsymbol l})b_{\boldsymbol q}+({e}^{-i\boldsymbol {qn}}-{e}^{-i\boldsymbol q \boldsymbol l})b_{\boldsymbol q}^{\dagger} \big]}_{B_{\boldsymbol n \boldsymbol l,\boldsymbol q;\mathrm{ac}}}\cdot \underbrace{ \big[({e}^{i\boldsymbol {qn}}-{e}^{i\boldsymbol q \boldsymbol l})b_{\boldsymbol q'}+({e}^{-i\boldsymbol {qn}}-{e}^{-i\boldsymbol q \boldsymbol l})b_{\boldsymbol q'}^{\dagger} \big]}_{B_{\boldsymbol n \boldsymbol l,\boldsymbol q';\mathrm{ac}}}\Bigg\}a_{\boldsymbol n}^{\dagger}a_{\boldsymbol n}
\end{split} 
\end{equation}

Where we define interaction matrix element $ U_{C_0}, B_{\boldsymbol n \boldsymbol l,\boldsymbol q;\mathrm{ac}}, B_{\boldsymbol n \boldsymbol l,\boldsymbol q';\mathrm{ac}} $, $ U^{C}_{\boldsymbol q;\mathrm{ac}}(\boldsymbol l-\boldsymbol n), U^{C}_{\boldsymbol q\boldsymbol q^{\prime};\mathrm{ac}}(\boldsymbol l-\boldsymbol n) $ such as,

\begin{equation}\label{eq-1.23}
\begin{aligned}
U^{C}_{\boldsymbol q;\mathrm{ac}}(\boldsymbol l-\boldsymbol n)&=-\frac{\hbar^{1/2}}{(2MN\omega_ {\boldsymbol q})^{1/2}}\sum_{i}e_{\boldsymbol q i}\frac{\partial }{\partial (\boldsymbol l_{i}-\boldsymbol n_{i})} \int {d}^{3}r|\phi(\boldsymbol r)|^{2}U(\boldsymbol r-\boldsymbol l+\boldsymbol n) \\
U^{C}_{\boldsymbol q\boldsymbol q^{\prime};\mathrm{ac}}(\boldsymbol l-\boldsymbol n) &=\frac{\hbar}{(4M^{2}N^{2}\omega_{\boldsymbol q}\omega_{\boldsymbol q^{\prime}})^{1/2}}\sum_{i,j}e_{\boldsymbol q i}e_{\boldsymbol q'j}\frac{\partial^{2} }{{\partial (\boldsymbol l_{i}-\boldsymbol n_{i})\partial (\boldsymbol l_{j}-\boldsymbol n_{j})}} \int {d}^{3}r|\phi(\boldsymbol r)|^{2}U(\boldsymbol r-\boldsymbol l+\boldsymbol n) \\
B_{\boldsymbol n \boldsymbol l,\boldsymbol q;\mathrm{ac}}&=({e}^{i\boldsymbol {qn}}-{e}^{i\boldsymbol q \boldsymbol l})b_{\boldsymbol q}+({e}^{-i\boldsymbol {qn}}-{e}^{-i\boldsymbol q \boldsymbol l})b_{\boldsymbol q}^{\dagger} \\
B_{\boldsymbol n \boldsymbol l,\boldsymbol q';\mathrm{ac}}&=({e}^{i\boldsymbol {qn}}-{e}^{i\boldsymbol q \boldsymbol l})b_{\boldsymbol q'}+({e}^{-i\boldsymbol {qn}}-{e}^{-i\boldsymbol q \boldsymbol l})b_{\boldsymbol q'}^{\dagger} \\
U_{C_0} & = \int {d}^{3}r|\phi(\boldsymbol r)|^{2}U(\boldsymbol r-\boldsymbol l+\boldsymbol n) \\
U^{C}_{\boldsymbol q;\mathrm{ac}}(\boldsymbol n^{\prime}-\boldsymbol n)& = - U^{C}_{\boldsymbol q;\mathrm{ac}}(\boldsymbol n - \boldsymbol n^{\prime}),\ \mathrm{from \ the \ inversion \ symmetric \ of \ lattice} \\
U^{C}_{\boldsymbol q\boldsymbol q^{\prime};\mathrm{ac}}(\boldsymbol n^{\prime}-\boldsymbol n)& = U^{C}_{\boldsymbol q\boldsymbol q^{\prime};\mathrm{ac}}(\boldsymbol n - \boldsymbol n^{\prime}),\ \mathrm{from \ the \ inversion \ symmetric \ of \ lattice} \\
U^{C}_{\boldsymbol q\boldsymbol q^{\prime};\mathrm{ac}}(\boldsymbol n^{\prime}-\boldsymbol n)& = U^{C}_{\boldsymbol q^{\prime}\boldsymbol q;\mathrm{ac}}(\boldsymbol n^{\prime}-\boldsymbol n),\ \mathrm{from \ the \ inversion \ symmetric \ of \ lattice}
\end{aligned}
\end{equation}

Therefore \cref{eq-1.22} read as,

\begin{equation}\label{eq-1.24}
H_{\mathrm{C}}=\sum_{\boldsymbol n}\sum_{\boldsymbol l}\Bigg\{\underbrace{U_{C_0}}_{\mathrm{Static~Part}} +\underbrace{\sum_{\boldsymbol q}U^{C}_{\boldsymbol q;ac}(\boldsymbol l-\boldsymbol n)B_{\boldsymbol n \boldsymbol l,\boldsymbol q;ac}+\frac{1}{2}\sum_{\boldsymbol q,\boldsymbol q'} U^{C}_{\boldsymbol q\boldsymbol q^{\prime};ac}(\boldsymbol l-\boldsymbol n)B_{\boldsymbol n \boldsymbol l,\boldsymbol q;ac}B_{\boldsymbol n \boldsymbol l,\boldsymbol q';ac}}_{\mathrm{Dynamic~Part;~which~is~further~discard~compare~to~dynamic~part~of~H_{\mathrm{J}}}}\Bigg\}a_{\boldsymbol n}^{\dagger}a_{\boldsymbol n} 
\end{equation}

From the argument of Friedman's work \textit{et al.}, \cite{friedman_electron-phonon_1965} $H_{\mathrm{C}}$ have less physical relevance compare to $H_{\mathrm{J}}$. Therefore, in the subsequent mathematical formulation, we will neglect the $H_{\mathrm{C}}$ part of Hamiltonian for purely mathematical simplification of the framework; however, the inclusion of $H_{\mathrm{C}}$ is not challenging from the theoretical point of view. We stated $H_{\mathrm{C}}$ here for the sake of completeness of the Hamiltonian. The electron-phonon interaction is non-diagonal in the space coordinates in the resonance Hamiltonian $H_{\mathrm{J}}$ part of electronic Hamiltonian $H_{\mathrm{E}}$. Furthermore, this interaction describes the internal fluctuation of electronic band states against each other due to the Bosonic field,

\begin{equation}\label{eq-1.25}
\begin{split}
H_{\mathrm{J}}=\sum_{\boldsymbol n,\boldsymbol n'} \Bigg\{&\underbrace{\int {d}^{3}\boldsymbol r\phi^{*}(\boldsymbol r)U(\boldsymbol r)\phi(\boldsymbol r-{\boldsymbol n^{\prime}}+{\boldsymbol n})}_{U_{J}(\boldsymbol n^{\prime}-\boldsymbol n)}\\& \boldsymbol{+}\sum_{\boldsymbol q}\underbrace{\frac{-\hbar^{1/2}}{(2MN\omega_ {\boldsymbol q})^{1/2}}\sum_{i}e_{\boldsymbol q i}\frac{\partial} {\partial (\boldsymbol n^{\prime}_{i}-\boldsymbol n_{i})}\int {d}^{3}r\phi^{*}(\boldsymbol r)U(\boldsymbol r)\phi(\boldsymbol r-\boldsymbol n^{\prime}+\boldsymbol n)}_{U^{J}_{\boldsymbol q;\mathrm{ac}}(\boldsymbol n^{\prime}-\boldsymbol n)}\cdot \\& \underbrace{\big[({e}^{i\boldsymbol {qn}}-{e}^{i\boldsymbol q \boldsymbol n^{\prime}})b_{\boldsymbol q}+({e}^{-i\boldsymbol {qn}}-{e}^{-i\boldsymbol q \boldsymbol n^{\prime}})b_{\boldsymbol q}^{\dagger}\big]}_{B_{\boldsymbol n \boldsymbol n^{\prime},\boldsymbol q;\mathrm{ac}}}\\
&\boldsymbol{+}\frac{1}{2}\sum_{\boldsymbol q,\boldsymbol q'}\underbrace{\frac{\hbar}{(4M^{2}N^{2}\omega_{\boldsymbol q}\omega_{\boldsymbol q^{\prime}})^{1/2}}\sum_{i,j}e_{\boldsymbol q i}e_{\boldsymbol q'j}
\frac{\partial^{2}} {\partial (\boldsymbol n^{\prime}_{i}-\boldsymbol n_{i})\partial (\boldsymbol n^{\prime}_{j}-\boldsymbol n_{j})}\cdot \int {d}^{3}r\phi^{*}(\boldsymbol r)U(\boldsymbol r)\phi(\boldsymbol r-\boldsymbol n^{\prime}+\boldsymbol n)}_{U^{J}_{\boldsymbol q\boldsymbol q^{\prime};\mathrm{ac}}(\boldsymbol n^{\prime}-\boldsymbol n)}\\& \cdot \underbrace{\big[({e}^{i\boldsymbol {qn}}-{e}^{i\boldsymbol q \boldsymbol n^{\prime}})b_{\boldsymbol q}+({e}^{-i\boldsymbol {qn}}-{e}^{-i\boldsymbol q \boldsymbol n^{\prime}})b_{\boldsymbol q}^{\dagger}\big]}_{B_{\boldsymbol n \boldsymbol n^{\prime},\boldsymbol q;ac}}\cdot \underbrace{\big[({e}^{i\boldsymbol {qn}}-{e}^{i\boldsymbol q \boldsymbol n^{\prime}})b_{\boldsymbol q'}+({e}^{-i\boldsymbol {qn}}-{e}^{-i\boldsymbol q \boldsymbol n^{\prime}})b_{\boldsymbol q'}^{\dagger}\big]}_{B_{\boldsymbol n \boldsymbol n^{\prime},\boldsymbol q';ac}}\Bigg\}a_{\boldsymbol n}^{\dagger}a_{\boldsymbol n'}
\end{split}
\end{equation}

Where we define interaction matrix element $ U_{J}(\boldsymbol n^{\prime}-\boldsymbol n), B_{\boldsymbol n \boldsymbol n^{\prime},\boldsymbol q;ac},B_{\boldsymbol n \boldsymbol n^{\prime},\boldsymbol q';ac} $, $U^{J}_{\boldsymbol q;ac}(\boldsymbol n^{\prime}-\boldsymbol n), U^{J}_{\boldsymbol q\boldsymbol q^{\prime};ac}(\boldsymbol n^{\prime}-\boldsymbol n)$ such as,

\begin{equation}\label{eq-1.26}
\begin{aligned}
U^{J}_{\boldsymbol q;ac}(\boldsymbol n^{\prime}-\boldsymbol n)&=-\frac{\hbar^{1/2}}{(2MN\omega_ {\boldsymbol q})^{1/2}}\sum_{i}e_{\boldsymbol q i}\frac{\partial} {\partial (\boldsymbol n^{\prime}_{i}-\boldsymbol n_{i})}\int {d}^{3}r\phi^{*}(\boldsymbol r)U(\boldsymbol r)\phi(\boldsymbol r-\boldsymbol n^{\prime}+\boldsymbol n) \\
U^{J}_{\boldsymbol q\boldsymbol q^{\prime};ac}(\boldsymbol n^{\prime}-\boldsymbol n)& =\frac{\hbar}{(4M^{2}N^{2}\omega_{\boldsymbol q}\omega_{\boldsymbol q^{\prime}})^{1/2}}\sum_{i,j}e_{\boldsymbol q i}e_{\boldsymbol q'j}
\frac{\partial^{2}} {\partial (\boldsymbol n^{\prime}_{i}-\boldsymbol n_{i})\partial (\boldsymbol n^{\prime}_{j}-\boldsymbol n_{j})}\int {d}^{3}r\phi^{*}(\boldsymbol r)U(\boldsymbol r)\phi(\boldsymbol r-\boldsymbol n^{\prime}+\boldsymbol n) \\ 
B_{\boldsymbol n \boldsymbol n^{\prime},\boldsymbol q;ac}&=({e}^{i\boldsymbol {qn}}-{e}^{i\boldsymbol q \boldsymbol n^{\prime}})b_{\boldsymbol q}+({e}^{-i\boldsymbol {qn}}-{e}^{-i\boldsymbol q \boldsymbol n^{\prime}})b_{\boldsymbol q}^{\dagger} \\
B_{\boldsymbol n \boldsymbol n^{\prime},\boldsymbol q';ac}&=({e}^{i\boldsymbol {qn}}-{e}^{i\boldsymbol q \boldsymbol n^{\prime}})b_{\boldsymbol q'}+({e}^{-i\boldsymbol {qn}}-{e}^{-i\boldsymbol q \boldsymbol n^{\prime}})b_{\boldsymbol q'}^{\dagger} \\
U_{J}(\boldsymbol n^{\prime}-\boldsymbol n)& = \int {d}^{3}\boldsymbol r\phi^{*}(\boldsymbol r)U(\boldsymbol r)\phi(\boldsymbol r-{\boldsymbol n^{\prime}}+{\boldsymbol n})\\
U_{J}(\boldsymbol n^{\prime}-\boldsymbol n)& = U_{J}(\boldsymbol n - \boldsymbol n^{\prime}), \mathrm{from \ the \ inversion \ symmetric \ of \ lattice}\\
U^{J}_{\boldsymbol q;ac}(\boldsymbol n^{\prime}-\boldsymbol n)& = - U^{J}_{\boldsymbol q;ac}(\boldsymbol n - \boldsymbol n^{\prime}),\ \mathrm{from \ the \ inversion \ symmetric \ of \ lattice} \\
U^{J}_{\boldsymbol q\boldsymbol q^{\prime};ac}(\boldsymbol n^{\prime}-\boldsymbol n)& = U^{J}_{\boldsymbol q\boldsymbol q^{\prime};ac}(\boldsymbol n - \boldsymbol n^{\prime}),\ \mathrm{from \ the \ inversion \ symmetric \ of \ lattice} \\
U^{J}_{\boldsymbol q\boldsymbol q^{\prime};\mathrm{ac}}(\boldsymbol n^{\prime}-\boldsymbol n)& = U^{J}_{\boldsymbol q^{\prime}\boldsymbol q;\mathrm{ac}}(\boldsymbol n^{\prime}-\boldsymbol n),\ \mathrm{from \ the \ inversion \ symmetric \ of \ lattice}
\end{aligned}
\end{equation}

Therefore \cref{eq-1.25} read as,
\begin{equation}\label{eq-1.27}
H_{\mathrm{J}}=\sum_{\boldsymbol n,\boldsymbol n'} \Bigg\{\underbrace{U_{J}(\boldsymbol n^{\prime}-\boldsymbol n)}_{\mathrm{Static~Part}}+\underbrace{\sum_{\boldsymbol q}U^{J}_{\boldsymbol q;\mathrm{ac}}(\boldsymbol n^{\prime}-\boldsymbol n)B_{\boldsymbol n \boldsymbol n^{\prime},\boldsymbol q;\mathrm{ac}}+\frac{1}{2}\sum_{\boldsymbol q,\boldsymbol q'}U^{J}_{\boldsymbol q\boldsymbol q^{\prime};\mathrm{ac}}(\boldsymbol n^{\prime}-\boldsymbol n)B_{\boldsymbol n \boldsymbol n^{\prime},\boldsymbol q;\mathrm{ac}}B_{\boldsymbol n \boldsymbol n^{\prime},\boldsymbol q';\mathrm{ac}}}_{\mathrm{Dynamic~Part;~Which~is~retained~compare~to~dynamic~part~of~H_{\mathrm{C}}}}\Bigg\}a_{\boldsymbol n}^{\dagger}a_{\boldsymbol n'}
\end{equation}

Similarly, polar optical phonon interaction with the electronic Hamiltonian can also expand in terms of the Coulomb integral $U_C$ and resonance integral $U_J$ up to second-order deviations terms with the electronic Hamiltonian and subsequently expanding the Hamiltonian. \cite{lang_kinetic_1963,schnakenberg_quasiteilchen-spektrum_1966} As hitherto concerned, in a practical organic device, additional imperfections such as lattice point defects, charged or uncharged impurity centers, adsorbent atoms, molecules in the nearby layer, and the surface may influence the charge transport. All those interactions with electrons depend upon the crystal lattice's finite distance point. Therefore depend upon the phonon field due to the distance fluctuation between diverse lattice points. However, these interactions can be modeled as  $H_{\mathrm{C}}$ or $H_{\mathrm{J}}$ type in the relevant lattice scenario.

The complete system Hamiltonian of electron-phonon gas incorporating one and two-phonon acoustic phonon interaction is, \cite{lang_kinetic_1962,lang_calculation_1968,bryksin_influence_1968}
\begin{equation}\label{eq-1.28}
H_{\mathrm{ac-ph}}=H_{\mathrm{S}}+H_{\mathrm{D}}   
\end{equation}

Where we separate the Hamiltonian in term of static $ H_{\mathrm{S}} $ and dynamic part $ H_{\mathrm{D}} $ for the Green's function perturbation expansion.

The static $ H_{\mathrm{S}} $ part of the complete system Hamiltonian of electron-phonon gas is,

\begin{equation}\label{eq-1.29}	
H_{\mathrm{S}}=(\varepsilon_{0}-\mu+\underbrace{U_{C_0})\sum_{\boldsymbol n}a_{\boldsymbol n}^{\dagger}a_{\boldsymbol n}}_{\mathrm{static \ part \ of} H_{\mathrm{C}}}+\underbrace{\sum_{\boldsymbol n,\boldsymbol n'}U_J(\boldsymbol n^{\prime}-\boldsymbol n)a_{\boldsymbol n}^{\dagger}a_{\boldsymbol n'}}_{\mathrm{static \ part \ of} \ H_{\mathrm{J}}} +\underbrace{\sum_{\boldsymbol q}\omega_{\boldsymbol q}b_{\boldsymbol q}^{\dagger}b_{\boldsymbol q}}_{H_{\mathrm{P}}}
\end{equation}

The dynamic $ H_{\mathrm{D}} $ part of the complete system Hamiltonian of electron-phonon gas is,

\begin{equation}\label{eq-1.30}
\begin{aligned}		
H_{\mathrm{D}}&=\underbrace{\sum_{\boldsymbol n,\boldsymbol n'}\Bigg\{\overbrace{\sum_{\boldsymbol q}U^{J}_{\boldsymbol q;\mathrm{ac}}(\boldsymbol n^{\prime}-\boldsymbol n)B_{\boldsymbol n \boldsymbol n^{\prime},\boldsymbol q;\mathrm{ac}}+\frac{1}{2}\sum_{\boldsymbol q,\boldsymbol q^{\prime}}U^{J}_{\boldsymbol q\boldsymbol q^{\prime};\mathrm{ac}}(\boldsymbol n^{\prime}-\boldsymbol n)B_{\boldsymbol n \boldsymbol n^{\prime},\boldsymbol q;\mathrm{ac}}B_{\boldsymbol n \boldsymbol n^{\prime},\boldsymbol q^{\prime};\mathrm{ac}}}^{H_{\mathrm{D};\boldsymbol n \boldsymbol n^{\prime}}} \Bigg\} a_{\boldsymbol n}^{\dagger}a_{\boldsymbol n^{\prime}}}_{\mathrm{dynamic \ part \ of} H_{\mathrm{J}}} \\
\end{aligned}  
\end{equation}

We will consider the $H_{\mathrm{D}}$ dynamic part of the complete system Hamiltonian of electron-phonon organic molecular gas as a perturbation to the static $ H_{\mathrm{S}} $ part due to Bosonic field and introduce the electronic Green's function $G(\boldsymbol n, t;\boldsymbol n^{\prime},t^{\prime})$ perturbation expanded in terms of $H_{\mathrm{D}}$. 

\section*{\texorpdfstring{Green's function $G(\boldsymbol n, t;\boldsymbol n^{\prime}, t^{\prime})$}{Green's function $G(n,t;n^{\prime},t^{\prime})$}}

The one-particle electronic Green's function propagator defined as, \cite{schwinger_greens-I_1951,schwinger_greens-II_1951}

\begin{equation}\label{eq-1.31}
G(\boldsymbol n, t;\boldsymbol n^{\prime}, t^{\prime};U)=-\frac{i}{\hbar}\frac{\Big\langle T^{c}\big\{a_{\boldsymbol n}(t)a_{\boldsymbol n'}^{\dagger}(t^{\prime})S\big\}\Big\rangle_{0}}{\langle S\rangle_{0}}
\end{equation}

Where $ S $ is scattering action functional from scattering matrix theory and inverse temperature $ \beta=\frac{1}{kT} $,

\begin{equation}\label{eq-1.32}
S=T^{c}\exp\Bigg\{-\frac{i}{\hbar}\int_{0}^{-i\hbar\beta}{d}\tau \sum_{\boldsymbol n, \boldsymbol n^{\prime}} U_{\boldsymbol n \boldsymbol n^{\prime}}(\tau) H_{D;\boldsymbol n \boldsymbol n^{\prime}}(\tau)\Bigg\}_{0}  
\end{equation}

The phenomenological explanation of the above equation can be interpreted for the $t>t'$ as the probability that a Fermion created in the quantum system at time $t'$ and at place $\boldsymbol n^{\prime}$ moves to another time $t$ at another place $\boldsymbol n$ is represented by Green's function $G(\boldsymbol n, t;\boldsymbol n^{\prime}, t^{\prime}; U)$. Where $T^{c}$ is the time ordering operator always moving the operator with the earlier time argument, and average values expressed as index $0$ and operators' time dependence is with respect to static part $H_{\mathrm{S}}$. Time-dependent $c$-number potential $U_{\boldsymbol n \boldsymbol n^{\prime}}(\tau)$ will become eventually zero. At the finite non-zero temperature, the quantum system is no longer in the ground state. Therefore bracket $\langle\cdots\rangle$ represents the grand canonical ensemble of thermodynamic average. The quantum device is in contact with a reservoir that has a temperature $\mathrm{T}$, and with the reservoir, the device might exchange heat as well as Fermions. To represent the zero and finite temperature equilibrium and non-equilibrium quantum system interaction, real-time and imaginary Green's functions as well as advanced $G^{A}$ and retarded $G^{R}$ Green's functions are defined. As well to completely describe the system, two Green's functions, the greater $G^{>}$ and the lesser $G^{<}$ Green's functions, are also defined as, \cite{martin_theory_1959,baym_conservation_1961,kadanoff_theory_1961,keldysh_diagram_1964}

\begin{equation}\label{eq-B6}
\begin{aligned}
G^{R}(\boldsymbol n, t;\boldsymbol n^{\prime}, t^{\prime}) &= -\frac{i}{\hbar}\theta(t-t^{\prime})\Big\langle\big[a_{\boldsymbol n}(t)a_{\boldsymbol n'}^{\dagger}(t^{\prime})\big]_{+}\Big\rangle \\
G^{A}(\boldsymbol n, t;\boldsymbol n^{\prime}, t^{\prime}) & = \frac{i}{\hbar}\theta(t^{\prime}-t)\Big\langle\big[a_{\boldsymbol n}(t)a_{\boldsymbol n'}^{\dagger}(t^{\prime})\big]_{+}\Big\rangle
\end{aligned}
\end{equation}

\begin{equation}\label{eq-B8}
\begin{aligned}
G^{<}(\boldsymbol n, t;\boldsymbol n',t') &=\frac{i}{\hslash}{\Big\langle a_{\boldsymbol n'}^{\dagger}(t^{\prime})a_{\boldsymbol n}(t) \Big\rangle}\\
G^{>} (\boldsymbol n, t;\boldsymbol n' , t') & =-\frac{i}{\hslash}\Big\langle a_{\boldsymbol n}(t)a_{\boldsymbol n'}^{\dagger}(t^{\prime})\Big\rangle
\end{aligned}
\end{equation}

The lesser $G^{<}$, greater $G^{>}$, advanced $G^{A}$, retarded $G^{R}$, and $G$ Green's functions are not uniquely independent but interrelated by following relationships, \cite{martin_theory_1959,baym_conservation_1961,kadanoff_theory_1961,keldysh_diagram_1964}
\begin{equation}\label{eq-B10}
\begin{aligned}
G(\boldsymbol n, t;\boldsymbol n', t') & = \theta(t-t')G^{>}(\boldsymbol n, t;\boldsymbol n', t')+\theta(t'-t)G^{<}(\boldsymbol n, t;\boldsymbol n', t') \\
G^{R,A}(\boldsymbol n, t;\boldsymbol n', t') & = \pm\theta(\pm t\mp t')\big[G^{>}(\boldsymbol n, t;\boldsymbol n', t')-G^{<}(\boldsymbol n, t;\boldsymbol n', t')\big] \\
G^{R}(\boldsymbol n, t;\boldsymbol n', t')&-G^{A}(\boldsymbol n, t;\boldsymbol n', t')  = G^{>}(\boldsymbol n, t;\boldsymbol n', t')-G^{<}(\boldsymbol n, t;\boldsymbol n', t')
\end{aligned}
\end{equation}

These equalities hold for both equilibrium and non-equilibrium pictures. The fluctuation-dissipation theorem linked all these properties in equilibrium. One subtle difference between the equilibrium and non-equilibrium pictures is in the derivation of the perturbation assumption. In equilibrium, zero-temperature Green's functions scenario system guarantees to return its initial state after an asymptotically large time. However, In the non-equilibrium picture, this is not true, as at time $t$ equal to $+\infty$, the final state will be very distinct from the initial state at time $t$ equal to $-\infty$. As a consequence, operator expectation values are built through the Feynman diagrams technique for contour integration, \cite{feynman_space-time_1948,schwinger_on_1951,mattuck_guide_1976} and Wick's decomposition for non-equilibrium situations. \cite{wick_the_1950,binder_nonequilibrium_1995} By using the \cref{eq-B8}, which holds for non-equilibrium, The non-equilibrium Green's function \cref{eq-1.31} is,

\begin{equation}\label{eq-B14}
G(\boldsymbol n, t;\boldsymbol n^{\prime}, t^{\prime}) = \theta(t, t^{\prime})G^{>}(\boldsymbol n, t;\boldsymbol n^{\prime}, t^{\prime})+\theta(t^{\prime}, t)G^{<}(\boldsymbol n, t;\boldsymbol n^{\prime}, t^{\prime})	
\end{equation}

Where on the contour the definition of the function $\theta(t, t^{\prime})$ is,

\begin{equation}\label{eq-B15}
\theta(t,\ t^{\prime}) = \left\{\begin{array}{l}
0,\ \mathrm{if}\ t\ \mathrm{is}\ \mathrm{earlier}\ \mathrm{on}\ \mathrm{a}\ \mathrm{contour}\ \mathrm{than}\ t^{\prime} \\
1,\ \mathrm{if}\ t\ \mathrm{is}\ \mathrm{later}\ \mathrm{on}\ \mathrm{a}\ \mathrm{contour}\ \mathrm{than}\ t^{\prime}
\end{array}\right.	
\end{equation}

The equation of motion for the electron in  Green's function terms \cref{eq-1.31} from the Bruevich, Tiablikov, Bogolyubov \textit{et al.} treatment \cite{bonch-bruevich_green_1962} and by Kadanoff and Baym \textit{et al.} method of functional derivatives, \cite{martin_theory_1959,baym_conservation_1961,kadanoff_quantum_1962} is,

\begin{equation}\label{eq-1.33}
\begin{aligned} 
&\Bigg[i\hbar\frac{\partial}{\partial t}-(\varepsilon_{0}-\mu+U_{C_0})\Bigg]G(\boldsymbol n,t;\boldsymbol n^{\prime},t^{\prime};U)=\delta_{\boldsymbol n \boldsymbol n^{\prime}}\delta(t-t^{\prime})+\frac{i}{\hbar}\sum_{\boldsymbol n,\boldsymbol n^{\prime}}\frac{1}{\langle S\rangle_{0}}\Big\langle T^{c}\{a_{\boldsymbol n}(t)H_{D;\boldsymbol n \boldsymbol n^{\prime}}(t)a_{\boldsymbol n^{\prime}}^{\dagger}(t^{\prime})S\}\Big\rangle
\end{aligned}
\end{equation}

By introducing intermediate dummy index $ \boldsymbol m  $ between the  $ \boldsymbol n  $ and $ \boldsymbol n^{\prime}  $ on the complex path and expanding the $ H_{\mathrm{D}} $ term,

\begin{equation}\label{eq-1.34}
\begin{aligned} 
\Bigg[i\hbar\frac{\partial}{\partial t}-(\varepsilon_{0}-\mu+U_{C_0})\Bigg]G(\boldsymbol n, t;\boldsymbol n^{\prime},t^{\prime};U)&=\delta_{\boldsymbol n \boldsymbol n^{\prime}}\delta(t-t^{\prime}) +\sum_{\boldsymbol m}{U_{J}}(\boldsymbol m-\boldsymbol n)G(\boldsymbol m, t;\boldsymbol n^{\prime}, t^{\prime};U)\\&+\sum_{\boldsymbol m}\Bigg[H_{\mathrm{D}}(\boldsymbol m, \boldsymbol n, t;U)+i\frac{\delta}{\delta U_{\boldsymbol {mn}}(t)}\Bigg]G(\boldsymbol m, t;\boldsymbol n^{\prime}, t^{\prime};U)
\end{aligned}
\end{equation}

Where,
\begin{equation}\label{eq-1.35}
H_{\mathrm{D}}(\boldsymbol m, \boldsymbol n, t;U)=-\frac{i}{\hbar}\frac{\Big\langle T^{c} \{H_{\mathrm{D};\boldsymbol m \boldsymbol n}(t)S\}\Big\rangle_0}{\langle S\rangle_0}  
\end{equation}

By transferring the terms towards left-hand side to make in Dyson's equation form, \cite{dyson_the_1949,schwinger_greens-I_1951,schwinger_greens-II_1951,hedin_new_1965}

\begin{equation}\label{eq-1.36}
\begin{aligned} 
\Bigg[i\hbar\frac{\partial}{\partial t}-(\varepsilon_{0}-\mu+U_{C_0})\Bigg]G(\boldsymbol n, t;\boldsymbol n^{\prime},t^{\prime};U)& -\sum_{\boldsymbol m}\int_{0}^{-i\hbar\beta}\mathrm{d}\tau\Sigma(\boldsymbol n,t;\boldsymbol m,\tau; U)G(\boldsymbol m,\tau; \boldsymbol n^{\prime}, t^{\prime};U)\\&=\delta_{\boldsymbol n \boldsymbol n^{\prime}}\delta(t-t^{\prime})\\
\end{aligned}
\end{equation}

After integrating \cref{eq-1.34} and \cref{eq-1.35}, \cite{bonch-bruevich_green_1962}

\begin{equation}\label{eq-1.37}
\begin{aligned}
G(\boldsymbol n,t;\boldsymbol n^{\prime},t^{\prime};U)&={G}_{0}(\boldsymbol n, {t};\boldsymbol n^{\prime}, {t}^{\prime})+\sum_{\boldsymbol m,\boldsymbol m^{\prime}}\int_{0}^{-i\hbar\beta}{d}\tau G_{0}(\boldsymbol n, {t};\boldsymbol m, \tau)U_{J}(\boldsymbol m^{\prime}-\boldsymbol m)\\&
\cdot\Bigg[H_{\mathrm{D}}(\boldsymbol m', \boldsymbol m, \tau;U)+i\frac{\delta}{\delta U_{\boldsymbol m' \boldsymbol m}(\tau)}\Bigg]G(\boldsymbol m', \tau;\boldsymbol n', {t}';U)   
\end{aligned}	
\end{equation}

Comparing \cref{eq-1.37} with Dyson's self-energy $\Sigma$,

\begin{equation}\label{eq-1.38}
\begin{aligned}
G(\boldsymbol n,{t};\boldsymbol n^{\prime}, t^{\prime};U)=G_{0}(\boldsymbol n, t;\boldsymbol n^{\prime}, t^{\prime})+\sum_{\boldsymbol m,\boldsymbol m^{\prime}}\int_{0}^{-i\hbar\beta}{d}\tau \int_{0}^{-i\hbar\beta}{d}\tau^{\prime}&G_{0}(\boldsymbol n, t;\boldsymbol m, \tau)\Sigma(\boldsymbol m,\tau;\boldsymbol m^{\prime},\tau^{\prime};U)\cdot\\&G(\boldsymbol m^{\prime}, \tau^{\prime};\boldsymbol n^{\prime}, t^{\prime};U)
\end{aligned}
\end{equation}

Where $G_{0}$ is free particle bare propagator correlation functions independent of $ U $ at $ H_{\mathrm{D};\boldsymbol m \boldsymbol n}=0 $ for the corresponding $H_{\mathrm{S}}$ in the $(\boldsymbol k, E)$ representation. The $G$ dressed particle propagator is a single particle picture incorporating all the many-body interactions. Also, By applying lattice translations invariance property and time-shift property, Green's function and self-energy in lattice index differences arguments, and by using the temporal periodicity properties,\cite{keldysh_diagram_1964, hybertsen_electron_1986} 

\begin{equation}\label{eq-1.39}
\begin{aligned}
G_{0}(\boldsymbol n, t;\boldsymbol n^{\prime}, t^{\prime})&=\delta_{\boldsymbol n \boldsymbol n^{\prime}} G_{0}(t, t^{\prime}) \\
G(\boldsymbol n,t;\boldsymbol n^{\prime}, t^{\prime})&\equiv G(\boldsymbol n-\boldsymbol n^{\prime}, t-t^{\prime})\\
\Sigma(\boldsymbol n,t;\boldsymbol n^{\prime}, t^{\prime})&\equiv \Sigma(\boldsymbol n-\boldsymbol n^{\prime}, t-t^{\prime})\\
G(\boldsymbol n, t\pm i\beta)&=-G(\boldsymbol n, t) \\
{G}_{0}^{<}(t)&=i\big\langle a_{\boldsymbol n}^{\dagger}(0)a_{\boldsymbol n}(t)\big\rangle_{0}=i\frac{e^{-i(\varepsilon_{0}-\mu+U_{C_0}) t}}{{e}^{\beta(\varepsilon_{0}-\mu+U_{C_0})}+1}\\
{G}_{0}^{>}(t)&=-i\big\langle a_{\boldsymbol n}(t)a_{\boldsymbol n}^{\dagger}(0)\big\rangle_{0}=-i\frac{{e}^{\beta(\varepsilon_{0}-\mu+U_{C_0})}e^{-i(\varepsilon_{0}-\mu+U_{C_0}) t}}{{e}^{\beta(\varepsilon_{0}-\mu+U_{C_0})}+1}\\
G_{0}(\boldsymbol k, E)&=\frac{1}{\Big[E-{\varepsilon}(\boldsymbol k)\Big]}  
\end{aligned}
\end{equation}

The self-energy $\Sigma$ can be determined by solving integro-functional equation from \cref{eq-1.37} and \cref{eq-1.38}, \cite{bonch-bruevich_green_1962,schnakenberg_electron-phonon_1969}

\begin{equation}\label{eq-1.40}
\begin{aligned}
\Sigma(\boldsymbol n, {t};\boldsymbol n^{\prime}, t^{\prime};U)&=U_J(\boldsymbol n^{\prime}-\boldsymbol n){H_D}(\boldsymbol n^{\prime}, \boldsymbol n, t;U)\delta(t-{t}^{\prime})\\&
+i\sum_{\boldsymbol m,\boldsymbol m^{\prime}} U_J(\boldsymbol m-\boldsymbol n)\int_{0}^{-i\hbar\beta}{d}\tau G(\boldsymbol m, {t};\boldsymbol m^{\prime}, \tau;U)\frac{\delta\Sigma(\boldsymbol m^{\prime},\tau;\boldsymbol n^{\prime},{t}^{\prime};U)}{\delta U_{\boldsymbol m\boldsymbol n}(t)} 
\end{aligned}
\end{equation}

The real-part of self-energy $\Sigma$ will provide the energy eigenvalues incorporating the effect of scattering. Furthermore, the inverse of the imaginary-part of self-energy $\Sigma$ will provide the lifetime of such scattering effect. For limit $U\rightarrow 0$, and expanding up to second order of Green's Function, from \cref{eq-1.40} the self-energy $\Sigma_{2}$ is,

\begin{equation}\label{eq-1.41}
\Sigma_{2}(\boldsymbol n,t;\boldsymbol n^{\prime},t^{\prime})=\sum_{\boldsymbol m,\boldsymbol m^{\prime}}U_J(\boldsymbol m-\boldsymbol n)U_J(\boldsymbol n'-\boldsymbol m^{\prime})G_{0}(\boldsymbol m,t;\boldsymbol m^{\prime},t^{\prime})\big\langle T^C\{{H}_{D;\boldsymbol {mn}}(t){H}_{D; \boldsymbol n^{\prime}\boldsymbol m^{\prime}}(t^{\prime})\}\big\rangle_{0} 
\end{equation}

We will first formally expand Green's function $G(\boldsymbol n, t;\boldsymbol n^{\prime}, t^{\prime})$ up to second-order and then evaluate the self-energy $\Sigma_{1,2}$ terms by summing up the first order and second-order perturbation term to the $ {H}_{\mathrm{D}} $ to set up the non-equilibrium Green's function framework for transport calculation. 

\section*{\texorpdfstring{Green's function $G(\boldsymbol n, t;\boldsymbol n^{\prime}, t^{\prime})$ Perturbation expansion}{Green's function $G(n,t;n^{\prime},t^{\prime})$ Perturbation expansion}}

By definition of electronic Green's function $G(\boldsymbol n, t;\boldsymbol n^{\prime}, t^{\prime})$ \cref{eq-1.31} and expanding the $H_{\mathrm{D}}$ part up-to second order for index $z=2$ as, \cite{bonch-bruevich_green_1962,bonch-bruevich_green_1962}

\begin{equation}\label{eq-1.42}
\begin{aligned}  
{G}(\boldsymbol n, t;\boldsymbol n^{\prime}, t^{\prime})&=-\frac{i}{\hbar}\frac{\Big\langle T^{c}\big\{a_{\boldsymbol n}(t)a_{\boldsymbol n'}^{\dagger}(t^{\prime})S\big\}\Big\rangle_{0}}{\langle S\rangle_{0}} \\
&=-\frac{i}{\hbar\langle S\rangle_{0}}\sum^{\infty}_{z=0}\frac{(-i)^{z}}{z!}\int_{0}^{-i\hbar\beta}{dt_1}\ldots\int_{0}^{-i\hbar\beta}{d}t_{z}\sum_{{\boldsymbol n}_{1},{\boldsymbol n}_{1}^{\prime}}\ldots\sum_{{\boldsymbol n}_{z},{\boldsymbol n}_{z}^{\prime}}\Big\langle T^{c}\big\{a_{\boldsymbol n}(t)a_{\boldsymbol n'}^{\dagger}({t}^{\prime}) \cdot\\ &
a_{\boldsymbol n_{1}}^{\dagger}(t_{1})a_{\boldsymbol n'_{1}}(t_{1})\ldots
a_{\boldsymbol n_{z}}^{\dagger}(t_{{z}})a_{\boldsymbol n_{z}^{\prime}}(t_{z})\big\}\Big\rangle_{0}\cdot\Big\langle T^{c}\big\{H_{\mathrm{D};\boldsymbol n_{1}\boldsymbol n_{1}^{\prime}}(t_{1})\ldots H_{\mathrm{D};\boldsymbol n_{z}\boldsymbol n_{z}^{\prime}}(t_{{z}})\big\}\Big\rangle_{0}
\end{aligned}
\end{equation}

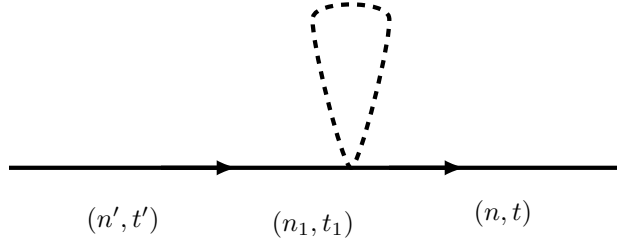
\begin{figure}[H]
\centering   
\begin{tikzpicture}
\draw[-latex,ultra thick]  (0,0)--node[below] {$(n^{\prime},t^{\prime})$} (3,0);
\draw[-latex,ultra thick] (2,0)--node[below] {$(n_{1},t_{1})$}  (6,0);
\draw[ultra thick]  (5,0)--node[below] {$(n,t)$}  (8,0);
\draw [ultra thick,dashed] plot[smooth cycle, tension=.4] coordinates {(4.5,0) (4,2) (5,2)};
\end{tikzpicture}
\caption{Self-energy contribution; first-order Green's function expansion; linked time-independent interaction}
\label{fig-1-0}
\end{figure}

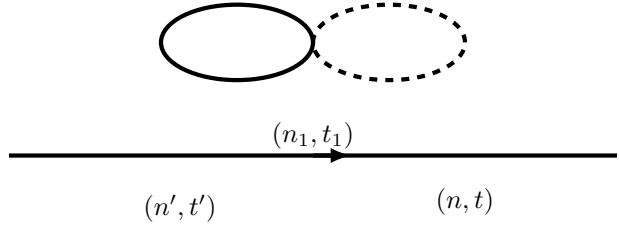
\begin{figure}[H]
\centering   
\begin{tikzpicture}
\draw[-latex,ultra thick]  (0,0)-- node[below] {$(n^{\prime},t^{\prime})$} (4.5,0);
\draw[ultra thick]  (4,0)--node[below] {$(n,t)$}  (8,0);
\draw[ultra thick]  (3,1.5) ellipse (1cm and .5cm); 
\draw[ultra thick,dashed]  (5,1.5) ellipse (1cm and .5cm);
\node at (4,1) [below]{$(n_{1},t_{1})$};
\end{tikzpicture}
\caption{Self-energy contribution; first-order Green's function expansion; unlinked interaction}
\label{fig-1-1}
\end{figure}

We have expanded Green's function up to $z=2$ second order to incorporate various scattering processes. In the case of  $z=1$ numerator term of \cref{eq-1.42} expand, it represented as Feynman integral diagrams. These graph sets are linked and unlinked first order Green's function expansion without any complex time dependence. For the case of first-order $z=1$ with incorporating, two phonon interactions process the numerator term of \cref{eq-1.42}, with the help of \cref{eq-1.37} and \cref{eq-1.40} is expanded and have two contribution term the first term corresponds to the graphical representation of linked contribution \cref{fig-1-0} and given as \cref{eq-1.43}, and the second term corresponds to the graphical representation of \cref{fig-1-1} and essential an unlinked contribution,

\begin{equation}\label{eq-1.43}
\int_{0}^{-i\hbar\beta}{d}t_{1}\sum_{\boldsymbol n_{1},\boldsymbol n'_{1}}G_{0}(\boldsymbol n-\boldsymbol n_{1},t-t_{1})G_{0}(\boldsymbol n_{1}^{\prime}-\boldsymbol n', t_{1}-t^{\prime})\cdot2\sum_{\boldsymbol q}U^{J}_{\boldsymbol q\boldsymbol q;\mathrm{ac}}(\boldsymbol n_{1}^{\prime}-\boldsymbol n_{1})\sin^{2}\bigg[\frac{\boldsymbol q(\boldsymbol n_{1}^{\prime}-\boldsymbol n_{1})}{2}\bigg]\coth(\frac{\beta\omega_{\boldsymbol q}}{2})
\end{equation}

\cref{fig-1-0} and \cref{fig-1-1} are two phonon processes; however, they are the time-independent contribution, and consequently, the sum of all the orders will provide a minor change in the $\varepsilon(\boldsymbol k)$  band and therefore safe to omit for further treatment. 

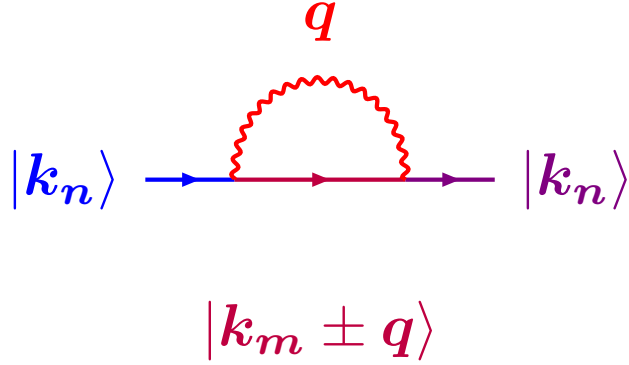
\begin{figure}[H]
\centering 
\feynmandiagram [scale=1.5,transform shape]{
a [blue,ultra thick,particle=\(|\boldsymbol{k_n}\rangle\)]  -- [blue,ultra thick, fermion] b -- [purple,ultra thick,fermion, edge label'=\(|\boldsymbol{k_m}\pm\boldsymbol q\rangle\)] c -- [violet,ultra thick,fermion] d [violet,ultra thick,particle=\(|\boldsymbol{k_n}\rangle\)] -- [fermion],
b  -- [red,ultra thick, boson, half left, looseness=2, edge label=\(\boldsymbol q\)]  c -- [red,ultra thick,boson,  half left, looseness=2],};
\caption{1-phonon interaction, second-order contribution linked irreducible time time-dependent graph to Green's function and self-energy}
\label{fig-2}
\end{figure}

With the assumption of the low density of electrons or holes in narrow bandwidth organic semiconductor and neglecting the influence of unlinked graph on the interaction,  for the irreducible linked contribution up to second-order $z=2$ expansion, thus have the numerical contributions as one phonon process interacting with an electron, for the case of second-order $z=2$ with incorporating, one phonon interactions process with electron the numerator term of \cref{eq-1.42} with the help of \cref{eq-1.37} and \cref{eq-1.40} expanded as in \cref{fig-2}, 

\begin{equation}\label{eq-1.44}
\begin{aligned}  
\int_{0}^{-i\hbar\beta}{dt_1}\int_{0}^{-i\hbar\beta}{d}{t}_{2}&\sum_{\boldsymbol n_{1},\boldsymbol n'_{1}}\sum_{\boldsymbol n_{2},\boldsymbol n'_{2}} G_{0}(\boldsymbol n-\boldsymbol n_{1}, {t}-t_{1})G_{0}(\boldsymbol n_{1}^{\prime}-\boldsymbol n_{2}, {t}_{1}-t_{2})G_{0}(\boldsymbol n'_{2}-\boldsymbol n^{\prime},t_{2}-t^{\prime})\cdot \\&\sum_{\boldsymbol q}U^{J}_{\boldsymbol q;\mathrm{ac}}(\boldsymbol n_{1}^{\prime}-\boldsymbol n_{1})U^{J}_{\boldsymbol q;\mathrm{ac}}(\boldsymbol n_{2}^{\prime}-\boldsymbol n_{2})({e}^{i\boldsymbol q\boldsymbol n_{1}}-{e}^{i\boldsymbol q\boldsymbol n_{1}^{\prime}})({e}^{-i\boldsymbol {qn}_2}-{e}^{-i\boldsymbol {qn}_{2}^{\prime}})\cdot\\&
\underbrace{\frac{-i}{\hbar}\Big[\Big\langle T^{c}\big\{b_{\boldsymbol q}({t_{1}-t_{2}})b_{\boldsymbol q}^{\dagger}\big\}\Big\rangle_{0}+\Big\langle T^{c}\big\{b_{\boldsymbol q}b_{\boldsymbol q}^{\dagger}({t_{1}-t_{2}})\big\}\Big\rangle_{0}\Big]}_{D(\boldsymbol q,t_{1}-t_{2})}
\end{aligned} 
\end{equation}

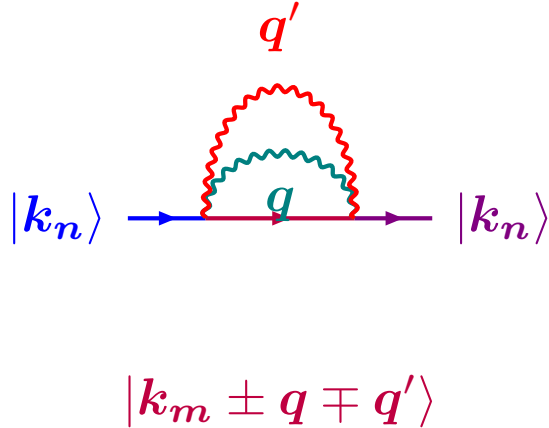
\begin{figure}[H]
\centering 
\feynmandiagram [scale=1.4,transform shape]{
a [blue,ultra thick,particle=\(|\boldsymbol{k_n}\rangle\)]  -- [blue,ultra thick,fermion] b -- [purple,ultra thick,fermion, edge label'=\(|\boldsymbol{k_m}\pm\boldsymbol q\mp\boldsymbol q^{\prime}\rangle\)] c -- [violet,ultra thick,fermion] d [violet,ultra thick,particle=\(|\boldsymbol{k_n}\rangle\)] -- [fermion],
b  -- [teal,ultra thick, boson, half left, looseness=1.5, edge label'=\(\boldsymbol q\)]  c -- [teal,ultra thick, boson,  half left, looseness=1.5],
b  -- [red,ultra thick,boson, half left, looseness=3,edge label=\(\boldsymbol q^{\prime}\)]  c -- [boson,  half left, looseness=3],};
\caption{Two-phonon interaction, second-order contribution linked irreducible time-dependent graph to Green's function and self-energy}
\label{fig-3}
\end{figure}

For the case of second-order $z=2$ with incorporating, two phonon interactions process with electron the numerator term of \cref{eq-1.42} with the help of \cref{eq-1.37} and \cref{eq-1.40} expanded as in  \cref{fig-3}, 

\begin{equation}\label{eq-1.45}
\begin{aligned} 
\int_{0}^{-i\hbar\beta}{dt_1}\int_{0}^{-i\hbar\beta}{d}{t}_{2}&\sum_{\boldsymbol n_{1},\boldsymbol n'_{1}}\sum_{\boldsymbol n_{2},\boldsymbol n'_{2}} G_{0}(\boldsymbol n-\boldsymbol n_{1}, {t}-t_{1})G_{0}(\boldsymbol n_{1}^{\prime}-\boldsymbol n_{2},{t}_{1}-t_{2})G_{0}(\boldsymbol n'_{2}-\boldsymbol n^{\prime},t_{2}-t^{\prime}) \cdot \\ \frac{1}{2}&\sum_{\boldsymbol q,\boldsymbol q^{\prime}}U^{J}_{\boldsymbol q\boldsymbol q^{\prime};\mathrm{ac}}(\boldsymbol n_{1}^{\prime}-\boldsymbol n_{1})U^{J}_{\boldsymbol q\boldsymbol q^{\prime};\mathrm{ac}}(\boldsymbol n_{2}^{\prime}-\boldsymbol n_{2})({e}^{i\boldsymbol q{\boldsymbol n}_{1}}-{e}^{{i}\boldsymbol q\boldsymbol n_{1}^{\prime}})({e}^{-i\boldsymbol {qn}_2}-{e}^{-i\boldsymbol {qn}_2^{\prime}})\cdot\\&({e}^{i\boldsymbol q^{\prime}\boldsymbol n_{1}}-{e}^{i\boldsymbol q^{\prime}\boldsymbol n_{1}^{\prime}})({e}^{-i\boldsymbol q^{\prime}\boldsymbol n_2}-{e}^{-i\boldsymbol q^{\prime}\boldsymbol n_{2}^{\prime}})\cdot
\\&
\underbrace{\frac{-i}{\hbar}\Big[\Big\langle T^{c}\big\{b_{\boldsymbol q}({t_{1}-t_{2}})b_{\boldsymbol q}^{\dagger}\big\}\Big\rangle_{0}+\Big\langle T^{c}\big\{b_{\boldsymbol q}b_{\boldsymbol q}^{\dagger}({t_{1}-t_{2}})\big\}\Big\rangle_{0}\Big]}_{D(\boldsymbol q,t_{1}-t_{2})}\cdot\\& 
\underbrace{\frac{-i}{\hbar}\Big[\Big\langle T^{c}\big\{b_{\boldsymbol q^{\prime}}({t_{1}-t_{2}})b_{\boldsymbol q^{\prime}}^{\dagger}\big\}\Big\rangle_{0}+\Big\langle T^{c}\big\{b_{\boldsymbol q^{\prime}}b_{\boldsymbol q^{\prime}}^{\dagger}({t_{1}-t_{2}})\big\}\Big\rangle_{0}\Big]}_{D(\boldsymbol q^{\prime},t_{1}-t_{2})} 
\end{aligned}
\end{equation}

Where phonon scattering probability through the phonon Green's function  $ D(\boldsymbol q, t) $ is, 

\begin{equation}\label{eq-1.46}
D(\boldsymbol q, t)=-\frac{i}{\hbar}\Big[\Big\langle T^{c}\big\{b_{\boldsymbol q}({t})b_{\boldsymbol q}^{\dagger}\big\}\Big\rangle_{0}+\Big\langle T^{c}\big\{b_{\boldsymbol q}b_{\boldsymbol q}^{\dagger}({t})\big\}\Big\rangle_{0}\Big]
\end{equation}

The Migdal theorem truncates Dyson's self-energy computation due to the boson field's lowest order phonon interaction. \cite{migdal_interaction_1958} Though it is computationally favorable for device simulation, it fails to incorporate multi-phonon processes and phonon lifetime relaxation. The more accurate spectral function is possible through the retarded cumulant expansion formalism, where electronic Green's function expanded in terms of cumulant function. \cite{nery_quasiparticles_2018}

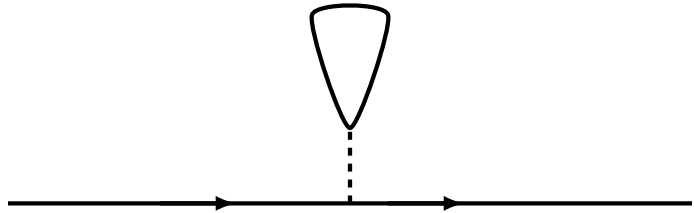
\begin{figure}[H]
\centering 
\begin{tikzpicture}
\draw[-latex,ultra thick]  (0,0)--node[below] {} (3,0);
\draw[-latex,ultra thick] (2,0)--node[below] {}  (6,0);
\draw[ultra thick]  (5,0)--node[below] {}  (9,0);
\draw[ultra thick,dashed]  (4.5,0)--  (4.5,1);
\draw [ultra thick] plot[smooth cycle, tension=.4] coordinates {(4.5,1) (4,2.5) (5,2.5)};
\end{tikzpicture}
\caption{One-phonon self-energy linked time-independent contribution; second-order Green's function expansion}
\label{fig-4-2}
\end{figure}

\begin{figure}[H]
\centering 
\begin{tikzpicture}
\draw[-latex,ultra thick] (0,0)--node[below] {} (2,0);
\draw[-latex,ultra thick] (1,0)--node[below] {} (5,0);
\draw[-latex,ultra thick] (4,0)--node[below] {} (7,0);
\draw[ultra thick] (6,0)--node[below] {} (9,0);
\draw [ultra thick,dashed] plot[smooth cycle, tension=.5] coordinates {(3,0) (2.5,2) (3.5,2)};
\draw [ultra thick,dashed] plot[smooth cycle, tension=.5] coordinates {(6,0) (5.5,2) (6.5,2)};
\end{tikzpicture}
\caption{Two-phonon self-energy linked time-independent contribution; second-order Green's function expansion, iterative process of \cref{fig-1-0} and reducible to the case of first-order $z=1$ as \cref{eq-1.43}}
\label{fig-4-4}
\end{figure}
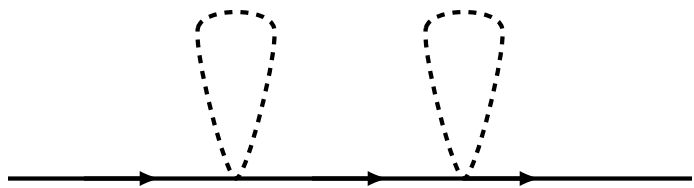

We have assumed that phonons baths remain in equilibrium during the interaction with electron motion. Moreover, the London effect or phonon drag due to electron and electron digging due to phonon in small polaron transport scenarios did not consider. Therefore, the phonon correlation in equilibrium is with $H_{\mathrm{S}}$ only. For completeness, we will mention here that in the case of second-order $z=2$ expansion, there are a total of twelve linked and unlinked contributions, out of which seven are unlinked and not represented here. Five are linked contributions out of which two time-dependents are mathematically expressed and represented in \cref{eq-1.44}, \cref{fig-2} and \cref{eq-1.45}, \cref{fig-3} respectively and the remaining three time-independent contributions shown in \cref{fig-4-2}, \cref{fig-4-3}, and \cref{fig-4-4}. The \cref{fig-4-2} is one-phonon linked contribution but time-independent in nature. The \cref{fig-4-3} and \cref{fig-4-4} are two-phonon linked contribution but time-independent in nature and \cref{fig-4-4} is iterative process of \cref{fig-1-0} and reducible to the case of first-order $z=1$ as discussed earlier in \cref{eq-1.43}. Moreover, a time-independent contribution from \cref{fig-4-2}, \cref{fig-4-3}, minuscule correction to the electronic band and neglected by the same argument discussed after \cref{eq-1.43}.

\begin{figure}[H]
\centering 
\begin{tikzpicture}
\draw[-latex,ultra thick]  (0,0)--node[below] {} (3,0);
\draw[-latex,ultra thick] (2,0)--node[below] {}  (6,0);
\draw[ultra thick]  (5,0)--node[below] {}  (9,0);
\draw [ultra thick](4.5,1.5) .. controls (5,2) and (5,2) .. (4.5,2.5);
\draw [ultra thick,dashed](4.5,0) .. controls (5,0.5) and (5,0.5) .. (4.5,1.5);
\draw [ultra thick,dashed](4.5,0) .. controls (4,0.5) and (4,0.5) .. (4.5,1.5);
\draw [ultra thick](4.5,2.5) .. controls (4,2) and (4,2) .. (4.5,1.5);
\end{tikzpicture}
\caption{Two-phonon self-energy linked time-independent contribution; second-order Green's function expansion}
\label{fig-4-3}
\end{figure}
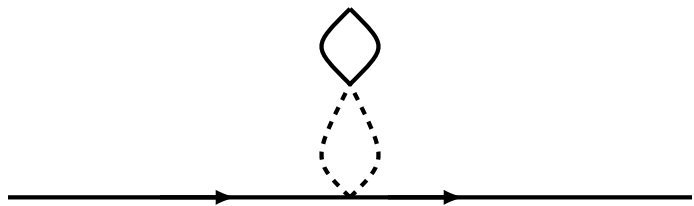

\begin{figure}[H]
\centering
\begin{tikzpicture}
\draw[-latex,ultra thick] (0,0)--node[below] {} (1.5,0);
\draw[-latex,ultra thick] (1,0)--node[below] {} (4,0);
\draw[-latex,ultra thick] (3.5,0)--node[below] {} (7.5,0);
\draw[-latex,ultra thick] (7,0)--node[below] {} (9.5,0);
\draw[ultra thick] (9,0)--node[below] {} (10,0);
\draw [ultra thick,dashed](2.5,0) arc (180:0:1.5);
\draw [ultra thick,dashed](5.5,0) arc (180:0:1.5);
\end{tikzpicture}
\caption{Third-order linked time-independent irreducible Self-energy contribution;  Green's function expansion; mixed one-phonon \& two-phonon interaction}
\label{fig-5}
\end{figure}
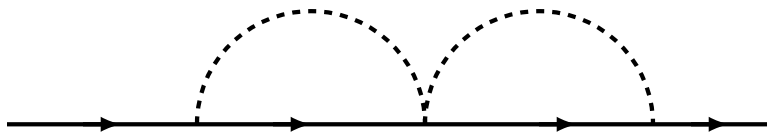

In the third-order $z=3$ expansion, the electron-phonon interactions, which contain one electron and combined processes of both one and two phonon interactions, will happen as shown in the \cref{fig-5}. In such a scenario, for electronic transport, adding the scattering rate through the Matthiessen rule for one, two, and three-phonon interaction is invalid. \cite{augustus1858xx,jacoboni_monte_1983,datta_quantum_2005,lundstrom_fundamentals_2009} As we have to compute and count mixed one and two-phonon scattering rates also. Therefore, we have neglected this third-order and higher-order interactions scattering in our hamiltonian and restricted it to second-order processes with one and two-phonon interactions only.

\begin{figure}[H]
\centering  
\begin{tikzpicture}[thick,scale=0.6, every node/.style={scale=0.6}]
\draw[-latex,ultra thick] (0,0)--node[below] {} (2,0);
\draw[-latex,ultra thick] (1,0)--node[below] {} (4,0);
\draw[-latex,ultra thick] (3,0)--node[below] {} (6,0);
\draw[ultra thick] (5.5,0)--node[below] {} (7.5,0);
\draw [ultra thick,dashed](2.5,0) arc (180:0:1.2);
\draw [ultra thick,dashed](1,0) arc (180:0:2.75);
\end{tikzpicture}
\begin{tikzpicture}[thick,scale=0.6, every node/.style={scale=0.6}]
\draw[-latex,ultra thick] (0,0)--node[below] {} (2,0);
\draw[-latex,ultra thick] (1,0)--node[below] {} (4,0);
\draw[-latex,ultra thick] (3,0)--node[below] {} (6,0);
\draw[ultra thick] (5.5,0)--node[below] {} (7.5,0);
\draw [ultra thick,dashed](3,0) arc (180:0:1.75);
\draw [ultra thick,dashed](1,0) arc (180:0:1.75);
\end{tikzpicture}
\caption{Linked irreducible self-energy contribution; two-phonon scattering in the lowest order non-simultaneous process}
\label{fig-8}
\end{figure}
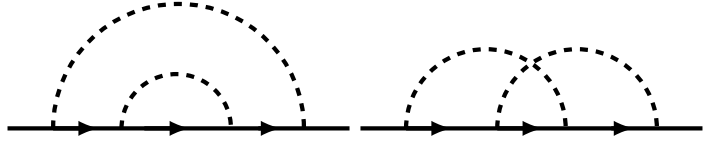

In  \cref{fig-8}, We have also illustrated linked, irreducible self-energy contribution in the lowest order where two-phonon non-simultaneous interaction contributes to the Green's function expansion. We have neglected all such non-simultaneous interactions.

The scattering self-energy $\Sigma$ incorporating the third and fourth-order interaction processes can be expanded based upon Konstantinov and Perel graphical method. \cite{konstantinov_diagram_1961} Lang, Firsov and Bryksin, \cite{lang_kinetic_1962,lang_calculation_1968,bryksin_influence_1968} first applied the third and fourth-order scattering processes in small polaron model. For expansion reference, we will give here the fourth-order self-energy $\Sigma_{4}$ in the limit $U\rightarrow 0$ from the \cref{eq-1.40}

\begin{equation}\label{eq-1.47}
\begin{aligned} 
\Sigma_{4}(1,2^{\prime})&=\sum_{1^{\prime},2}U_J(1)U_J(2)G_{0}(1', 2)\cdot\Big\langle T^{c}\big\{{H}_{D}(1){H}_{D}(2)\big\}\Big\rangle_{0}\\&
+\sum_{1^{\prime},2}\sum_{3,3^{\prime}} U_J(1)U_J(2)U_J(3)\int_{0}^{-i\hbar\beta}{d}t_{3}G_{0}(1', 3)G_{0}(3', 2)\cdot\Big\langle T^{c}\big\{{H}_{D}(1){H}_{D}(2){H}_{D}(3)\big\}\Big\rangle_{0}\\& +\sum_{1^{\prime},2}\sum_{3,3^{\prime}}\sum_{4,4^{\prime}}U_J(1)U_J(2)U_J(3)U_J(4)\int_{0}^{-i\hbar\beta}{d}t_{3}\int_{0}^{-{i}\hbar\beta}{d}{t}_{4} G_{0}(1^{\prime},3)G_{0}(3^{\prime},4)G_{0}(4^{\prime},2)\\&
\cdot\Bigg[\Big\langle T^{c}\big\{{H}_{D}(1){H}_{D}(2){H}_{D}(3){H}_{D}(4)\big\}\Big\rangle_{0}-\Big\langle T^{c}\big\{{H}_{D}(1){H}_{D}(3)\big\}\Big\rangle_{0}\Big\langle T^{c}\big\{{H}_{D}(2){H}_{D}(4)\big\}\Big\rangle_{0}\Bigg]
\end{aligned}
\end{equation}

We have used the following abbreviations in the \cref{eq-1.47},

\begin{equation}\label{eq-1.48}
\begin{aligned}
\sum_{1^{\prime},2}&=\sum_{\boldsymbol n_{1}^{\prime},\boldsymbol n_{2}}\\ 
U_J(1)&=U_J(\boldsymbol n_{1}^{\prime}-\boldsymbol n_{1})\\ G_{0}(1^{\prime},2)&=G_{0}(\boldsymbol n_{1}^{\prime}, t_{1};\boldsymbol n_{2}, {t}_{2})\\
{H}_{D}(1)&={H}_{D;\boldsymbol n_{1}^{\prime}\boldsymbol n_{1}}(t_{1})
\end{aligned}
\end{equation}

In the second-order perturbation, the \cref{eq-1.44} is exact expansion as up to second-order no additional integral diagram term will contribute the Green's function, and all the possible interaction is incorporated in the model. By using Wick's theorem, \cite{wick_the_1950,binder_nonequilibrium_1995} linked-graph theorem techniques \cite{fanchenko_generalized_1983} and Kadanoff and Baym method of functional derivatives, \cite{kadanoff_quantum_2018}  to the Green's function \cref{eq-1.42} expansion and summing all the topologically distinct, irreducible connected graphs, in the second-order, self-energy contribution of one-phonon \cref{eq-1.44} and two-phonon \cref{eq-1.45}.

\section*{\texorpdfstring{Self-energy $ \Sigma(\boldsymbol n,t) $}{Self-energy $ Sigma(n,t)$}}

The interaction self-energy expansion $ \Sigma_{1,2}(\boldsymbol n,t) $ with the help of \cref{eq-1.37} and \cref{eq-1.40} as expanded in \cref{eq-1.45} is,\cite{bogolyubov_kinetic_1978,bogolyubov_generalized_1979}

\begin{equation}\label{eq-1.49}
\begin{aligned} 
\Sigma_{1,2}(\boldsymbol n-\boldsymbol n',t-{t}^{\prime}) & =\sum_{\boldsymbol m,\boldsymbol m'}G_{0}(\boldsymbol m-\boldsymbol m',t-t^{\prime})\Bigg\{\sum_{\boldsymbol q}U^{J}_{\boldsymbol q;\mathrm{ac}}(\boldsymbol m-\boldsymbol n)U^{J}_{\boldsymbol q;\mathrm{ac}}(\boldsymbol n^{\prime}-\boldsymbol m^{\prime})({e}^{i\boldsymbol q\boldsymbol n}-{e}^{i\boldsymbol {qm}})\cdot\\&({e}^{-i\boldsymbol q\boldsymbol m^{\prime}}-{e}^{-{i}\boldsymbol q{\boldsymbol n}^{\prime}}) D(\boldsymbol q,t-t^{\prime}) -\frac{1}{2}\sum_{\boldsymbol q,\boldsymbol q^{\prime}}U^{J}_{\boldsymbol q\boldsymbol q^{\prime};\mathrm{ac}}(\boldsymbol m-\boldsymbol n)U^{J}_{\boldsymbol q\boldsymbol q^{\prime};\mathrm{ac}}(\boldsymbol n^{\prime}-\boldsymbol m^{\prime})\cdot\\& ({e}^{i\boldsymbol q\boldsymbol n}-{e}^{i\boldsymbol q\boldsymbol m})({e}^{-i\boldsymbol q \boldsymbol m^{\prime}}-{e}^{i\boldsymbol q{\boldsymbol n}^{\prime}})({e}^{i\boldsymbol q^{\prime}\boldsymbol n}-{e}^{i\boldsymbol q^{\prime}\boldsymbol m})({e}^{-{i}\boldsymbol q^{\prime}\boldsymbol m^{\prime}}-{e}^{-i\boldsymbol q^{\prime}\boldsymbol n^{\prime}})\cdot\\&  D(\boldsymbol q, t-t^{\prime})D(\boldsymbol q^{\prime}, t-t^{\prime})\Bigg\}
\end{aligned}
\end{equation}

It is convenient to work by taking Fourier transformation of Green's function $ G(\boldsymbol n,t) $ and self-energy $ \Sigma(\boldsymbol n,t) $ from the imaginary time domain to the wavefunction, energy domain representation $(\boldsymbol k, E)$ by

\begin{equation}\label{eq-1.50}
\begin{aligned} 
G(\boldsymbol n,t)&=\frac{1}{N}\sum_{\boldsymbol k}{e}^{i\boldsymbol {kn}}\frac{i}{\beta}\sum_{E}{e}^{-\frac{iEt}{\hbar}}G(\boldsymbol k, E)  \\
G(\boldsymbol k, E)&=\int_{0}^{-i\hbar\beta}{d}t{e}^{\frac{iEt}{\hbar}} \sum_{\boldsymbol n}{e}^{-i\boldsymbol {kn}}G(\boldsymbol n,t) \\
\Sigma(\boldsymbol n,t) & =\frac{1}{N}\sum_{\boldsymbol k}{e}^{i\boldsymbol {kn}}\frac{i}{\beta}\sum_{E}{e}^{-\frac{iEt}{\hbar}}\Sigma(\boldsymbol k, E)  \\
E&= \frac{(2l+1)\pi i}{\beta},\ \mathrm{where} \ l=0,\ \pm 1,\ \pm 2,\ldots
\end{aligned}
\end{equation}

Phonon Green's function $ D(\boldsymbol q, \zeta) $ with the Boson complex frequency $ \boldsymbol q $ and energy $ \zeta $,

\begin{equation}\label{eq-1.51}
\begin{aligned} 
D(\boldsymbol q,t) &=\frac{i}{\beta}\sum_{\zeta}{e}^{-\frac{i\zeta t}{\hbar}}D(\boldsymbol q, \zeta) \\  \zeta&=\frac{2l\pi i}{\beta}, \ \mathrm{where} \ l=0, \pm 1, \pm 2,\ldots
\end{aligned}
\end{equation}
\begin{equation}\label{eq-1.52}
\begin{aligned}
{G}_{0}(\boldsymbol k, E)&=\frac{1}{[E-\varepsilon(\boldsymbol k)]}\\ D(\boldsymbol q, \zeta)&=\frac{1}{[\zeta-\hbar\omega_{\boldsymbol q}]}-\frac{1}{[\zeta+\hbar\omega_{\boldsymbol q}]}
\end{aligned}
\end{equation}\\

\subsection*{\texorpdfstring{Acoustic Phonon Self-energy $ \Sigma_{\mathrm{ac}}(k, E) $}{Acoustic Phonon Self-energy $Sigma {\mathrm{ac}}(k,E) $}}

The self-energy $ \Sigma_{\mathrm{ac}}(\boldsymbol k, E) $ from \cref{eq-1.49} is, \cite{baumann_quantum_1962,baumann_quantum_1963}

\begin{equation}\label{eq-1.53}
\begin{split}  
\Sigma_{\mathrm{ac}}(\boldsymbol k, E) &=\underbrace{\sum_{\boldsymbol q}|U^{J}_{\boldsymbol q;\mathrm{ac}}(\boldsymbol k, \boldsymbol q)|^{2}\cdot\Bigg\{\overbrace{\frac{\mathcal{N}_{\boldsymbol q}+1-\mathcal{F}(\varepsilon(\boldsymbol k-\boldsymbol q))}{E-\varepsilon(\boldsymbol k-\boldsymbol q)-\hbar\omega_{\boldsymbol q}}}^{\text{1-Acoustic Phonon; Emission}}+\overbrace{\frac{\mathcal{N}_{\boldsymbol q}+\mathcal{F}(\varepsilon(\boldsymbol k-\boldsymbol q))}{E-\varepsilon(\boldsymbol k-\boldsymbol q)+\hbar\omega_{\boldsymbol q}}}^{\text{1-Acoustic Phonon; Absorption}}\Bigg\}}_{\Sigma_{\mathrm{ac;ph-1}}(\boldsymbol k, E)} \\&
\boldsymbol{+}\underbrace{\begin{aligned}\frac{1}{2}\sum_{\boldsymbol q,\boldsymbol q^{\prime}}|U^{J}_{\boldsymbol q\boldsymbol q^{\prime};\mathrm{ac}}(\boldsymbol k, \boldsymbol q, \boldsymbol q^{\prime})|^{2}
\cdot \Bigg\{\overbrace{\frac{\mathcal{F}(\varepsilon(\boldsymbol k-\boldsymbol q-\boldsymbol q^{\prime}))}{\mathcal{F}(\varepsilon(\boldsymbol k-\boldsymbol q-\boldsymbol q^{\prime})+\hbar\omega_{\boldsymbol q}+\hbar\omega_{\boldsymbol q^{\prime}})} \cdot\frac{\mathcal{N}_{\boldsymbol q}\mathcal{N}_{\boldsymbol q^{\prime}}}{E-\varepsilon(\boldsymbol k-\boldsymbol q-\boldsymbol q^{\prime})-\hbar\omega_{\boldsymbol q}-\hbar\omega_{\boldsymbol q^{\prime}}}}^{\text{2-Acoustic Phonon; Both Simultaneous Emission}}\\+\overbrace{\frac{\mathcal{F}(\varepsilon(\boldsymbol k-\boldsymbol q-\boldsymbol q^{\prime}))}{\mathcal{F}(\varepsilon(\boldsymbol k-\boldsymbol q-\boldsymbol q^{\prime})-\hbar\omega_{\boldsymbol q}-\hbar\omega_{\boldsymbol q^{\prime}})}\cdot\frac{(\mathcal{N}_{\boldsymbol q}+1)(\mathcal{N}_{\boldsymbol q^{\prime}}+1)}{E-\varepsilon(\boldsymbol k-\boldsymbol q-\boldsymbol q^{\prime})+\hbar\omega_{\boldsymbol q}+\hbar\omega_{\boldsymbol q^{\prime}}}}^{\text{2-Acoustic Phonon; Both Simultaneous Absorption}} \\ +\overbrace{\frac{\mathcal{F}(\varepsilon(\boldsymbol k-\boldsymbol q-\boldsymbol q^{\prime}))}{\mathcal{F}(\varepsilon(\boldsymbol k-\boldsymbol q-\boldsymbol q^{\prime})+\hbar\omega_{\boldsymbol q}-\hbar\omega_{\boldsymbol q^{\prime}})}\cdot\frac{\mathcal{N}_{\boldsymbol q}(\mathcal{N}_{\boldsymbol q^{\prime}}+1)}{E-\varepsilon(\boldsymbol k-\boldsymbol q-\boldsymbol q^{\prime})-\hbar\omega_{\boldsymbol q}+\hbar\omega_{\boldsymbol q^{\prime}}}}^{\text{2-Acoustic Phonon; Both Simultaneous Emission \& Absorption}}\\+\overbrace{\frac{\mathcal{F}(\varepsilon(\boldsymbol k-\boldsymbol q-\boldsymbol q^{\prime}))}{\mathcal{F}(\varepsilon(\boldsymbol k-\boldsymbol q-\boldsymbol q^{\prime})-\hbar\omega_{\boldsymbol q}+\hbar\omega_{\boldsymbol q^{\prime}})}\cdot\frac{(\mathcal{N}_{\boldsymbol q}+1)\mathcal{N}_{\boldsymbol q^{\prime}}}{E-\varepsilon(\boldsymbol k-\boldsymbol q-\boldsymbol q^{\prime})+\hbar\omega_{\boldsymbol q}-\hbar\omega_{\boldsymbol q^{\prime}}}}^{\text{2-Acoustic Phonon; Both Simultaneous Absorption \& Emission}}\Bigg\}\end{aligned}}_{\Sigma_{\mathrm{ac;ph-2}}(\boldsymbol k, E)}\\
&= \Sigma_{\mathrm{ac;ph-1}}(\boldsymbol k, E) + \Sigma_{\mathrm{ac;ph-2}}(\boldsymbol k, E)
\end{split}
\end{equation}

Where $ \Sigma_{\mathrm{ac;ph-1}}(\boldsymbol k, E) $ are self-energy due to one acoustic phonon interaction, $ \Sigma_{\mathrm{ac;ph-2}}(\boldsymbol k, E) $ are self-energy due to two simultaneous acoustic phonon interaction with the electron. There are six contribution term from the one and two-phonon process, where matrix element $ U^{J}_{\boldsymbol q;\mathrm{ac}}(\boldsymbol k, \boldsymbol q) $, $ U^{J}_{\boldsymbol q\boldsymbol q^{\prime};\mathrm{ac}}(\boldsymbol k,\boldsymbol q,\boldsymbol q^{\prime}) $, $ \mathcal{N}_{\boldsymbol q} $, $ \mathcal{F}(\varepsilon) $, and $ \varepsilon(\boldsymbol k^{\prime}) $ defined as,

\begin{equation}\label{eq-1.54}
\begin{aligned}
U^{J}_{\boldsymbol q;\mathrm{ac}}(\boldsymbol k, \boldsymbol q)&=\sum_{\boldsymbol n^{\prime}-\boldsymbol n}{e}^{i\boldsymbol {k}(\boldsymbol n^{\prime}-\boldsymbol n)}U^{J}_{\boldsymbol q;\mathrm{ac}}(\boldsymbol n^{\prime}-\boldsymbol n)-\sum_{\boldsymbol n^{\prime}-\boldsymbol n}{e}^{i {(\boldsymbol k- \boldsymbol q)(\boldsymbol n^{\prime}-\boldsymbol n)}}U^{J}_{\boldsymbol q;\mathrm{ac}}(\boldsymbol n^{\prime}-\boldsymbol n) \\
U^{J}_{\boldsymbol q\boldsymbol q^{\prime};\mathrm{ac}}(\boldsymbol k,\boldsymbol q,\boldsymbol q^{\prime})&=\sum_{\boldsymbol n^{\prime}-\boldsymbol n}{e}^{i{\boldsymbol k}(\boldsymbol n^{\prime}-\boldsymbol n)}U^{J}_{\boldsymbol q\boldsymbol q^{\prime};\mathrm{ac}}(\boldsymbol n^{\prime}-\boldsymbol n)-\sum_{\boldsymbol n^{\prime}-\boldsymbol n}{e}^{i{(\boldsymbol k-\boldsymbol q)}(\boldsymbol n^{\prime}-\boldsymbol n)}U^{J}_{\boldsymbol q\boldsymbol q^{\prime};\mathrm{ac}}(\boldsymbol n^{\prime}-\boldsymbol n)\\&-\sum_{\boldsymbol n^{\prime}-\boldsymbol n}{e}^{i{(\boldsymbol k-\boldsymbol q^{\prime})}(\boldsymbol n^{\prime}-\boldsymbol n)}U^{J}_{\boldsymbol q\boldsymbol q^{\prime};\mathrm{ac}}(\boldsymbol n^{\prime}-\boldsymbol n)+\sum_{\boldsymbol n^{\prime}-\boldsymbol n}{e}^{i{(\boldsymbol k-\boldsymbol q-\boldsymbol q^{\prime})}(\boldsymbol n^{\prime}-\boldsymbol n)}U^{J}_{\boldsymbol q\boldsymbol q^{\prime};\mathrm{ac}}(\boldsymbol n^{\prime}-\boldsymbol n)\\&
\mathrm{Bose-Einstein \ distribution \ function} \,\, \mathcal{N}_{\boldsymbol q}=\frac{1}{{e}^{\beta\hbar\omega_{\boldsymbol q}}-1}\\&
\mathrm{Fermi-Dirac \ distribution \ function} \,\, \mathcal{F}(\varepsilon(\boldsymbol k^{\prime}))=\frac{1}{{e}^{\beta\varepsilon(\boldsymbol k^{\prime})}+1} \\&
\varepsilon(\boldsymbol k^{\prime})=\varepsilon(\boldsymbol k-\boldsymbol q-\boldsymbol q^{\prime})\\&
\varepsilon(\boldsymbol k)=\underbrace{\varepsilon_{0}-\mu+U_{C_0}+\sum_{\boldsymbol n}U_J(\boldsymbol n){e}^{i\boldsymbol {kn}}}_{\text{electronic part of static} H_{s} \text{Hamiltonian}}
\end{aligned}
\end{equation}

The coupling matrix elements in the acoustic case $ U^{J}_{\boldsymbol q;\mathrm{ac}}(\boldsymbol k, \boldsymbol q) $, $ U^{J}_{\boldsymbol q\boldsymbol q^{\prime};\mathrm{ac}}(\boldsymbol k,\boldsymbol q,\boldsymbol q^{\prime}) $, are deformation-potential type interaction.

In the \cref{eq-1.54}, electronic energy eigenvalue of static $H_{s}$ Hamiltonian is for the isolated molecular crystal. When contacted through the electrode in the external circuit, an infinitesimally slight broadening energy is added to the hamiltonian to include the effect of contact. We will explicitly include this effect in the treatment of the non-equilibrium Green's function framework. There is a slight shift in the energy spectrum due to the coupling of molecular orbitals with the electrodes. For the exact solution in the many-body quantum system, electron-phonon self-energies through Dyson's equation for the phonon Green's function $D(\boldsymbol{q}; \zeta)$ is solved in a coupled way with electron Green's functions which is very expansive. \cite{wagner_expansions_1991} The first-order phonon renormalization process can be neglected at a price to miss to capture a possible phonon lifetime reduction. According to the Migdal theorem, \cite{migdal_interaction_1958} phonon induced renormalization process of the electron-phonon vertex scales with the ratio of electron mass to ion mass. Therefore it is safe to omit the renormalization process at the first level. \cite{fetter_quantum_1971} Therefore, We have assumed the phonon bath is in thermal equilibrium and full phonon Green's function $D(\boldsymbol{q};\zeta)$ approximated to the non-interacting free phonon Green's functions $D^{0}(\boldsymbol{q};\zeta)$. The Bose distribution for the phonons is $\mathcal{N}_{\boldsymbol{q}}$ with phonon frequency $\omega_{\boldsymbol{q}}$. For the narrow bandwidth, low density of state organic semiconductor crystal at the room temperature, the Fermi-Dirac distribution function $ \mathcal{F} (\varepsilon(\boldsymbol k)) $, is slowly varying in energy within the full span of k-space and electron and holes in these low density of state crystal reasonably approximate as following the Boltzmann statics and $\mathcal{F}_{ k}\ll 1$ and $1-\mathcal{F}_{ k}\approx 1$. 
Moreover variation in $ \mathcal{F} (\varepsilon(\boldsymbol k^{\prime})) $ due to various contribution by the $ \boldsymbol k^{\prime} $, $ \omega_{\boldsymbol q} $ and $ \omega_{\boldsymbol q^{\prime}} $ interaction with in one-phonon and two-phonon can calculated by taking summation of all $ \boldsymbol k^{\prime} $ in the modified Bessel function.

\subsection*{\texorpdfstring{Electronic Green's function $ G(k, E) $}{Electronic Green's function $G(k,E)$}}

The electronic Green's function $ G(\boldsymbol k, E) $ is,
\begin{equation}\label{eq-1.57}
\begin{aligned} 
G(\boldsymbol k, E)&=\bigg[\frac{1}{E-\varepsilon(\boldsymbol k)-\Sigma(\boldsymbol k, E) \pm i\eta}\bigg]_{\eta\rightarrow 0^+} =\bigg[\int_{-\infty}^{+\infty}\frac{\mathrm{d}\omega}{2\pi}\frac{A(\boldsymbol k,\omega)}{E-\omega \pm i\delta} \bigg]_{\delta\rightarrow 0^+}
\end{aligned}  
\end{equation}

Where in the energy domain constant $ \eta $ tend to zero and force the integrands to zero for reaching $ \pm\infty $ and similarly in spectral-domain constant $ \delta $ tend to zero and force the integrands to peak at zero frequency. Broadening function $\Gamma(\boldsymbol k,\omega)$ from imaginary part of $\Sigma(\boldsymbol k, E) $ in spectral representation defined as,

\begin{equation}\label{eq-1.58}
\Gamma(\boldsymbol k, \omega)=i\Big[\Sigma(\boldsymbol k, \omega+i\delta)-\Sigma(\boldsymbol k, \omega-i\delta)\Big]_{\delta\rightarrow 0^+}   
\end{equation}

In transport calculation, the real part of self-energy $\Sigma$ provides a minor correction in $\varepsilon(\boldsymbol k)$ bandstructure through renormalization of the chemical potential that can be neglected for the low density of state, single-band organic crystal, while the imaginary part of $\Sigma$ will provide coupling strength and relaxation times from the \cref{eq-1.57} near the real $z$-axis for $\omega=\varepsilon(\boldsymbol k)$, Green's function is,

\begin{equation}\label{eq-1.59}
G(\boldsymbol k, \omega\pm i\delta)=\Bigg[\frac{1}{\omega \pm i\delta-\varepsilon(\boldsymbol k)\pm \Big[\frac{i\Gamma(\boldsymbol k, \omega)}{2}\Big]}\Bigg]_{\delta\rightarrow 0^+}   
\end{equation}

And the spectral function $A(\boldsymbol k, \omega)$ which is related to the observed angle-resolved photoemission spectroscopy signal of dressed Green's function $G(\boldsymbol k, \omega)$ incorporating all the electron-phonon interaction for $\omega\approx\varepsilon(\boldsymbol k)$,

\begin{equation}\label{eq-1.60}
\begin{aligned}
A(\boldsymbol k, \omega)&=i\Big[G(\boldsymbol k, \omega+i\delta)-G(\boldsymbol k, \omega-i\delta)\Big]_{\delta\rightarrow 0^+}
\\&=\frac{\Gamma(\boldsymbol k,\omega)}{\Big[\omega-\varepsilon(\boldsymbol k)\Big]^{2}+\Big[\frac{\Gamma(\boldsymbol k,\omega)}{2}\Big]^{2}} \\&
\approx\frac{\Gamma(\boldsymbol k,\varepsilon(\boldsymbol k))}{\Big[\omega-\varepsilon(\boldsymbol k)\Big]^{2}+\Big[\frac{\Gamma(\boldsymbol k,\varepsilon(\boldsymbol k))}{2}\Big]^{2}}
\end{aligned} 
\end{equation}

By Dyson's self-energy equation, all the electron-phonon interaction contained in self-energy $\Sigma(\boldsymbol k, E) $ and connected through broadening function $\Gamma(\boldsymbol k, \omega)$ for $\omega\approx\varepsilon(\boldsymbol k)$ from \cref{eq-1.53} as,

\begin{equation}\label{eq-1.61}
\begin{split}  
\Gamma(\boldsymbol k, \varepsilon(\boldsymbol k)) &=\underbrace{\begin{aligned}\frac{2\pi}{\hbar}\sum_{\boldsymbol q}|U^{J}_{\boldsymbol q;\mathrm{ac}}(\boldsymbol k, \boldsymbol q)|^{2}\Bigg\{&\overbrace{\Big[\mathcal{N}_{\boldsymbol q}+1-\mathcal{F}(\varepsilon(\boldsymbol k-\boldsymbol q))\Big]\cdot\delta\Big(\varepsilon(\boldsymbol k)-\varepsilon(\boldsymbol k-\boldsymbol q)-\hbar\omega_{\boldsymbol q}\Big)}^{\text{1-Acoustic Phonon; Emission}}\\&+\overbrace{\Big[\mathcal{N}_{\boldsymbol q}+\mathcal{F}(\varepsilon(\boldsymbol k-\boldsymbol q))\Big]\cdot\delta\Big(\varepsilon(\boldsymbol k)-\varepsilon(\boldsymbol k-\boldsymbol q)+\hbar\omega_{\boldsymbol q}\Big)}^{\text{1-Acoustic Phonon;  Absorption}}\Bigg\}\end{aligned}}_{\Gamma_{\mathrm{ac;ph-1}}(\boldsymbol k, \varepsilon(\boldsymbol k))}\\ 
&\boldsymbol{+}\underbrace{\begin{aligned} \frac{2\pi}{\hbar}&\sum_{\boldsymbol q,\boldsymbol q^{\prime}}|U^{J}_{\boldsymbol q\boldsymbol q^{\prime};\mathrm{ac}}(\boldsymbol k,\boldsymbol q,\boldsymbol q^{\prime})|^{2}\bigg[
\frac{\mathcal{F}(\varepsilon(\boldsymbol k-\boldsymbol q-\boldsymbol q^{\prime}))}{\mathcal{F}(\varepsilon(\boldsymbol k))}
\bigg]\cdot \\&\Bigg\{\overbrace{\mathcal{N}_{\boldsymbol q}\mathcal{N}_{\boldsymbol q'} \cdot\delta\Big(\varepsilon(\boldsymbol k)-\varepsilon(\boldsymbol k-\boldsymbol q-\boldsymbol q^{\prime})-\hbar\omega_{\boldsymbol q}-\hbar\omega_{\boldsymbol q^{\prime}}\Big)}^{\text{2-Acoustic Phonon; Both Simultaneous Emission}}\\& +\overbrace{(\mathcal{N}_{\boldsymbol q}+1)(\mathcal{N}_{\boldsymbol q'} +1)\cdot\delta\Big(\varepsilon(\boldsymbol k)-\varepsilon(\boldsymbol k-\boldsymbol q-\boldsymbol q^{\prime})+\hbar\omega_{\boldsymbol q}+\hbar\omega_{\boldsymbol q^{\prime}}\Big)}^{\text{2-Acoustic Phonon; Both Simultaneous  Absorption}} \\&+\overbrace{\mathcal{N}_{\boldsymbol q}(\mathcal{N}_{\boldsymbol q'} +1) \cdot\delta\Big(\varepsilon(\boldsymbol k)-\varepsilon(\boldsymbol k-\boldsymbol q-\boldsymbol q^{\prime})-\hbar\omega_{\boldsymbol q}+\hbar\omega_{\boldsymbol q^{\prime}}\Big)}^{\text{2-Acoustic Phonon; Both Simultaneous Emission \& Absorption}}\\&
+\overbrace{(\mathcal{N}_{\boldsymbol q}+1)\mathcal{N}_{\boldsymbol q'} \cdot\delta\Big(\varepsilon(\boldsymbol k)-\varepsilon(\boldsymbol k-\boldsymbol q-\boldsymbol q^{\prime})+\hbar\omega_{\boldsymbol q}-\hbar\omega_{\boldsymbol q^{\prime}}\Big)}^{\text{2-Acoustic Phonon; Both Simultaneous Absorption \& Emission}}\Bigg\}\end{aligned}}_{\Gamma_{\mathrm{ac;ph-2}}(\boldsymbol k, \varepsilon(\boldsymbol k))} \\
&= \Gamma_{\mathrm{ac;ph-1}}(\boldsymbol k, \varepsilon(\boldsymbol k))+ \Gamma_{\mathrm{ac;ph-2}}(\boldsymbol k, \varepsilon(\boldsymbol k))
\end{split}
\end{equation}

In the self-energy $ \Sigma(\boldsymbol k, E) $ \cref{eq-1.53} and broadening function $\Gamma(\boldsymbol k, \omega)$  \cref{eq-1.61} in the second-order expansion total of six phonon contribution, two contributions is due to single phonon absorption and emission interaction and other four are contribution due to combine emission or absorption of two phonons simultaneous. In the narrow bandwidth semiconductor, due to the low density of state in the energy window of the electron and high energy of acoustic and polar phonon, the possibility of one-phonon processes scattering are in the tiny region of $\boldsymbol k$ wave vectors space to satisfy the energy conservation. For the two-phonon processes, the probability of interaction in which both phonons emit energy $\delta(\varepsilon(\boldsymbol k)-\varepsilon(\boldsymbol k-\boldsymbol q-\boldsymbol q^{\prime})-\hbar\omega_{\boldsymbol q}-\hbar\omega_{\boldsymbol q^{\prime}})$ or absorb energy $ \delta(\varepsilon(\boldsymbol k)-\varepsilon(\boldsymbol k-\boldsymbol q-\boldsymbol q^{\prime})+\hbar\omega_{\boldsymbol q}+\hbar\omega_{\boldsymbol q^{\prime}})$, simultaneously is even minuscule. By eliminating two terms in the aforementioned two phonon interactions and with this approximation $\Gamma(\boldsymbol k, \varepsilon(\boldsymbol k))$ is,

\begin{equation}\label{eq-1.62}
\begin{split}  
\Gamma(\boldsymbol k, \varepsilon(\boldsymbol k)) &=\underbrace{\begin{aligned}\frac{2\pi}{\hbar}\sum_{\boldsymbol q}|U^{J}_{\boldsymbol q;\mathrm{ac}}(\boldsymbol k, \boldsymbol q)|^{2}\Bigg\{&\overbrace{\Big[\mathcal{N}_{\boldsymbol q}+1-\mathcal{F}(\varepsilon(\boldsymbol k-\boldsymbol q))\Big]\cdot\delta\Big(\varepsilon(\boldsymbol k)-\varepsilon(\boldsymbol k-\boldsymbol q)-\hbar\omega_{\boldsymbol q}\Big)}^{\text{1-Acoustic Phonon; Emission}}\\&+\overbrace{\Big[\mathcal{N}_{\boldsymbol q}+\mathcal{F}(\varepsilon(\boldsymbol k-\boldsymbol q))\Big]\cdot\delta\Big(\varepsilon(\boldsymbol k)-\varepsilon(\boldsymbol k-\boldsymbol q)+\hbar\omega_{\boldsymbol q}\Big)}^{\text{1-Acoustic Phonon;  Absorption}}\Bigg\}\end{aligned}}_{\Gamma_{\mathrm{ac;ph-1}}(\boldsymbol k, \varepsilon(\boldsymbol k))}\\
&\boldsymbol{+}\underbrace{\begin{aligned} \frac{2\pi}{\hbar}&\sum_{\boldsymbol q,\boldsymbol q^{\prime}}|U^{J}_{\boldsymbol q\boldsymbol q^{\prime};\mathrm{ac}}(\boldsymbol k,\boldsymbol q,\boldsymbol q^{\prime})|^{2}\bigg[
\frac{\mathcal{F}(\varepsilon(\boldsymbol k-\boldsymbol q-\boldsymbol q^{\prime}))}{\mathcal{F}(\varepsilon(\boldsymbol k))}
\bigg]\cdot \\&\Bigg\{\overbrace{\mathcal{N}_{\boldsymbol q}(\mathcal{N}_{\boldsymbol q'} +1) \cdot\delta\Big(\varepsilon(\boldsymbol k)-\varepsilon(\boldsymbol k-\boldsymbol q-\boldsymbol q^{\prime})-\hbar\omega_{\boldsymbol q}+\hbar\omega_{\boldsymbol q^{\prime}}\Big)}^{\text{2-Acoustic Phonon; Both Simultaneous Emission \& Absorption}}\\&
+\overbrace{(\mathcal{N}_{\boldsymbol q}+1)\mathcal{N}_{\boldsymbol q'}  \cdot\delta\Big(\varepsilon(\boldsymbol k)-\varepsilon(\boldsymbol k-\boldsymbol q-\boldsymbol q^{\prime})+\hbar\omega_{\boldsymbol q}-\hbar\omega_{\boldsymbol q^{\prime}}\Big)}^{\text{2-Acoustic Phonon; Both Simultaneous  Absorption \& Emission}}\Bigg\}\end{aligned}}_{\Gamma_{\mathrm{ac;ph-2}}(\boldsymbol k, \varepsilon(\boldsymbol k))} \\
&= \Gamma_{\mathrm{ac;ph-1}}(\boldsymbol k, \varepsilon(\boldsymbol k))+ \Gamma_{\mathrm{ac;ph-2}}(\boldsymbol k, \varepsilon(\boldsymbol k))
\end{split}
\end{equation}

Where $ \Gamma_{\mathrm{ac;ph-1}}(\boldsymbol k, \varepsilon(\boldsymbol k)) $ are broadening function due to one acoustic phonon interaction, $ \Gamma_{\mathrm{ac;ph-2}}(\boldsymbol k, \varepsilon(\boldsymbol k)) $ are broadening function due to two simultaneous acoustic phonon interaction with the electron. We have made two approximations in the calculation first by restricting the scattering process to second order in perturbation expansion and second, $\omega \approx \varepsilon(\boldsymbol k)$ in the $\Gamma(\boldsymbol k, \omega)$ and $A(\boldsymbol k, \omega)$.

\subsection*{\texorpdfstring{Polar Optical Phonon Self-energy $ \Sigma_{\mathrm{op}}(k, E) $}{Polar Optical Phonon Self-energy $Sigma {\mathrm{op}}(k,E)$}}

High energy, high frequency polar optical phonon interact with the electron through the intramolecular lattice vibration to calculate the self-energy Interaction through optical phonon in a weak coupling regime. Intramolecular vibrations amplitude $y_{\boldsymbol n}$ at site $\boldsymbol n$ within the unit cell describes the unit cell's internal deformations. Therefore, molecular eigenstate energy $\varepsilon$ will also fluctuate in addition to $U_C$ and $U_J$ as previously described in the acoustic case \cref{eq-1.12}. In the weak coupling regime, amplitude $y_{\boldsymbol n}$ at site $\boldsymbol n$ only influences the molecular eigenstate energy $\varepsilon_{\boldsymbol n}$ at site $\boldsymbol n$ and nearby site $\boldsymbol n'$ not influence the eigenstate energy with this assumption of $\varepsilon_{\boldsymbol n}=\varepsilon(y_{\boldsymbol n})$, We have assumed at the first order simpler linear dependence, however more complex sub-linear or quadratic dependence can be treated in the proposed framework. In the intramolecular optical vibrations, we have assumed that fluctuations in $ \varepsilon_{\boldsymbol n}$ by $\varepsilon(y_{\boldsymbol n})$ is most robust, and the fluctuations in $U_J$ and  $U_C$ by optical phonon vibration $y_{\boldsymbol n}$ is negligible. Therefore the complete system Hamiltonian modify from \cref{eq-1.10} and \cref{eq-1.13} as,

\begin{equation}\label{eq-1.63}
H=\underbrace{\sum_{\boldsymbol n}\big[\varepsilon(y_{\boldsymbol n})-\mu\big]a_{\boldsymbol n}^{\dagger}a_{\boldsymbol n}}_{H_{\mathrm{C}}}+\underbrace{\sum_{\boldsymbol n,\boldsymbol n^{\prime}} U_J(\boldsymbol n^{\prime}-\boldsymbol n)a_{\boldsymbol n}^{\dagger}a_{\boldsymbol n^{\prime}}}_{H_{\mathrm{J}}} +\underbrace{\sum_{\boldsymbol q}\omega_{\boldsymbol q}b_{\boldsymbol q}^{\dagger}b_{\boldsymbol q}}_{H_{\mathrm{P}}}   
\end{equation}

From \cref{eq-1.63} by expanding up to second-order terms $\varepsilon(y_{\boldsymbol n})$ in powers of $y_{\boldsymbol n}$, Again, The complete system Hamiltonian of electron-phonon gas incorporating one and two-optical phonon processes interaction is,

\begin{equation}\label{eq-1.64}
H_{\mathrm{op-ph}}=H_{\mathrm{S}}+H_{\mathrm{D}}  
\end{equation}

Where we again separate the Hamiltonian in terms of static $ H_{\mathrm{S}} $ and the dynamic part $ H_{\mathrm{D}} $ as for the Green's function perturbation expansion.

The static $ H_{\mathrm{S}} $ part of the complete system Hamiltonian of electron-phonon gas is,

\begin{equation}\label{eq-1.65}
H_{\mathrm{S}}=\underbrace{\Big(\big[\varepsilon(\boldsymbol y) \big]_{\boldsymbol y=0}-\mu\Big)\sum_{\boldsymbol n}a_{\boldsymbol n}^{\dagger}a_{\boldsymbol n}}_{\mathrm{static \ part \ of}\ H_{\mathrm{C}}}+\underbrace{\sum_{\boldsymbol n,\boldsymbol n^{\prime}}U_J(\boldsymbol n^{\prime}-\boldsymbol n)a_{\boldsymbol n}^{\dagger}a_{\boldsymbol n^{\prime}}}_{\mathrm{static \ part \ of}\ H_{\mathrm{J}}}+\underbrace{\sum_{\boldsymbol q}\omega_{\boldsymbol q}b_{\boldsymbol q}^{\dagger}b_{\boldsymbol q}}_{H_{\mathrm{P}}}
\end{equation}

The dynamic $ H_{\mathrm{D}} $ part of the complete system Hamiltonian of electron-phonon gas is,

\begin{equation}\label{eq-1.66}
H_{\mathrm{D}}=\underbrace{\sum_{\boldsymbol n}\Bigg\{
\sum_{\boldsymbol q}U^{C}_{\boldsymbol q;\mathrm{op}}B_{\boldsymbol n,\boldsymbol q;op} 
+\frac{1}{2}\sum_{\boldsymbol q,\boldsymbol q^{\prime}} U^{C}_{\boldsymbol q\boldsymbol q^{\prime};\mathrm{op}}B_{\boldsymbol n,\boldsymbol q;op}B_{\boldsymbol n,\boldsymbol q^{\prime};op}\Bigg\}a_{\boldsymbol n}^{\dagger}a_{\boldsymbol n}}_{\mathrm{dynamic \ part \ of}\ H_{\mathrm{C}}}
\end{equation}

Where we define interaction matrix element $ U^{C}_{\boldsymbol q;\mathrm{op}}, U^{C}_{\boldsymbol q\boldsymbol q^{\prime};\mathrm{op}}, B_{\boldsymbol n,\boldsymbol q;\mathrm{op}}, B_{\boldsymbol n,\boldsymbol q^{\prime};\mathrm{op}} $ such as,

\begin{equation}\label{eq-1.67}
\begin{aligned}
U^{C}_{\boldsymbol q;\mathrm{op}} & =\frac{\hbar^{1/2}}{(2MN\omega_ {\boldsymbol q})^{1/2}}\sum_{i}e_{\boldsymbol q i} \Bigg[\frac{\partial}{\partial \boldsymbol y_{i}}\varepsilon(\boldsymbol y)\Bigg]_{\boldsymbol y=0} \\
U^{C}_{\boldsymbol q\boldsymbol q^{\prime};\mathrm{op}}& =\frac{\hbar}{(4M^{2}N^{2}\omega_{\boldsymbol q}\omega_{\boldsymbol q^{\prime}})^{1/2}}\sum_{i,j}e_{\boldsymbol q i}e_{\boldsymbol q^{\prime}j} \Bigg[\frac{\partial^{2}}{\partial \boldsymbol y_{i}\partial \boldsymbol y_{j}}\varepsilon(\boldsymbol y)\Bigg]_{\boldsymbol y=0} \\
B_{\boldsymbol n,\boldsymbol q;\mathrm{op}} & =[{e}^{i\boldsymbol q\boldsymbol n}b_{\boldsymbol q}+{e}^{-i\boldsymbol q\boldsymbol n}b_{\boldsymbol q}^{\dagger}]\\
B_{\boldsymbol n,\boldsymbol q^{\prime};\mathrm{op}} &=[{e}^{i\boldsymbol q^{\prime}\boldsymbol n}b_{\boldsymbol q^{\prime}} +{e}^{-i\boldsymbol q^{\prime}\boldsymbol n}b_{\boldsymbol q^{\prime}}^{\dagger}] 
\end{aligned}
\end{equation}

\begin{equation}\label{eq-1.68}
\varepsilon_{0}=\Bigg[\varepsilon(\boldsymbol y) \Bigg]_{\boldsymbol y=0} \ \ \varepsilon_{i}=\Bigg[\frac{\partial}{\partial \boldsymbol y_{i}}\varepsilon(\boldsymbol y)\Bigg]_{\boldsymbol y=0}\ \ \varepsilon_{ij}=\Bigg[\frac{\partial^{2}}{\partial \boldsymbol y_{i}\partial \boldsymbol y_{j}}\varepsilon(\boldsymbol y)\Bigg]_{\boldsymbol y=0}   
\end{equation}

The matrix elements $U^{C}_{\boldsymbol q;\mathrm{op}}$ and $U^{C}_{\boldsymbol q\boldsymbol q^{\prime};\mathrm{op}}$ coupling is a molecular dipole potential type of interaction. Hereinafter the acoustic phonon treatment for the Green's function from \cref{eq-1.42} and perturbation expansion treatment of Hamiltonian $H_{\mathrm{D}}$ from \cref{eq-1.66} and following the procedure of \cref{eq-1.53}. The self-energy $ \Sigma_{\mathrm{op}}(\boldsymbol k, E) $ for intramolecular vibrations is,

\begin{equation}\label{eq-1.69}
\begin{split}  
\Sigma(\boldsymbol k, E) &=\underbrace{\sum_{\boldsymbol q}|U^{C}_{\boldsymbol q;\mathrm{op}}|^{2}\cdot\Bigg\{\overbrace{\frac{\mathcal{N}_{\boldsymbol q}+1-\mathcal{F}(\varepsilon(\boldsymbol k-\boldsymbol q))}{E-\varepsilon(\boldsymbol k-\boldsymbol q)-\hbar\omega_{\boldsymbol q}}}^{\text{1-Optical Phonon; Emission}}+\overbrace{\frac{\mathcal{N}_{\boldsymbol q}+\mathcal{F}(\varepsilon(\boldsymbol k-\boldsymbol q))}{E-\varepsilon(\boldsymbol k-\boldsymbol q)+\hbar\omega_{\boldsymbol q}}}^{\text{1-Optical Phonon; Absorption}}\Bigg\}}_{\Sigma_{\mathrm{op;ph-1}}(\boldsymbol k, E)} \\&
\boldsymbol{+}\underbrace{\begin{aligned}\frac{1}{2}\sum_{\boldsymbol q,\boldsymbol q^{\prime}}|U^{C}_{\boldsymbol q\boldsymbol q^{\prime};\mathrm{op}}|^{2}
\cdot \Bigg\{\overbrace{\frac{\mathcal{F}(\varepsilon(\boldsymbol k-\boldsymbol q-\boldsymbol q^{\prime}))}{\mathcal{F}(\varepsilon(\boldsymbol k-\boldsymbol q-\boldsymbol q^{\prime})+\hbar\omega_{\boldsymbol q}+\hbar\omega_{\boldsymbol q^{\prime}})} \cdot\frac{\mathcal{N}_{\boldsymbol q}\mathcal{N}_{\boldsymbol q^{\prime}}}{E-\varepsilon(\boldsymbol k-\boldsymbol q-\boldsymbol q^{\prime})-\hbar\omega_{\boldsymbol q}-\hbar\omega_{\boldsymbol q^{\prime}}}}^{\text{2-Optical Phonon; Both Simultaneous Emission}}\\+\overbrace{\frac{\mathcal{F}(\varepsilon(\boldsymbol k-\boldsymbol q-\boldsymbol q^{\prime}))}{\mathcal{F}(\varepsilon(\boldsymbol k-\boldsymbol q-\boldsymbol q^{\prime})-\hbar\omega_{\boldsymbol q}-\hbar\omega_{\boldsymbol q^{\prime}})}\cdot\frac{(\mathcal{N}_{\boldsymbol q}+1)(\mathcal{N}_{\boldsymbol q^{\prime}}+1)}{E-\varepsilon(\boldsymbol k-\boldsymbol q-\boldsymbol q^{\prime})+\hbar\omega_{\boldsymbol q}+\hbar\omega_{\boldsymbol q^{\prime}}}}^{\text{2-Optical Phonon; Both Simultaneous Absorption}} \\ +\overbrace{\frac{\mathcal{F}(\varepsilon(\boldsymbol k-\boldsymbol q-\boldsymbol q^{\prime}))}{\mathcal{F}(\varepsilon(\boldsymbol k-\boldsymbol q-\boldsymbol q^{\prime})+\hbar\omega_{\boldsymbol q}-\hbar\omega_{\boldsymbol q^{\prime}})}\cdot\frac{\mathcal{N}_{\boldsymbol q}(\mathcal{N}_{\boldsymbol q^{\prime}}+1)}{E-\varepsilon(\boldsymbol k-\boldsymbol q-\boldsymbol q^{\prime})-\hbar\omega_{\boldsymbol q}+\hbar\omega_{\boldsymbol q^{\prime}}}}^{\text{2-Optical Phonon; Both Simultaneous Emission \& Absorption}}\\+\overbrace{\frac{\mathcal{F}(\varepsilon(\boldsymbol k-\boldsymbol q-\boldsymbol q^{\prime}))}{\mathcal{F}(\varepsilon(\boldsymbol k-\boldsymbol q-\boldsymbol q^{\prime})-\hbar\omega_{\boldsymbol q}+\hbar\omega_{\boldsymbol q^{\prime}})}\cdot\frac{(\mathcal{N}_{\boldsymbol q}+1)\mathcal{N}_{\boldsymbol q^{\prime}}}{E-\varepsilon(\boldsymbol k-\boldsymbol q-\boldsymbol q^{\prime})+\hbar\omega_{\boldsymbol q}-\hbar\omega_{\boldsymbol q^{\prime}}}}^{\text{2-Optical Phonon; Both Simultaneous Absorption \& Emission}}\Bigg\}\end{aligned}}_{\Sigma_{\mathrm{op;ph-2}}(\boldsymbol k, E)}\\
&= \Sigma_{\mathrm{op;ph-1}}(\boldsymbol k, E) + \Sigma_{\mathrm{op;ph-2}}(\boldsymbol k, E)
\end{split}
\end{equation}

Where $ \Sigma_{\mathrm{op;ph-1}}(\boldsymbol k, E) $ are self-energy due to one optical phonon interaction, $ \Sigma_{\mathrm{op;ph-2}}(\boldsymbol k, E) $ are self-energy due to two simultaneous optical phonon interaction with the electron. There are six contribution term from the one and two-phonon process, where matrix element $  \mathcal{N}_{\boldsymbol q} $, $ \mathcal{F}(\varepsilon) $, and $ \varepsilon(\boldsymbol k-\boldsymbol q-\boldsymbol q^{\prime}) $  defined as previously. For the two-phonon processes, the probability of interaction in which both phonons emit energy $\delta(\varepsilon(\boldsymbol k)-\varepsilon(\boldsymbol k-\boldsymbol q-\boldsymbol q^{\prime})-\hbar\omega_{\boldsymbol q}-\hbar\omega_{\boldsymbol q^{\prime}})$ or absorb energy $ \delta(\varepsilon(\boldsymbol k)-\varepsilon(\boldsymbol k-\boldsymbol q-\boldsymbol q^{\prime})+\hbar\omega_{\boldsymbol q}+\hbar\omega_{\boldsymbol q^{\prime}})$, simultaneously is even minuscule. By eliminating two terms in the aforementioned two phonon interactions and with this approximation, $\Gamma(\boldsymbol k, \varepsilon(\boldsymbol k))$ is,

\begin{equation}\label{eq-1.70}
\begin{split}  
\Gamma(\boldsymbol k, \varepsilon(\boldsymbol k)) &=\underbrace{\begin{aligned}\frac{2\pi}{\hbar}\sum_{\boldsymbol q}|U^{C}_{\boldsymbol q;\mathrm{op}}|^{2}\Bigg\{&\overbrace{\Big[\mathcal{N}_{\boldsymbol q}+1-\mathcal{F}(\varepsilon(\boldsymbol k-\boldsymbol q))\Big]\cdot\delta\Big(\varepsilon(\boldsymbol k)-\varepsilon(\boldsymbol k-\boldsymbol q)-\hbar\omega_{\boldsymbol q}\Big)}^{\text{1-Optical Phonon; Emission}}\\&+\overbrace{\Big[\mathcal{N}_{\boldsymbol q}+\mathcal{F}(\varepsilon(\boldsymbol k-\boldsymbol q))\Big]\cdot\delta\Big(\varepsilon(\boldsymbol k)-\varepsilon(\boldsymbol k-\boldsymbol q)+\hbar\omega_{\boldsymbol q}\Big)}^{\text{1-Optical Phonon; Absorption}}\Bigg\}\end{aligned}}_{\Gamma_{\mathrm{op;ph-1}}(\boldsymbol k, \varepsilon(\boldsymbol k))}\\ 
&\boldsymbol{+}\underbrace{\begin{aligned} \frac{2\pi}{\hbar}&\sum_{\boldsymbol q,\boldsymbol q^{\prime}}|U^{C}_{\boldsymbol q\boldsymbol q^{\prime};\mathrm{op}}|^{2}\bigg[
\frac{\mathcal{F}(\varepsilon(\boldsymbol k-\boldsymbol q-\boldsymbol q^{\prime}))}{\mathcal{F}(\varepsilon(\boldsymbol k))}
\bigg]\cdot \\&\Bigg\{\overbrace{\mathcal{N}_{\boldsymbol q}(\mathcal{N}_{\boldsymbol q'} +1) \cdot\delta\Big(\varepsilon(\boldsymbol k)-\varepsilon(\boldsymbol k-\boldsymbol q-\boldsymbol q^{\prime})-\hbar\omega_{\boldsymbol q}+\hbar\omega_{\boldsymbol q^{\prime}}\Big)}^{\text{2-Optical Phonon; Emission \& Absorption}}\\&
+\overbrace{(\mathcal{N}_{\boldsymbol q}+1)\mathcal{N}_{\boldsymbol q'}  \cdot\delta\Big(\varepsilon(\boldsymbol k)-\varepsilon(\boldsymbol k-\boldsymbol q-\boldsymbol q^{\prime})+\hbar\omega_{\boldsymbol q}-\hbar\omega_{\boldsymbol q^{\prime}}\Big)}^{\text{2-Optical Phonon; Absorption \& Emission}}\Bigg\}\end{aligned}}_{\Gamma_{\mathrm{op;ph-2}}(\boldsymbol k, \varepsilon(\boldsymbol k))} \\
&= \Gamma_{\mathrm{op;ph-1}}(\boldsymbol k, \varepsilon(\boldsymbol k))+ \Gamma_{\mathrm{op;ph-2}}(\boldsymbol k, \varepsilon(\boldsymbol k))
\end{split}
\end{equation}

\section*{Non-Equilibrium Green's Function formalism}

The non-equilibrium Green's function formalism provides a framework to calculate the non-equilibrium carrier statistics in the externally biased connected device as an ensemble average of single-particle correlation Green's Function. Keldysh, Kadanoff, and Baym developed the non-equilibrium Green's function formalism in 1960. \cite{baym_conservation_1961,keldysh_diagram_1964} The adaption of non-equilibrium Green's function formalism to semiconductor devices was first demonstrated by Datta, Lake, and Klimeck in 1997 and later in 2002 by Wacker. \cite{datta_electronic_1997,lake_single_1997,wacker_semiconductor_2002} The non-equilibrium Green's function formalism can study the time evolution of many-body quantum fields, either in thermodynamic equilibrium or non-equilibrium. \cite{danielewicz_quantum-I_1984,danielewicz_quantum-II_1984,lake_single_1997,gebauer_current_2004,burke_density_2005,frederiksen_inelastic_2007} These quantum fields are constituted by carriers such as electrons, phonons, and spin in semiconductor devices. In the non-equilibrium Green's function formalism, The Schr\"{o}dinger-Poisson equation solved with the open boundary conditions under the non-equilibrium Fermi contact potentials with the coupling to the contacts and energy dissipative scattering processes. \cite{landauer_spatial_1957,beenakker_quantum_1991,weinmann_quantum_1994,datta_nanoscale_2000,lundstrom_fundamentals_2009,anantram_modeling_2008,kubis_quantum_2009,luisier_quantum_2006,hirsbrunner_review_2019}  In the device simulation, the carrier densities $\mathfrak N(\boldsymbol n, t)$ and $\mathfrak J(\boldsymbol n, t)$ the current  densities are two most important physical observable quantities. By solving the equation of motion, lesser Green's functions $G^{<}(\boldsymbol n, t;\boldsymbol n^{\prime}, t^{\prime})$ are calculated which is only possible because Green's function $G(\boldsymbol n_{1}, t_{1};\boldsymbol n_{2}, t_{2})$ and the self-energy $\Sigma(\boldsymbol n_{1}, t_{1};\boldsymbol n_{2}, t_{2})$ has the same symmetry properties, and that is required for the calculation of carrier densities. This is shown by Craig et.al. \cite{craig_perturbation_1968} and later prove by Danielewicz et.al. \cite{danielewicz_quantum_1984} By using the Langreth theorem, \cite{haug_quantum_2008} expansion of eigenfunction, and for ($t-t^{\prime}$) time difference taking the Fourier transform of the Green's functions, The closed set of equations of motion is,

\begin{equation}\label{eq-B22}
\begin{aligned}
{\it ih} \frac{\mathrm{d}}{\mathrm{d}t}G_{\boldsymbol n \boldsymbol m}^{<}(\boldsymbol{k},tt^{\prime})-\sum_{\boldsymbol l}h_{\boldsymbol n \boldsymbol l}G_{\boldsymbol l \boldsymbol m}^{<}(\boldsymbol{k},tt^{\prime}) &= \sum_{\boldsymbol l}\int_{t_{0}}^{\infty}\mathrm{d}t_{1}\Sigma_{\boldsymbol n \boldsymbol l}^{R}(\boldsymbol{k},tt_{1})G_{\boldsymbol l \boldsymbol m}^{<}(\boldsymbol{k},t_{1}t^{\prime}) \\&
+\sum_{\boldsymbol l}\int_{t_{0}}^{\infty}\mathrm{d}t_{1}\Sigma_{\boldsymbol n \boldsymbol l}^{<}(\boldsymbol{k},tt_{1})G_{\boldsymbol l \boldsymbol m}^{A}(\boldsymbol{k},t_{1}t^{\prime})
\end{aligned}
\end{equation}

\begin{equation}\label{eq-B23}
\begin{aligned}
-ih\frac{\mathrm{d}}{\mathrm{d}t^{\prime}}G_{\boldsymbol n \boldsymbol m}^{<}(\boldsymbol{k},tt^{\prime})-\sum_{\boldsymbol l}G_{\boldsymbol n \boldsymbol l}^{<}(\boldsymbol{k},tt^{\prime})h_{\boldsymbol l \boldsymbol m} & = \sum_{\boldsymbol l}\int_{t_{0}}^{\infty}\mathrm{d}t_{1}G_{\boldsymbol n \boldsymbol l}^{R}(\boldsymbol{k},tt_{1})\Sigma_{\boldsymbol l \boldsymbol m}^{<}(\boldsymbol{k},t_{1}t^{\prime}) \\&
+\sum_{\boldsymbol l}\int_{t_{0}}^{\infty}\mathrm{d}t_{1}G_{\boldsymbol n \boldsymbol l}^{<}(\boldsymbol{k},tt_{1})\Sigma_{\boldsymbol l \boldsymbol m}^{A}(\boldsymbol{k},t_{1}t^{\prime})\
\end{aligned}
\end{equation}

As Hamiltonian is hermitian $h_{\boldsymbol m \boldsymbol l}^{*}=h_{\boldsymbol l \boldsymbol m}$ and where $h_{\boldsymbol n \boldsymbol m}=\int \mathrm{d}\boldsymbol{r}\phi_{\boldsymbol n}^{*}(\boldsymbol{r})H_{\mathrm{S}}(\boldsymbol{r})\phi_{m}(\boldsymbol{r})$, and $ H_{\mathrm{S}} $ is purely electronic part of static Hamiltonian from \cref{eq-1.29} as also subsequently mentioned in \cref{eq-1.54}.

Retarded and advanced Green's function $G^{R,A}$, and lesser self-energy $\Sigma^{<}$, and retarded and advanced self-energy $\Sigma^{R,A}$ is required to solve the close-set of the equations. These Green's functions and self-energy calculate by solving the equation of motion. Similarly, $R$'s can be replaced by $A' \mathrm{s}$ and by similarly fashion $G^{A}$ equations can be obtained,

\begin{equation}\label{eq-B24}
{i\hbar} \frac{\mathrm{d}}{\mathrm{d}t}G_{\boldsymbol n \boldsymbol m}^{R}(\boldsymbol{k},tt^{\prime})-\sum_{\boldsymbol l}h_{\boldsymbol n \boldsymbol l}G_{\boldsymbol l \boldsymbol m}^{R}(\boldsymbol{k},tt^{\prime}) = \delta_{\boldsymbol n \boldsymbol m}(tt^{\prime})+\sum_{\boldsymbol l}\int_{t_{0}}^{\infty}\mathrm{d}t_{1}\Sigma_{\boldsymbol nl}^{R}(\boldsymbol{k},tt_{1})G_{\boldsymbol l \boldsymbol m}^{R}(\boldsymbol{k},t_{1}t^{\prime})
\end{equation}
Where $ \delta_{\boldsymbol n \boldsymbol m}(tt^{\prime}) $ is defined as, 
\begin{equation}\label{eq-B25}
\delta_{\boldsymbol n \boldsymbol m}(tt^{\prime})=
\sum_{\boldsymbol l}\int_{t_{0}}^{\infty}\mathrm{d}t_{1}[G_{\boldsymbol n \boldsymbol l}^{R}(\boldsymbol{k},tt_{1})]^{-1}G_{\boldsymbol l \boldsymbol m}^{R}(\boldsymbol{k},t_{1}t^{\prime})
\end{equation}

After simplification by Langreth theorem. The lesser Green's function, the central equation is with the coupling between $G^{<}$ and $G^{R,A}$,

\begin{equation}\label{eq-B26}
G_{\boldsymbol n \boldsymbol m}^{<}(\boldsymbol{k},tt^{\prime}) =\sum_{\boldsymbol l,\boldsymbol v}\int_{t_{0}}^{\infty}\mathrm{d}t_{1}\int_{t_{0}}^{\infty}\mathrm{d}t_{2}G_{\boldsymbol n \boldsymbol l}^{R}(\boldsymbol{k},tt_{1})\Sigma_{\boldsymbol l \boldsymbol v}^{<}(\boldsymbol{k},t_{1}t_{2})G_{\boldsymbol v \boldsymbol m}^{A}(\boldsymbol{k},t_{2}t^{\prime})
\end{equation}

$G_{\boldsymbol n \boldsymbol m}(\boldsymbol{k},tt^{\prime})$ Green's functions depend upon time ($t$) and ($t^{\prime}$), but once non-equilibrium system reach to stationary state solution the Green's functions depend on the time difference ($t-t^{\prime}$).Through the Langreth theorem, ($t$) and ($t^{\prime}$) upon reaching the stationary state, no longer reside on the imaginary time contour. Furthermore, using the advantage of Fourier transform Green's functions for the time difference $t-t^{\prime}$ can be modified in the energy. Similar relationship hold for $G^{R,A}$,
\begin{equation}\label{eq-B27}
G_{\boldsymbol n \boldsymbol m}^{<}(\boldsymbol{k},E) = \int \mathrm{d}(t-t^{\prime})e^{iE(t-t^{\prime})/\hslash}G_{\boldsymbol n \boldsymbol m}^{<}(\boldsymbol{k},t-t^{\prime}) 
\end{equation}
Therefore, in the stationary state, equations of motion of the quantum system can be simplified as follow,
\begin{equation}\label{eq-B28}
\begin{aligned}
EG_{\boldsymbol n \boldsymbol m}^{<}(\boldsymbol{k},E)-\sum_{\boldsymbol l}h_{\boldsymbol n \boldsymbol l}G_{\boldsymbol l \boldsymbol m}^{<}(\boldsymbol{k},E) & = \sum_{\boldsymbol l}\Sigma_{\boldsymbol n \boldsymbol l}^{R}(\boldsymbol{k},E)G_{\boldsymbol l \boldsymbol m}^{<}(\boldsymbol{k},E)+\sum_{\boldsymbol l}\Sigma_{\boldsymbol n \boldsymbol l}^{<}(\boldsymbol{k},E)G_{\boldsymbol l \boldsymbol m}^{A}(\boldsymbol{k},E) \\ 
EG_{\boldsymbol n \boldsymbol m}^{<}(\boldsymbol{k},E)-\sum_{\boldsymbol l}G_{\boldsymbol n \boldsymbol l}^{<}(\boldsymbol{k},E)h_{\boldsymbol l \boldsymbol m} &  = \sum_{\boldsymbol l}G_{\boldsymbol n \boldsymbol l}^{R}(\boldsymbol{k},E)\Sigma_{\boldsymbol l \boldsymbol m}^{<}(\boldsymbol{k},E)+\sum_{\boldsymbol l}G_{\boldsymbol n \boldsymbol l}^{<}(\boldsymbol{k},E)\Sigma_{{\boldsymbol l \boldsymbol m}}^{A}(\boldsymbol{k},E) \\
G_{\boldsymbol n \boldsymbol m}^{<}(\boldsymbol{k},E) & = \sum_{\boldsymbol l,\boldsymbol v}G_{\boldsymbol n \boldsymbol l}^{R}(\boldsymbol{k},E)\Sigma_{lv}^{<}(\boldsymbol{k},E)G_{vm}^{A}(\boldsymbol{k},E) \\ 
G_{\boldsymbol n \boldsymbol m}^{<}(\boldsymbol{k},E) & = -[G_{\boldsymbol m\boldsymbol n}^{<}(\boldsymbol{k},E)]^{\dagger} \\
EG_{\boldsymbol n \boldsymbol m}^{R}(\boldsymbol{k},E)-\sum_{\boldsymbol l}h_{\boldsymbol n \boldsymbol l}G_{\boldsymbol l \boldsymbol m}^{R}(\boldsymbol{k},E) & = \delta_{\boldsymbol n \boldsymbol m}+\sum_{\boldsymbol l}\Sigma_{\boldsymbol n \boldsymbol l}^{R}(\boldsymbol{k},E)G_{\boldsymbol l \boldsymbol m}^{R}(\boldsymbol{k},E) \\ 
G_{\boldsymbol n \boldsymbol m}^{A}(\boldsymbol{k},E) & = [G_{\boldsymbol m\boldsymbol n}^{R}(\boldsymbol{k},E)]^{\dagger} \\ 
G_{\boldsymbol m\boldsymbol n}^{R}(\boldsymbol{k},E)-G_{\boldsymbol m\boldsymbol n}^{A}(\boldsymbol{k},E) & = G_{\boldsymbol m\boldsymbol n}^{>}(\boldsymbol{k},E)-G_{\boldsymbol m\boldsymbol n}^{<}(\boldsymbol{k},E) \\ 
A(\boldsymbol{k},E) & = i(G_{\boldsymbol m\boldsymbol n}^{R}(\boldsymbol{k},E)-G_{\boldsymbol m\boldsymbol n}^{A}(\boldsymbol{k},E)) \\
\Sigma^{R,<}(\boldsymbol{k},E) &=\Sigma^{R,<}_{\mathrm{ac;ph-1}}(\boldsymbol k, E) + \Sigma^{R,<}_{\mathrm{ac;ph-2}}(\boldsymbol k, E)+
\Sigma^{R,<}_{\mathrm{op;ph-1}}(\boldsymbol k, E) + \Sigma^{R,<}_{\mathrm{op;ph-2}}(\boldsymbol k, E)
\end{aligned}
\end{equation}

Where $G^{R}$, $G^{<}$, and for the different relevant interactions self-energies $\Sigma_{\boldsymbol n \boldsymbol m}^{R,<}(\boldsymbol{k}, E)$, have to calculate for the relevant lattice point index on the grid in this \cref{eq-B28} coupled system of equations.  $\Sigma_{\boldsymbol n \boldsymbol m}^{R,<}(\boldsymbol{k},E)$ is self-energies sum due to one phonon acoustic self-energy $ \Sigma^{R,<}_{\mathrm{ac;ph-1}}(\boldsymbol k, E) $, two phonon acoustic self-energy $ \Sigma^{R,<}_{\mathrm{ac;ph-2}}(\boldsymbol k, E) $, from \cref{eq-1.53}, and  one phonon optical self-energy $ \Sigma^{R,<}_{\mathrm{op;ph-1}}(\boldsymbol k, E) $, two phonon optical self-energy $ \Sigma^{R,<}_{\mathrm{op;ph-2}}(\boldsymbol k, E)  $ from \cref{eq-1.69}.  This couple set of equations is computational intensive, for a device with tight-binding model Hamiltonian $h$, matrix element is $N_{H}\times N_{H}$, if the device contain lattice $N_{L}$ points, $k$ wavevector $N_{k}$ points, and energy $N_{E}$ points. The functions $G^{R}, G^{<}, \Sigma^{R}$, and $\Sigma^{<}$ size to calculate and store is $N_{L}\times N_{L}\times N_{k}\times N_{E}\times N_{H}\times N_{H}$. The size of the matrices is $N\mathrm{x}N$, as $N$ denotes the total number of atoms in the device. The retarded Green's function computed by the Recursive Green's function (RGF) algorithm of complexity $ \mathcal{O}(N)$, \cite{haydock_recursive_1980,teichert_improved_2017,thouless_conductivity_1981,mackinnon_calculation_1985,lee_nonequilibrium_2002} This exploits the property of block tri-diagonal matrix structure with minimal computational resources compared to the massive matrix inversion operation of complexity $ \mathcal{O}(N^3)$. The algorithm to calculate the couple set of equation start with an initial value of $G^{<}$ and $G^{R}$. The system with no interactions is taken as the initial value of free Green's function $G^{0R}$ and $G^{0<}$. For all the lattice, $H$ $E$, $k$, points, self-energies $\Sigma^{<}$and $\Sigma^{R}$ derived by calculating the actual $G^{<}$ and $G^{R}$. The calculation of new $G^{R}$ and $G^{<}$ is performed by using the self-energies $\Sigma^{<}$ and $\Sigma^{R}$ values from the previous iteration. This loop continues until Jacobi iterations reach convergence. With this $G^{<}$, the actual carrier density is obtained and solved in the Poisson equation loop. The carrier-carrier interactions will be approximately treated on the mean-field level with the Hartree self-energies as part of the Poisson potential. The algorithm restarted with the newly calculated Poisson potential and continued to run till convergence was achieved. The carrier density and current density calculation are computationally expensive in Green's function formalism. \cite{anantram_modeling_2008,rahman_theory_2003,cauley_distributed_2011} Speed up can be achieved by parallelizing the self energies computation and neglecting some higher-order parts of the self-energies.

\subsection*{\texorpdfstring{Carrier Density $ \mathfrak{N}(\boldsymbol n, t) $}{Carrier Density $ \mathfrak{N}(n,t)$}}

In the non-equilibrium Green's functions formalism, the carrier density $\mathfrak{N} (\boldsymbol n, t)$ is,

\begin{equation}\label{eq-B42}
\mathfrak{N}(\boldsymbol n, t) = -i\hbar G(\boldsymbol n, t;\boldsymbol n^{\prime}, t^{\prime})
\end{equation}

In the stationary regime of non-equilibrium Green's functions solution eigenfunction expansion gives the carrier density per unit area $ {A} $ with \cref{eq-1.42}  and \cref{eq-1.60},

\begin{equation}\label{eq-B43}
\mathfrak{N}(\boldsymbol n, t) = -\frac{i}{A}\sum_{\boldsymbol{k}}\sum_{\boldsymbol{n}, \boldsymbol n^{\prime}}\int\frac{\mathrm{d}E}{2\pi}G_{\boldsymbol{n}, \boldsymbol{n^{\prime}}}(\boldsymbol k, E)
\phi_{\boldsymbol k}(\boldsymbol r-\boldsymbol n)\phi_{\boldsymbol k}^{*}(\boldsymbol r-\boldsymbol n^{\prime}) 
\end{equation}

\subsection*{\texorpdfstring{Current Density $\mathfrak J(\boldsymbol n, t)$}{Current Density $\mathfrak J(n,t)$}}

The current density calculation is more computationally expensive compared to carrier density. The current density $\mathfrak J(\boldsymbol n, t)$ is related to carrier density $\mathfrak{N}(\boldsymbol n, t)$ by the continuity equation,

\begin{equation}\label{eq-B44}
\frac{\mathrm{d}}{\mathrm{d}t}\mathfrak N (\boldsymbol n, t)+\mathrm{d}\mathrm{i}\mathrm{v}\mathfrak J(\boldsymbol n, t) = 0
\end{equation}
The carrier density $\mathfrak N(\boldsymbol n, t)$ is derived from the lesser Green's functions $G(\boldsymbol n, t;\boldsymbol n^{\prime}, t^{\prime})$ as,

\begin{equation}\label{eq-B45}
\frac{\mathrm{d}}{\mathrm{d}t}\mathfrak N(\boldsymbol n, t) = \lim_{t'\rightarrow t}(-i\hbar)e\bigg[\frac{\mathrm{d}}{\mathrm{d}t}G^{<}(\boldsymbol n, t,\boldsymbol n^{\prime}, t^{\prime})+\frac{\mathrm{d}}{\mathrm{d}t'}G^{<}(\boldsymbol n, t,\boldsymbol n^{\prime}, t^{\prime})\bigg]
\end{equation}

In the case of the $z$-directional current transport and assuming the  eigenfunctions centered around one lattice point, \cite{lake_single_1997} by using \cref{eq-B44}, 
\begin{equation}\label{eq-B46}
\begin{aligned}
\frac{\mathrm{d}}{\mathrm{d}t}\mathfrak N(\boldsymbol n, t) & = \frac{e}{A\triangle}\sum_{\boldsymbol{k}}\frac{\mathrm{d}}{\mathrm{d}t}\big\langle a_{\boldsymbol n,\boldsymbol{k}}^{\dagger}(t)a_{\boldsymbol n,\boldsymbol{k}}(t)\big\rangle, \; \mathrm{where} \mathrm{e} +\mathrm{ive} \; \mathrm{sign \; for} \; \mathrm{hole}, \;\mathrm{e} -\mathrm{ive} \; \mathrm{sign \; for} \; \mathrm{electron}\; \\
& =\lim_{t'\rightarrow t}(-i\hbar)\frac{e}{A\triangle}\sum_{\boldsymbol{k}}\sum_{\boldsymbol{n}, \boldsymbol n^{\prime}}\bigg[\frac{\mathrm{d}}{\mathrm{d}t}G_{\boldsymbol n \boldsymbol n^{\prime}}^{<}(\boldsymbol{k},tt^{\prime})+\frac{\mathrm{d}}{\mathrm{d}t'}G_{\boldsymbol n^{\prime} \boldsymbol n}^{<}(\boldsymbol{k},tt^{\prime})\bigg] \\
& = -\frac{\mathfrak J_{\boldsymbol n}(t)-\mathfrak J_{\boldsymbol n-1}(t)}{\triangle}
\end{aligned}
\end{equation}

Where $\mathfrak N(\boldsymbol n, t)$ is charge density and $\mathfrak J(\boldsymbol n, t)$  is the current density at place $\boldsymbol n$, $a_{\boldsymbol n,\boldsymbol{k}}(t)$ annihilates an electron at position $\boldsymbol n$, with state $\boldsymbol{k}$, at time $t$ with in a volume $V=A\Delta$, $e$ is negative charge for electrons and the positive charge for holes transport, $A$ the area in the {\it xy} plane, $\Delta=\boldsymbol n-(\boldsymbol {n-1})$, $a_{\boldsymbol n,\boldsymbol{k}}^{\dagger}$ creates an electron at position $\boldsymbol n$, with state $\boldsymbol{k}$, at time $t$ with in a volume $V=A\Delta$, 
$\mathfrak J(\boldsymbol n, t)$ is the current density between point $\boldsymbol n$ and $\boldsymbol {n+1}$, The current density $\mathfrak J(\boldsymbol n, t)$ is calculated by taking the two derivatives of the lesser Green's function $G_{\boldsymbol n \boldsymbol n^{\prime}}^{<}(\boldsymbol{k}, {\it tt}')$  from \cref{eq-B22} and \cref{eq-B23} and inserting into \cref{eq-B46} yield,

\begin{equation}\label{eq-B47}
\begin{aligned}
\frac{\mathrm{d}}{\mathrm{d}t}\mathfrak N(\boldsymbol n, t) & = -\frac{e}{A\triangle}\sum_{\boldsymbol{k}}\sum_{\boldsymbol m}\bigg[h_{\boldsymbol n \boldsymbol m}G_{\boldsymbol m \boldsymbol n}^{<}(\boldsymbol{k},tt)-G_{\boldsymbol n \boldsymbol m}^{<}(\boldsymbol{k},tt)h_{\boldsymbol m \boldsymbol n}\bigg] \\
& = -\frac{\mathfrak J_{\boldsymbol n}(t)-\mathfrak J_{\boldsymbol n-1}(t)}{\triangle}
\end{aligned}
\end{equation}

By decomposing \cref{eq-B47}, $\mathfrak J_{\boldsymbol n}$ and $\mathfrak J_{\boldsymbol n-1}$ are separated. An ansatz given by Caroli \textit{et al.} \cite{caroli_direct_1971} The current $\mathfrak J_{\boldsymbol n}$ between point $\boldsymbol n-1$ and points $\boldsymbol n$ defined as the difference between the flow of Fermions from right to left and from left to right. Therefore, for stationary as well as for non-stationary cases when scattering mechanisms are present, the current $\mathfrak J_{\boldsymbol n}(t)$ is given by,

\begin{equation}\label{eq-B48}
\mathfrak J_{\boldsymbol n}(t) = -\frac{e}{A}\sum_{\boldsymbol l\geq \boldsymbol n+1}\sum_{\boldsymbol m\leq \boldsymbol n}\sum_{\boldsymbol{k}}\bigg[h_{\boldsymbol l \boldsymbol m}G_{\boldsymbol m \boldsymbol l}^{<}(\boldsymbol{k},tt)-G_{\boldsymbol l \boldsymbol m}^{<}(\boldsymbol{k},tt)h_{\boldsymbol m \boldsymbol l}\bigg] 
\end{equation}

for $\mathfrak J_{\boldsymbol n}$ \cref{eq-B48} with similar expression for $\mathfrak J_{\boldsymbol n-1}$ satisfies the \cref{eq-B47}.

The current is everywhere the same in the stationary state of the device. Therefore, we can choose where to compute the current, assuming that contacts are big and in thermal equilibrium.

We have assumed that in-between active parts of the device and contacts, no scattering occurs. Current is calculated at the interface of the active region and contact. By choosing this place, the index $\boldsymbol l$ corresponds to points in the active region, and index $\boldsymbol m$ corresponds to the contact points. By using  \cref{eq-B27}, \cref{eq-B48} $\mathfrak J_{\boldsymbol n}$ is simplified as,

\begin{equation}\label{eq-B49}
\begin{aligned}
\mathfrak J_{\boldsymbol n} & = -\frac{e}{\hbar A}\sum_{\boldsymbol l\geq \boldsymbol n+1}\sum_{\boldsymbol m\leq \boldsymbol n}\sum_{\boldsymbol{k}}\int\frac{\mathrm{d}E}{2\pi}\bigg[h_{\boldsymbol l\boldsymbol m}G_{\boldsymbol m\boldsymbol l}^{<}(\boldsymbol{k},E)-G_{\boldsymbol l\boldsymbol m}^{<}(\boldsymbol{k},E)h_{\boldsymbol m\boldsymbol l}\bigg] \\
& = -\frac{e}{\hbar A}\sum_{\boldsymbol l\geq \boldsymbol n+1}\sum_{\boldsymbol m\leq \boldsymbol n}\sum_{\boldsymbol{k}}\int\frac{\mathrm{d}E}{2\pi}2\mathbb{R}e\bigg\{h_{\boldsymbol l\boldsymbol m}G_{\boldsymbol m\boldsymbol l}^{<}(\boldsymbol{k},E)\bigg\}
\end{aligned}
\end{equation}
Using \cref{eq-B28} in the above equation second line is evaluated. $G_{\boldsymbol m\boldsymbol l}^{<}(\boldsymbol{k},E)$ is simplified by using the appropriate boundary conditions. Where $\boldsymbol l$ belongs to the active part of the device, with define carriers Fermi distribution $\boldsymbol m$ belongs to any point in the contacts, and $\boldsymbol n$ belongs to the interface between both regions by using the corresponding boundary conditions for $h_{\boldsymbol m \boldsymbol l,\boldsymbol l \boldsymbol m}$ and with equilibrium contacts assumption. In the contacts, the Fermi distribution is in equilibrium and by using the fluctuation-dissipation theorem. The current density \cref{eq-B49} simplified to,

\begin{equation}\label{eq-B50}
\mathfrak J_{n}=\frac{e}{\hslash A}\sum_{\boldsymbol l\geq \boldsymbol n+1}\sum_{\boldsymbol l_{1}\geq \boldsymbol n+1}\sum_{\boldsymbol{k}}\int\frac{\mathrm{d}E}{2\pi}\Gamma_{\boldsymbol l \boldsymbol l_{1}}^{\mathrm{Contact}}(\boldsymbol{k},E)\bigg[f^{\mathrm{Contact}}(E)A_{\boldsymbol l_{1}\boldsymbol l}(\boldsymbol{k},E)+iG_{\boldsymbol l_{1}\boldsymbol l}^{<}(\boldsymbol{k},E)\bigg]
\end{equation}
Where $f^{Cont}(E)$ is contact Fermi-distribution function,
\begin{equation}\label{eq-B51}
\Gamma_{\boldsymbol l\boldsymbol l_{1}}^{\mathrm{Contact}}(\boldsymbol{k},E) = \sum_{\boldsymbol m\leq \boldsymbol n}\sum_{\boldsymbol m_{1}\leq \boldsymbol n}h_{\boldsymbol l\boldsymbol m}{A_{\boldsymbol m\boldsymbol m_{1}}}(\boldsymbol{k},E)h_{\boldsymbol m_{1}\boldsymbol l_{1}}
\end{equation}

In the device's active region, with many-body interactions and scattering processes, \cref{eq-B50} is still valid, and only one assumption made to derive the \cref{eq-B50} that self-energies between active region and contacts disappear. This equation corresponds to equation (5) of the Landauer formula for the current through an interacting electron region. \cite{meir_landauer_1992} The carrier density from  \cref{eq-B43}, and current density from \cref{eq-B50} is computed in the non-equilibrium Kadanoff-Keldysh formalism. Where $f^{Cont}(E)$ is contact Fermi-distribution function. \cite{martin_theory_1959,kadanoff_theory_1961,keldysh_diagram_1964}

Interacting current in the device, where the central part interacts with self-energy and contacts or leads are non-interacting, is defined by \cref{eq-B50}. The interacting part of the device has a set of expressions for $A_{l_{1}l}$ and $G_{l_{1}l}^{<}$ as,

\begin{equation}\label{eq-B53}
\begin{aligned}
G_{l_{1}l}^{<}& = \sum_{l_{2},l_{3}}G_{l_{1}l_{2}}^{R}\Big[if^{L}\Gamma_{l_{2}l_{3}}^{L}+if^{R}\Gamma_{l_{2}l_{3}}^{R}+\Sigma_{l_{2}l_{3}}^{<}\Big]G_{l_{3}l}^{A} \\
A_{l_{1}l} & = \sum_{l_{2},l_{3}}G_{l_{1}l_{2}}^{R}\Big[\Gamma_{l_{2}l_{3}}^{L}+\Gamma_{l_{2}l_{3}}^{R}+i\big[\Sigma_{l_{2}l_{3}}^{>}-\Sigma_{l_{2}l_{3}}^{<}\big]\Big]G_{l_{3}l}^{A}  
\end{aligned}
\end{equation}

The self-energies due to interactions e.g. carrier-phonon scattering are incorporated into $\Sigma_{l_{2}l_{3}}^{<}$ and $\Sigma_{l_{2}l_{3}}^{>}$ through one and two simultaneous acoustic phonon \cref{eq-1.53} and one and two simultaneous optical phonon \cref{eq-1.69} interatction from the aforementioned self-energies precipitation. In the \cref{eq-B53} $G_{l_{1}l}^{<}$ Green's function and spectral function $A_{l_{1}l}$ produce two-part current density $\mathfrak{J}$ as,

\begin{equation}\label{eq-B54}
\mathfrak{J} = \mathfrak{J}_{coh}+\mathfrak{J}_{in}
\end{equation}

In the presence of interaction, the expression $G_{l_{1}l_{2}}^{R}$ and $G_{l_{3}l}^{A}$ in the \cref{eq-B53} calculated for coherent current density case $\mathfrak{J}_{coh}$ by assuming no interaction self-energy in the active region. The \cref{eq-B50} simplified by appropriate boundary conditions by expressing lesser Green's function $G_{l_{1}l}^{<}(\boldsymbol{k}_{\mathrm{t}};E)$, and the spectral function $A_{l_{1}l}(\boldsymbol{k}_{\mathrm{t}};E)$ and using some algebra indices $m_{1}$ and $m_{2}$ to the left contact, non-interacting active part indices $l_{2}$ and $l_{3}$, and $m_{3}$ and $m_{4}$ for the right contact. Carrier distribution within the equilibrated contacts represent by  $g_{m_{1}m_{2}}^{<}$ and $g_{m_{3}m_{4}}^{<}$. Which enables the use of the fluctuation-dissipation theorem. Furthermore, recalling the definition \cref{eq-B51} leads to the following equations, which is two-terminal non-interacting Landauer formula,

\begin{equation}\label{eq-B52}
\mathfrak{J}_{coh} = \frac{e}{\hbar A}\sum_{ll_{1}}\sum_{l_{2}l_{3}}\sum_{\boldsymbol{k}_{\mathrm{t}}}\int\frac{\mathrm{d}E}{2\pi}\Big\{\Gamma_{ll_{1}}^{L}G_{l_{1}l_{2}}^{R}\Gamma_{l_{2}l_{3}}^{R}G_{l_{3}l}^{A}\Big\}(\boldsymbol{k}_{\mathrm{t}};E)\Big[f^{L}(E)-f^{R}(E)\Big] 
\end{equation}

Where in the non-interacting active part of the device indices $l, l_{1}, l_{2}, l_{3}$ run covering all the points. 

Further, The interaction current density $\mathfrak{J}_{in}$, where the interacting part of the device has a different set of expressions from \cref{eq-B54} and placing \cref{eq-B53} into \cref{eq-B50} as,

\begin{equation}\label{eq-B55}
\begin{aligned}
\mathfrak{J}_{in} & = \frac{ie}{\hbar A}\sum_{ll_{1}}\sum_{l_{2}l_{3}}\sum_{\boldsymbol{k}_{\mathrm{t}}}\int\frac{\mathrm{d}E}{2\pi}\Gamma_{ll_{1}}^{L}(\boldsymbol{k}_{\mathrm{t}};E)G_{l_{1}l_{2}}^{R}(\boldsymbol{k}_{\mathrm{t}};E)\Big\{\Sigma_{l_{2}l_{3}}^{>}(\boldsymbol{k}_{\mathrm{t}};E)f^{L}(E)  \\
&+\Sigma_{l_{2}l_{3}}^{<}(\boldsymbol{k}_{\mathrm{t}};E)\big[1-f^{L}(E)\big]\Big\}G_{l_{3}l}^{A}(\boldsymbol{k}_{\mathrm{t}};E) \\
& = \frac{ie}{\hbar A}\sum_{ll_{1}l_{2}l_{3}}\sum_{m_{1}m_{2}}\sum_{\boldsymbol{k}_{\mathrm{t}}}\int\frac{\mathrm{d}E}{2\pi}h_{lm_{1}}A_{m_{1}m_{2}}(\boldsymbol{k}_{\mathrm{t}};E)h_{m_{2}l_{1}}G_{l_{1}l_{2}}^{R}(\boldsymbol{k}_{\mathrm{t}};E)G_{l_{3}l}^{A}(\boldsymbol{k}_{\mathrm{t}};E) \\
&
\times\Big\{\Sigma_{l_{2}l_{3}}^{>}(\boldsymbol{k}_{\mathrm{t}};E)if^{L}(E)+\Sigma_{l_{2}l_{3}}^{<}(\boldsymbol{k}_{\mathrm{t}};E)i\big[1-f^{L}(E)\big]\Big\} \\
& = \frac{e}{\hbar A}\sum_{ll_{1}l_{2}l_{3}}\sum_{m_{1}m_{2}}\sum_{\boldsymbol{k}_{\mathrm{t}}}\int\frac{\mathrm{d}E}{2\pi}h_{lm_{1}}h_{m_{2}l_{1}}G_{l_{1}l_{2}}^{R}(\boldsymbol{k}_{\mathrm{t}};E)G_{l_{3}l}^{A}(\boldsymbol{k}_{\mathrm{t}};E) \\
& 
\times\Big\{\Sigma_{l_{2}l_{3}}^{>}(\boldsymbol{k}_{\mathrm{t}};E)g_{m_{1}m_{2}}^{<}(\boldsymbol{k}_{\mathrm{t}};E)-\Sigma_{l_{2}l_{3}}^{<}(\boldsymbol{k}_{\mathrm{t}};E)g_{m_{1}m_{2}}^{>}(\boldsymbol{k}_{\mathrm{t}};E)\Big\}
\end{aligned}
\end{equation}

Where in left lead or contact, indices $m_{1}, m_{2}$ are situated and within the interacting central part of the device, indices $l, l_{1}, l_{2}, l_{3}$ run covering all the points. By using the fluctuation-dissipation theorem, the last equality evaluated in \cref{eq-B55}. In the self-consist born approximation, the carrier density from \cref{eq-B43} and current density from \cref{eq-B50} is computed.

The algorithm flow is as follows, at the start of the simulation, at the first step  $G_{0}^{<,>}$  and $G_{0}^{R, A}$ non-interacting Green's functions evaluated to calculate the first iterated self-energies $\Sigma^{<,>}$ and $\Sigma^{R, A}$. In the second step, $G^{R, A}$ matrix equation calculate to get the actual values of $\Sigma^{<,>}$ and $\Sigma^{R, A}$ self-energies from the \cref{eq-1.53} and \cref{eq-1.69}. Actual self-energies use for the computation of $G^{<,>}$. In the third step, updated values of $G^{<,>}$ and $G^{R,A}$  adopt to estimate new scattering self-energies $\Sigma^{<,>}$ and $\Sigma^{R,A}$ from the \cref{eq-1.53} and \cref{eq-1.69}. The scattering self-energies utilizes to determine the new Hartree potential, which through $V(z)$ part directly updates the Hamiltonian $H_{0}$. It is equivalent to finding the solution of potential in Poisson's equation with the carrier density in \cref{eq-B43} and iteratively updating the device potential. The self-consistent iterative loop between the self-energies and Green's functions will run continuously until convergence. Once the convergence achieves, the algorithm proceeds in the last step. In the fourth step, definitive device potential obtained from self-consistent self-energies and Green's functions loop is used in \cref{eq-B50} to calculate the current density.

\subsection*{Boltzmann Transport Equation: Semi-classical Treatment}

For the Boltzmann transport equation calculation in the molecular crystal device, as we introduce lattice site index in tight-binding model and later used in the non-equilibrium Green's function formalism to connect one lattice site to another through the translation vector, for the Boltzmann transport equation solution,\cite{boltzmann_theorie_van_1896,boltzmann_theorie_der_1896} we will here reintroduce the generalized band index $ \mathrm n $ and $ \mathrm m $, as discussed in paragraph after \cref{eq-1.12}, the author here want to caution the reader these electronic band branch index $ \mathrm n $, $\mathrm m $, and phonon branch index $\mathrm v $ are in phase space and connected through the respective electron momentum $ \boldsymbol{k} $ and lattice phonon momentum $ \boldsymbol{q} $ through the Fourier transformation in the reciprocal vector space of the first Brillouin zone. However, as discussed hitherto and illustrated in the figures \cref{fig-15},\cref{fig-16},\cref{fig-17}, the electronic band is narrow in energy bandwidth due to the peculiar nature of organic molecular crystal. Therefore, few electronic band branches are available to scatter than classical solid-state semiconductor materials. It will indicate that mobility should be higher; however, the near-flat electronic dispersion curve gives rise to a slow-moving electron in the crystal. Also, as discussed and illustrated in the figures \cref{fig-31},\cref{fig-32}, the phonon spectra are in the high energy range compared to the electronic dispersion curve. Furthermore, the electron-phonon interaction is only possible when more than one phonon interacts to conserve the energy and momentum of the scattering event. We have investigated up to two phonon processes. However, as earlier mentioned, it is possible at microscopic level three to four phonon interaction governing underlying dynamics of electronic transport in molecular crystal and organic polymer.

The statement of Boltzmann transport equation is, an electronic distribution function $ \mathscr{F}(\boldsymbol{r}, \boldsymbol{k_n}, t) $ in the phase-space variables with electron momentum $ \boldsymbol{k_n} $ at $ \mathrm n^{th} $ band index and spatial coordinate $ \boldsymbol{r} $, it's time evolution of the electron occupations is equal to the rate of change of $ \mathscr{F}(\boldsymbol{r}, \boldsymbol{k_n}, t) $ due to various collisions interaction, and by applying the chain rule of derivatives,

\begin{equation}\label{eq-BTE-13}
\underbrace{\frac{\partial  \mathscr{F}(\boldsymbol{r}, \boldsymbol{k_n}, t) }{\partial t}}_{\text {Time evolution of  electron occupations}}+\underbrace{\boldsymbol v_{\boldsymbol {k_n}} \cdot \nabla_{\boldsymbol r}  \mathscr{F}(\boldsymbol{r}, \boldsymbol{k_n}, t) +\frac{1}{\hbar} \boldsymbol{F}_{e} \cdot \nabla_{\boldsymbol k_n}  \mathscr{F}(\boldsymbol{r}, \boldsymbol{k_n}, t)}_{\text {Drift terms due to external electric fields}} =\underbrace{\hat{\mathcal{C}}  \mathscr{F}(\boldsymbol{r}, \boldsymbol{k_n}, t)}_{\text {Collisions
terms}} = \left.\frac{d \mathscr{F}}{d t}\right|_{\text {coll }}
\end{equation}

Where $ \boldsymbol v_{\boldsymbol {k_n}} =  d\boldsymbol r/dt $ is the band velocity and $ \boldsymbol{F}_{e}=  d\boldsymbol {k_n}/dt $ the Lorentz force from Newtonian dynamics on the species, for the electron this force is due to externally applied electric field for conductance. The collision or scattering operator $ \hat{\mathcal{C}} \mathscr{F}(\boldsymbol{r}, \boldsymbol {k_n}, t) $ is the difference of in-scattering rate and out-scattering rate from a phase-space state $\mathrm n $ and $\mathrm m $,

\begin{equation}\label{eq-BTE-14-1}
\begin{aligned}
\left.\frac{d \mathscr{F}}{d t}\right|_{\text {coll }}&=\hat{\mathcal{C}} \mathscr{F}(\boldsymbol{r}, \boldsymbol{k_n}, t)=
\underbrace{\sum_{\boldsymbol {k_n}} S(\boldsymbol {k_m}, \boldsymbol {k_n}) \mathcal{F}(\boldsymbol {k_n})[1-\mathcal{F}(\boldsymbol {k_m})]}_{\text { in-scattering flux }}-
\underbrace{\sum_{\boldsymbol {k_m}} S(\boldsymbol {k_n}, \boldsymbol {k_m}) \mathcal{F}(\boldsymbol {k_m})[1-\mathcal{F}(\boldsymbol {k_n})]}_{\text { out-scattering flux }} \\
&=\text {In-scattering rate; each interaction type}-\text {Out-scattering rate; each interaction type}\\
\left.\frac{d \mathscr{F}}{d t}\right|_{\text {coll }}& =\left.\frac{d\mathscr{F}}{d t}\right|_{\mathrm{ac;ph-1 }}+\left.\frac{d\mathscr{F}}{d t}\right|_{\mathrm{ac;ph-2 }}+\left.\frac{d\mathscr{F}}{d t}\right|_{\mathrm{op;ph-1 }}+\left.\frac{d\mathscr{F}}{d t}\right|_{\mathrm{op;ph-2 }}
\end{aligned}
\end{equation}

\begin{figure}[H]
\centering  
\begin{tikzpicture}
\draw  [ultra thick,blue]plot[smooth, tension=.7] coordinates {(-2,-0.5) (0.5,-1.5) (3,-0.5)};
\draw  [ultra thick,magenta]plot[smooth, tension=.7] coordinates {(1.5,-0.5) (3,-2) (4,-0.5)};
\node (v1) at (-1,-1.05) {};
\node (v2) at (3.5,-1.5) {};
\draw [-latex,ultra thick, cyan](v1) -- (v2);
\node [below] at (-2.0237,-0.6) {$|\boldsymbol{k_n}\rangle$ };
\node [below] at (-1,-1) {$\varepsilon(\boldsymbol {k_n})$};
\node [above] at (4,-1.7) {$|\boldsymbol{k_m}-\boldsymbol q\rangle$};
\node [below] at (3.5,-1.3) {$\varepsilon(\boldsymbol {k_m}-\boldsymbol q)$};
\node at (0.5,0) {};
\node at (0.6,1.5) {$\textbf{1-Phonon\ Emission}$};
\draw[ultra thick,red,decorate,decoration={coil,segment length=4pt}] (v1) -- (0.7,0.5);
\draw[ball color=blue] (v1) circle (.1);
\draw[ball color=magenta] (v2) circle (.1);
\node [above] at (0.7,0.4) {$|\boldsymbol{q}\rangle$};
\node [below] at (1.1,0.7) {$\hbar\omega_{\boldsymbol q}$};
\end{tikzpicture}
\begin{tikzpicture}
\draw  [ultra thick,blue]plot[smooth, tension=.7] coordinates {(-2.5,-0.5) (-0.5,-1.5) (2.5,-0.5)};
\draw  [ultra thick,magenta]plot[smooth, tension=.7] coordinates {(-0.5,1) (2,-1) (5,0.5)};
\node (v1) at (-0.5,-1.5) {};
\node (v2) at (0.5,-0.1) {};
\draw [-latex,ultra thick, cyan](v1) -- (v2);
\node [below] at (-1.2,-1.2) {$|\boldsymbol{k_n}\rangle$};
\node [below] at (-0.5,-1.5) {$\varepsilon(\boldsymbol {k_n})$};
\node at (1,1.5) {$\textbf{1-Phonon\ Absorption}$};
\draw[ultra thick,green,decorate,decoration={coil,segment length=4pt}] (v1) -- (-1.5,0.5);
\draw[ball color=blue] (v1) circle (.1);
\draw[ball color=magenta] (v2) circle (.1);
\node [above] at (1,-0.3) {$|\boldsymbol{k_m}-\boldsymbol q\rangle$};
\node [below] at (1.2,0.3) {$\varepsilon(\boldsymbol {k_m}-\boldsymbol q)$};
\node [below] at (-1.5,1) {$|\boldsymbol{q}\rangle$};
\node [below] at (-0.5,0) {$\hbar\omega_{\boldsymbol q}$};
\end{tikzpicture}
\caption{One-Phonon Emission and One-Phonon Absorption Process}
\label{fig-0-1}
\end{figure}
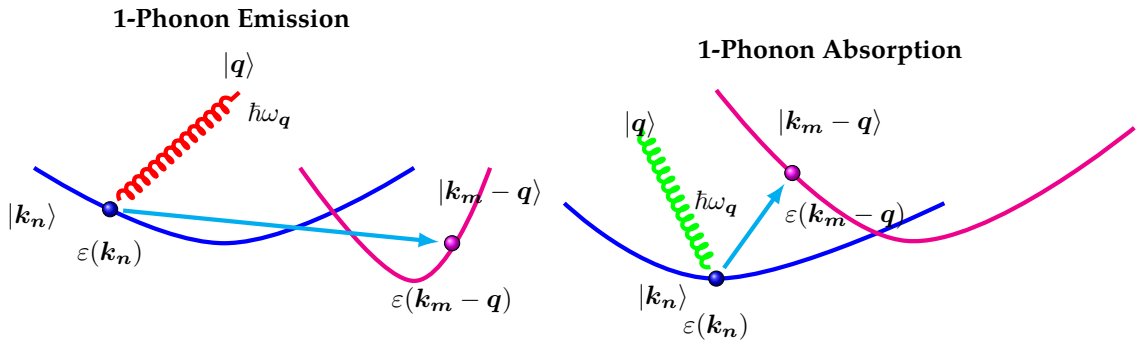

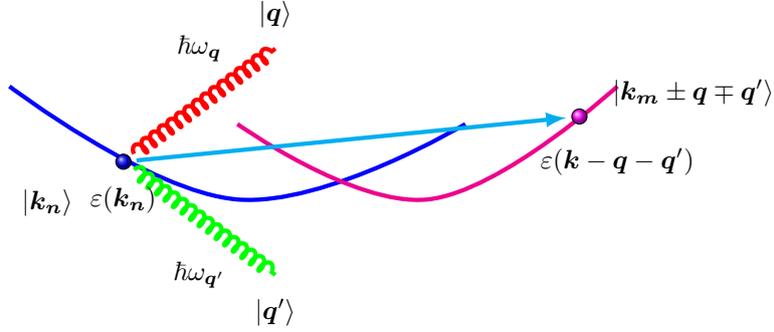
\begin{figure}[H]
\centering  
\begin{tikzpicture}
\draw  [ultra thick,blue]plot[smooth, tension=.7] coordinates {(-2.5,0) (0.5,-1.5) (3.5,-0.5)};
\draw  [ultra thick,magenta]plot[smooth, tension=.7] coordinates {(0.5,-0.5) (3,-1.5) (5.5,0)};
\node (v1) at (-1,-1) {};
\node (v2) at (5,-0.4) {};
\draw [-latex,ultra thick, cyan](v1) -- (v2);
\node [below] at (-2,-1) {$|\boldsymbol{k_n}\rangle$};
\node at (0.5,0) {};
\node at (1,1.5) {$\textbf{2-Phonon;\ Emission\ \& Absorption}$};
\draw[ultra thick,red,decorate,decoration={coil,segment length=4pt}] (v1) -- (1,0.5);
\draw[ball color=blue] (v1) circle (.1);
\draw[ball color=magenta] (v2) circle (.1);
\node [above] at (6.5,-1.3) {$|\boldsymbol{k_m}\pm\boldsymbol q\mp\boldsymbol q^{\prime}\rangle$};
\node at (1,-2.5) {};
\draw[ultra thick,green,decorate,decoration={coil,segment length=4pt}] (v1) -- (1,-2.5);
\node [above] at(1,0.5) {$|\boldsymbol q \rangle$};
\node [below] at(1,-2.5) {$|\boldsymbol q^{\prime}\rangle$};
\node at (0,-2.5) {$\hbar\omega_{\boldsymbol q^{\prime}}$};
\node at (0,0.5) {$\hbar\omega_{\boldsymbol q}$};
\node at (5.5,-1) {$\varepsilon(\boldsymbol k-\boldsymbol q-\boldsymbol q^{\prime})$};
\node at (-1,-1.5) {$\varepsilon(\boldsymbol {k_n})$};
\end{tikzpicture}
\caption{Two-Phonon Simultaneous Emission and Absorption Process}
\label{fig-0-2}
\end{figure}

In the semi-classical transport regime, the physical description of equation \cref{eq-BTE-14-1} is that in the phase space Bloch electron wave are propagating in the organic molecular crystal from $| \boldsymbol {k_n} \rangle $ state to $| \boldsymbol {k_m} \rangle $ state under the influence of scattering probability $S(\boldsymbol {k_m}, \boldsymbol {k_n})$ whose amplitude strength is governed by the relevant phonon interaction type. Further the rate calculation is performed by weighted multiplication of Fermi distribution occupation function $ \mathcal{F}(\boldsymbol {k_n}) $  while assuming all the incoming $| \boldsymbol {k_n} \rangle $ state are initially filled. To successfully migrate to $| \boldsymbol {k_m} \rangle $ state it is also weighted multiplied by Fermi distribution vacancy function $ [1-\mathcal{F}(\boldsymbol {k_m})] $ to enforce that initially all $| \boldsymbol {k_m} \rangle $ state are empty and to be occupied by incoming electron with $| \boldsymbol {k_n} \rangle $ state. Next, summing over all $| \boldsymbol {k_n} \rangle $ give the incoming scattering flux and similar calculation is performed for the outgoing flux from $| \boldsymbol {k_m} \rangle $ to $| \boldsymbol {k_n} \rangle $ state and net rate is calculated by taking the difference of these two flux per unit time. Scattering probability $S(\boldsymbol {k_m}, \boldsymbol {k_n})$ is calculated for one phonon emission and absorption interaction and two phonons simultaneous emission and absorption process for acoustic and optical scattering-type as illustrated in the figures \cref{fig-0-1}, \cref{fig-0-2}. The total net scattering rate is the sum of all these individual scattering types. The most dominant one will govern the carrier dynamics and macroscopic measurable transport properties, i.e., conductivity and mobility of organic molecular crystal.

Recalling, broadening amplitude $ \Gamma_{\mathrm{ac;ph-1}}(\boldsymbol k, \varepsilon(\boldsymbol k)) $ and $ \Gamma_{\mathrm{ac;ph-2}}(\boldsymbol k, \varepsilon(\boldsymbol k)) $ from the Green's function perturbation treatment due to one acoustic phonon interaction and simultaneous two acoustic phonon interaction with the electron propagator from the \cref{eq-1.61} and \cref{eq-1.62}, for reference, we state here again one last time these equations as,

\begin{equation}\label{eq-1.62-1}
\begin{split}  
\Gamma_{\mathrm{ac;ph-1}}(\boldsymbol k, \varepsilon(\boldsymbol k)) =\frac{2\pi}{\hbar}\sum_{\boldsymbol q}|U^{J}_{\boldsymbol q;\mathrm{ac}}(\boldsymbol k, \boldsymbol q)|^{2}&\Bigg\{\overbrace{\Big[\mathcal{N}_{\boldsymbol q}+1-\mathcal{F}(\varepsilon(\boldsymbol k-\boldsymbol q))\Big]\cdot\delta\Big(\varepsilon(\boldsymbol k)-\varepsilon(\boldsymbol k-\boldsymbol q)-\hbar\omega_{\boldsymbol q}\Big)}^{\text{1-Acoustic Phonon; Emission}}\\&+\overbrace{\Big[\mathcal{N}_{\boldsymbol q}+\mathcal{F}(\varepsilon(\boldsymbol k-\boldsymbol q))\Big]\cdot\delta\Big(\varepsilon(\boldsymbol k)-\varepsilon(\boldsymbol k-\boldsymbol q)+\hbar\omega_{\boldsymbol q}\Big)}^{\text{1-Acoustic Phonon;  Absorption}}\Bigg\}
\end{split}
\end{equation}

\begin{equation}\label{eq-1.62-3}
\begin{split}  
\Gamma_{\mathrm{ac;ph-2}}(\boldsymbol k, \varepsilon(\boldsymbol k)) &=
\frac{2\pi}{\hbar}\sum_{\boldsymbol q,\boldsymbol q^{\prime}}|U^{J}_{\boldsymbol q\boldsymbol q^{\prime};\mathrm{ac}}(\boldsymbol k,\boldsymbol q,\boldsymbol q^{\prime})|^{2}\cdot \Bigg\{\overbrace{\mathcal{N}_{\boldsymbol q}(\mathcal{N}_{\boldsymbol q'} +1)\cdot\delta\Big(\varepsilon(\boldsymbol k)-\varepsilon(\boldsymbol k-\boldsymbol q-\boldsymbol q^{\prime})-\hbar\omega_{\boldsymbol q}+\hbar\omega_{\boldsymbol q^{\prime}}\Big)}^{\text{2-Acoustic Phonon; Both Simultaneous Emission \& Absorption}}
\Bigg\}
\end{split}
\end{equation}

With the treatment as mentioned earlier in \cref{eq-BTE-14-1}, summing the scattering probability amplitude over all inward coming
$| \boldsymbol {k_n} \rangle $ and outward leaving $| \boldsymbol {k_m} \rangle$ state with holding energy-momentum conservation for each electronic interaction with respective one phonon $( \boldsymbol {q} =\boldsymbol {q}_\mathrm {v})  $ i.e. from the $ \mathrm {v}^{th} $ branch and two phonon processes $( \boldsymbol {q} =\boldsymbol {q}_\mathrm {v}, \boldsymbol {q^{\prime}} = \boldsymbol {q}_\mathrm {v^{\prime}} )$ i.e., from the $ \mathrm {v}^{th} $ and $ \mathrm {v^{\prime}}^{th} $ branches. In this notation we have drop writing  $\mathrm {v}$ and $\mathrm {v^{\prime}}$ in the every subscript of $ \boldsymbol {q}$ to keep notation trackable and clean. However, we want to emphasize here that summation over $ \boldsymbol {q}$ restrict to the first Brillouin zone of the crystal, and summation must perform over the various mode of the phonon. Such a summation is performed in the  \cref{eq-1.12} and therefore, in subsequent treatment, we have skipped the writing  $\mathrm {v}$ and $\mathrm {v^{\prime}}$ in every subscript of phonon wavevector $ \boldsymbol {q}$. Further multiplying with electronic state occupation or vacancy Fermi distribution function of relevant band indexes $ n $ and $ m $. As $ \mathcal{F}(\boldsymbol {k_n})[1-\mathcal{F}(\boldsymbol {k_m}-\boldsymbol q)] $ and $ \mathcal{F}(\boldsymbol {k_m}-\boldsymbol q)[1-\mathcal{F}(\boldsymbol {k_n})] $ for net transition rate in between respective band indexes. Furthermore, treating by subtracting from the time-reversal conjugate as the prescription of \cref{eq-BTE-14-1} for the calculation of net scattering rate. In the Boltzmann statistics for Fermion, further simplification is achieved with the assumption initially for all inward moving flux at state $|\boldsymbol {k_m}\rangle $ from state $| \boldsymbol {k_n} \rangle $ the all $ \mathcal{F}(\boldsymbol {k_m}) $ state are empty and consequently there occupancy is zero, and similarly for all outward leaving flux to state $| \boldsymbol {k_n} \rangle $ from state $|\boldsymbol {k_m}\rangle $ all initial occupancy state are empty and therefore $ \mathcal{F}(\boldsymbol {k_n}) $ is zero, with these assumption one phonon acoustic scattering rate is,

\begin{equation}\label{eq-1.62-2}
\begin{split}  
\left.\frac{d\mathscr{F}}{d t}\right|_{\mathrm{ac;ph-1 }}=& \frac{2\pi}{\hbar}\sum_{\boldsymbol {k_m}-\boldsymbol q}\sum_{\boldsymbol q}|U^{J}_{\boldsymbol q;\mathrm{ac}}(\boldsymbol {k_n}, \boldsymbol q)|^{2}\Bigg\{\underbrace{ \delta\Big(\varepsilon(\boldsymbol {k_n})-\varepsilon(\boldsymbol {k_m}-\boldsymbol q)-\hbar\omega_{\boldsymbol q}\Big)}_{\text{Energy Conservation; Phonon Emission}}\cdot \\& \bigg[\underbrace{\Big(\mathcal{N}_{\boldsymbol q}+1\Big)\mathcal{F}(\boldsymbol {k_n})\Big(1-\mathcal{F}(\boldsymbol {k_m}-\boldsymbol q)\Big)}_{\text{Acoustic Phonon Emission}}  -\underbrace{\mathcal{N}_{\boldsymbol q}\mathcal{F}(\boldsymbol {k_m}-\boldsymbol q)\Big(1-\mathcal{F}(\boldsymbol {k_n})\Big)}_{\text{Time-reversal Conjugate; Phonon Emission}}\bigg]\\&  +\bigg[\underbrace{\mathcal{N}_{\boldsymbol q}\mathcal{F}(\boldsymbol {k_n})\Big(1-\mathcal{F}(\boldsymbol {k_m}-\boldsymbol q)\Big)}_{\text{Acoustic Phonon Absorption}} -\underbrace{\Big(\mathcal{N}_{\boldsymbol q}+1\Big)\mathcal{F}(\boldsymbol {k_m}-\boldsymbol q)\Big(1-\mathcal{F}(\boldsymbol {k_n})\Big)}_{\text{Time-reversal Conjugate; Phonon Absorption}}\bigg]\cdot\\& \underbrace{\delta\Big(\varepsilon(\boldsymbol {k_n})-\varepsilon(\boldsymbol {k_m}-\boldsymbol q)+\hbar\omega_{\boldsymbol q}\Big)}_{\text{Energy Conservation; Phonon Absorption}}\Bigg\}
\end{split}
\end{equation}

Where energy conservation hold through $ \varepsilon(\boldsymbol {k_n})=\varepsilon(\boldsymbol {k_m}-\boldsymbol q)+\hbar\omega_{\boldsymbol q} $ for emission process interaction with the electron and energy conservation hold through $ \varepsilon(\boldsymbol {k_n})=\varepsilon(\boldsymbol {k_m}-\boldsymbol q)-\hbar\omega_{\boldsymbol q} $ for absorption process interaction with the electron and momentum is conserved in the interaction through $ \boldsymbol{k_m} = \boldsymbol{k_n} +\boldsymbol{q} $.
Following the same mathematical prescription to count the scattering rate in the Boltzmann equation as discussed above, by weighted multiplying with $ \mathcal{F}(\boldsymbol {k_n})[1-\mathcal{F}(\boldsymbol {k_m}-\boldsymbol q-\boldsymbol q^{\prime})] $ and $ \mathcal{F}(\boldsymbol {k_m}-\boldsymbol q-\boldsymbol q^{\prime})[1-\mathcal{F}(\boldsymbol {k_n})] $ for the net transition rate in between respective band indexes. Two-phonon acoustic scattering rate is,

\begin{equation}\label{eq-1.62-4}
\begin{split}  
&\left.\frac{d\mathscr{F}}{d t}\right|_{\mathrm{ac;ph-2 }} =\frac{2\pi}{\hbar}\sum_{\boldsymbol {k_m}-\boldsymbol q}\sum_{\boldsymbol {k_m}-\boldsymbol q^{\prime}}\sum_{\boldsymbol q,\boldsymbol q^{\prime}}|U^{J}_{\boldsymbol q\boldsymbol q^{\prime};\mathrm{ac}}(\boldsymbol {k_n},\boldsymbol q,\boldsymbol q^{\prime})|^{2}\cdot\Bigg\{\bigg[\underbrace{\mathcal{N}_{\boldsymbol q}(\mathcal{N}_{\boldsymbol q'} +1)\mathcal{F}(\boldsymbol {k_n})\Big(1-\mathcal{F}(\boldsymbol {k_m}-\boldsymbol q-\boldsymbol q^{\prime})\Big)}_{\text{Acoustic Phonon Emission \& Absorption}} \\&-\underbrace{\mathcal{N}_{\boldsymbol q}(\mathcal{N}_{\boldsymbol q'} +1)\mathcal{F}(\boldsymbol {k_m}-\boldsymbol q-\boldsymbol q^{\prime})\Big(1-\mathcal{F}(\boldsymbol {k_n})\Big)}_{\text{Time-reversal Conjugate; Phonon Emission \& Absorption}}\bigg] \cdot\underbrace{\delta\Big(\varepsilon(\boldsymbol {k_n})-\varepsilon(\boldsymbol {k_m}-\boldsymbol q-\boldsymbol q^{\prime})-\hbar\omega_{\boldsymbol q}+\hbar\omega_{\boldsymbol q^{\prime}}\Big)}_{\text{Energy Conservation; Phonon Emission \& Absorption}}\Bigg\}
\end{split}
\end{equation}

Where energy conservation hold through $ \varepsilon(\boldsymbol {k_n})=\varepsilon(\boldsymbol {k_m}-\boldsymbol q-\boldsymbol q^{\prime})+\hbar\omega_{\boldsymbol q}-\hbar\omega_{\boldsymbol q^{\prime}} $ for two phonon simultaneous emission and absorption interaction with the electron and momentum is conserved in the interaction through $ \boldsymbol{k_m} = \boldsymbol{k_n} +\boldsymbol{q}+\boldsymbol{q}^{\prime} $. Now again recalling, broadening amplitude $ \Gamma_{\mathrm{op;ph-1}}(\boldsymbol k, \varepsilon(\boldsymbol k)) $ and $ \Gamma_{\mathrm{op;ph-2}}(\boldsymbol k, \varepsilon(\boldsymbol k)) $ from the Green's function perturbation treatment due to one optical phonon interaction and simultaneous two optical phonon interaction with the electron propagator from the \cref{eq-1.69} and \cref{eq-1.70}, for reference we state here one last time these equation as,

\begin{equation}\label{eq-1.70-1}
\begin{split}  
\Gamma_{\mathrm{op;ph-1}}(\boldsymbol k, \varepsilon(\boldsymbol k)) =\frac{2\pi}{\hbar}\sum_{\boldsymbol q}|U^{C}_{\boldsymbol q;\mathrm{op}}|^{2}&\Bigg\{\overbrace{\Big[\mathcal{N}_{\boldsymbol q}+1-\mathcal{F}(\varepsilon(\boldsymbol k-\boldsymbol q))\Big]\cdot\delta\Big(\varepsilon(\boldsymbol k)-\varepsilon(\boldsymbol k-\boldsymbol q)-\hbar\omega_{\boldsymbol q}\Big)}^{\text{1-Optical Phonon; Emission}}\\&+\overbrace{\Big[\mathcal{N}_{\boldsymbol q}+\mathcal{F}(\varepsilon(\boldsymbol k-\boldsymbol q))\Big]\cdot\delta\Big(\varepsilon(\boldsymbol k)-\varepsilon(\boldsymbol k-\boldsymbol q)+\hbar\omega_{\boldsymbol q}\Big)}^{\text{1-Optical Phonon; Absorption}}\Bigg\}
\end{split}
\end{equation}

\begin{equation}\label{eq-1.70-2}
\begin{split}  
\Gamma_{\mathrm{op;ph-2}}(\boldsymbol k, \varepsilon(\boldsymbol k)) =
\frac{2\pi}{\hbar}\sum_{\boldsymbol q,\boldsymbol q^{\prime}}|U^{C}_{\boldsymbol q\boldsymbol q^{\prime};\mathrm{op}}|^{2}\cdot \Bigg\{\overbrace{\mathcal{N}_{\boldsymbol q}(\mathcal{N}_{\boldsymbol q'} +1) \cdot\delta\Big(\varepsilon(\boldsymbol k)-\varepsilon(\boldsymbol k-\boldsymbol q-\boldsymbol q^{\prime})-\hbar\omega_{\boldsymbol q}+\hbar\omega_{\boldsymbol q^{\prime}}\Big)}^{\text{2-Optical Phonon; Emission \& Absorption}}
\end{split}\Bigg\}
\end{equation}

Following the same mathematical prescription to count the scattering rate in the Boltzmann equation as discussed above for the acoustic phonon case in the above paragraph, the one and two optical phonon scattering rates are,

\begin{equation}\label{eq-1.70-3}
\begin{split}  
\left.\frac{d\mathscr{F}}{d t}\right|_{\mathrm{op;ph-1 }}=& \frac{2\pi}{\hbar}\sum_{\boldsymbol {k_m}-\boldsymbol q}\sum_{\boldsymbol q}|U^{C}_{\boldsymbol q;\mathrm{op}}|^{2}\Bigg\{ \underbrace{\delta\Big(\varepsilon(\boldsymbol {k_n})-\varepsilon(\boldsymbol {k_m}-\boldsymbol q)-\hbar\omega_{\boldsymbol q}\Big)}_{\text{Energy Conservation; Phonon Emission}}\cdot \\& \bigg[\underbrace{\Big(\mathcal{N}_{\boldsymbol q}+1\Big)\mathcal{F}(\boldsymbol {k_n})\Big(1-\mathcal{F}(\boldsymbol {k_m}-\boldsymbol q)\Big)}_{\text{Optical Phonon Emission}}  -\underbrace{\mathcal{N}_{\boldsymbol q}\mathcal{F}(\boldsymbol {k_m}-\boldsymbol q)\Big(1-\mathcal{F}(\boldsymbol {k_n})\Big)}_{\text{Time-reversal Conjugate; Phonon Emission}}\bigg]\\&  +\bigg[\underbrace{\mathcal{N}_{\boldsymbol q}\mathcal{F}(\boldsymbol {k_n})\Big(1-\mathcal{F}(\boldsymbol {k_m}-\boldsymbol q)\Big)}_{\text{Optical Phonon Absorption}} -\underbrace{\Big(\mathcal{N}_{\boldsymbol q}+1\Big)\mathcal{F}(\boldsymbol {k_m}-\boldsymbol q)\Big(1-\mathcal{F}(\boldsymbol {k_n})\Big)}_{\text{Time-reversal Conjugate; Phonon Absorption}}\bigg]\cdot\\& \underbrace{\delta\Big(\varepsilon(\boldsymbol {k_n})-\varepsilon(\boldsymbol {k_m}-\boldsymbol q)+\hbar\omega_{\boldsymbol q}\Big)}_{\text{Energy Conservation; Phonon Absorption}}\Bigg\}
\end{split}
\end{equation}

\begin{equation}\label{eq-1.70-4}
\begin{split}  
&\left.\frac{d\mathscr{F}}{d t}\right|_{\mathrm{op;ph-2 }} =\frac{2\pi}{\hbar}\sum_{\boldsymbol {k_m}-\boldsymbol q}\sum_{\boldsymbol {k_m}-\boldsymbol q^{\prime}}\sum_{\boldsymbol q,\boldsymbol q^{\prime}}|U^{C}_{\boldsymbol q\boldsymbol q^{\prime};\mathrm{op}}|^{2}\cdot\Bigg\{\bigg[\underbrace{\mathcal{N}_{\boldsymbol q}(\mathcal{N}_{\boldsymbol q'} +1)\mathcal{F}(\boldsymbol {k_n})\Big(1-\mathcal{F}(\boldsymbol {k_m}-\boldsymbol q-\boldsymbol q^{\prime})\Big)}_{\text{Optical Phonon Emission \& Absorption}} \\&-\underbrace{\mathcal{N}_{\boldsymbol q}(\mathcal{N}_{\boldsymbol q'} +1)\mathcal{F}(\boldsymbol {k_m}-\boldsymbol q-\boldsymbol q^{\prime})\Big(1-\mathcal{F}(\boldsymbol {k_n})\Big)}_{\text{Time-reversal Conjugate; Phonon Emission \& Absorption}}\bigg] \cdot\underbrace{\delta\Big(\varepsilon(\boldsymbol {k_n})-\varepsilon(\boldsymbol {k_m}-\boldsymbol q-\boldsymbol q^{\prime})-\hbar\omega_{\boldsymbol q}+\hbar\omega_{\boldsymbol q^{\prime}}\Big)}_{\text{Energy Conservation; Phonon Emission \& Absorption}}\Bigg\}
\end{split}
\end{equation}

In the above equation \cref{eq-1.62-2}, \cref{eq-1.62-4},   one phonon coupling matrix 
$|U^{J}_{\boldsymbol q;\mathrm{ac}}(\boldsymbol {k_n}, \boldsymbol q)|$, and two phonon coupling matrix $|U^{J}_{\boldsymbol q\boldsymbol q^{\prime};\mathrm{ac}}(\boldsymbol {k_n},\boldsymbol q,\boldsymbol q^{\prime})|$, for acoustic phonon interaction is calculated from \cref{eq-1.54} using the the prescription of \cref{eq-1.26} in it. Similarly, in the equation \cref{eq-1.70-3}, \cref{eq-1.70-4}  one phonon coupling matrix 
$|U^{C}_{\boldsymbol q;\mathrm{op}}|$, and two phonon coupling matrix $|U^{C}_{\boldsymbol q\boldsymbol q^{\prime};\mathrm{op}}|$, for optical phonon interaction is calculated from \cref{eq-1.67}, \cref{eq-1.68} using the the prescription of \cref{eq-1.23} in it. Here we want to state again 
$ U^{J}_{\boldsymbol q;\mathrm{ac}}(\boldsymbol k, \boldsymbol q) $, $ U^{J}_{\boldsymbol q\boldsymbol q^{\prime};\mathrm{ac}}(\boldsymbol k,\boldsymbol q,\boldsymbol q^{\prime}) $, are deformation-potential type of electron-phonon interaction and 
$U^{C}_{\boldsymbol q;\mathrm{op}}$ and $U^{C}_{\boldsymbol q\boldsymbol q^{\prime};\mathrm{op}}$ coupling is a molecular dipole potential type of interaction.
The advantage of using  \cref{eq-1.54}, \cref{eq-1.67}, \cref{eq-1.68} is that they are already in the energy, momentum domain, and electronic momentum and boson momenta can operate with each other on equal footing. This advantage was achieved due to the earlier application of Green's function description to transfer the entire electron-phonon gas system using the second quantization language in Heisenberg representation. Nevertheless, we want to emphasize that the coupling matrixes mentioned above can also be calculated using the density functional perturbation theory, which incorporates the lattice dynamics effect in the Kohn–Sham framework. However, Kohn–Sham equation is essentially a one-electron Schr\"{o}dinger equation in real space. Therefore, as admitted by Holstein, treatment of Bosonic dilation perturbation and electronic wavefunction is difficult to book-keep beyond one unit cell. To overcome these bottlenecks additional double Fourier transformation is needed to keep molecular crystal Boson and Fermion on the same footing. In practice, to overcome this as reported for the one phonon interaction for semiconductor calculation in the EPW \cite{ponce_epw_2016} and Perturbo code \cite{jin_perturbo_2021}. The adaptive coarse grid is used and later interpolated to the fine grid with a large cut-off in real space, and further sampling is required to handle Bosonic perturbation in the unit cell within the Bloch basis.  

The Boltzmann transport equation \cref{eq-BTE-14-1} solved in the semiconductor domain under the assumption of relaxation time approximation; however, such an approximation is harsh for the non-parabolic band and narrow bandwidth organic crystal. The solution of the Boltzmann equation obtains through various approximations. In the low-field transport regime, externally applied electric field drift the system out of Fermi-Dirac equilibrium distribution.
Furthermore, scattering self-energy from electron-phonon interaction works to restore the equilibrium distribution and the dynamic equation reach a steady-state transport regime. Drift in the distribution function due to applied electric field is assumed small, which is an approximation. Another method is to linearize the Boltzmann equation in the lower first order by expanding the occupation/vacancy Fermi-distribution function  $ \mathcal{F}(\boldsymbol {k_n}) $ or $ [1-\mathcal{F}(\boldsymbol {k_m})] $ around the zero-field solution $ \mathcal{F}^{0}(\boldsymbol {k_n}) $ or $ [1-\mathcal{F}^{0}(\boldsymbol {k_m})] $ of the relaxation time approach. Furthermore, use this initial solution as a starting point to solve the linearized system of equations in the iterative method $ \mathcal{F}^{i+1}(\boldsymbol {k_n}) $ or $ [1-\mathcal{F}^{i+1}(\boldsymbol {k_m})] $. However, we want to reiterate that the relaxation time approximation holds only for the wide bandwidth semiconductor domain—the computational challenge in calculating the scattering coupling matrix for each time step of the Boltzmann transport equation. Moreover, we have to compute the scattering coupling matrix within the current time step for the population distribution of electrons. The coupling matrix for various scattering mechanisms can be efficiently calculated in an explicit time-step approach in the iterative method by the first-order Euler method or using the fourth-order Runge-Kutta technique. Further reduction in the memory storage and computation time is achieved by selecting an energy window and bands of interest for the transport regime and retaining relevant k-points sampling. Also, the non-equilibrium electron's distribution function is further approximately modeled as Lorentzian, Gaussian, and Fermi-Dirac distribution of equilibrium state. These techniques will reduce the calculation overhead and faster implement the Boltzmann transport equation. Using the computation strategies mentioned beforehand ultrafast electron-phonon interaction dynamics up to hundreds of pico-second with a resolution of femtosecond time scales can be computed. Nevertheless, after evaluating the scattering rates from the bottom-up quantum treatment, the solution of the Boltzmann equation accomplishes through stochastic methods in the Monte-Carlo framework without any approximation of the low-field transport regime. However, these calculations are again computationally demanding, and sampling is used to deduce the scattering rate efficiently. In the semi-classical treatment through the Boltzmann equation, we have to assume that the transport in the organic molecular crystal is in the weak scattering regime, infrequent scattering in the crystal. Therefore there is enough time between scattering events so that propagating state has sharply defined energy and does not consider collisional broadening effects in the range of transport regime.

\section*{Discussion}
Organic molecular crystals and polymers have a wide variety of crystal structures and properties. Therefore, before moving ahead and applying any transport theory, a quantum mechanical viewpoint and subsequent treatment of molecular crystals Hamiltonian are essential for defining the theoretical calculation and describing the experimental phenomena. For the inorganic and organic material with high polarity in the crystal structures, Fr\"{o}hlich Hamiltonian should be a good starting point, which is remarkably successful with ionic lattice crystal of inorganic material, e.g., $KCl, CsI, SrTiO_3, RbCl $. Fr\"{o}hlich coupling strength is a valuable parameter to start the treatment of the framework. Holstein molecular-crystal treatment is widely adopted in the polaron community for the polar organic molecular crystal material. Most organic molecules and polymers fall in this domain. Though historically this molecular crystal inspired by Fr\"{o}hlich, however, it is still a different Hamiltonian construction. In the weak interaction coupling regime of organic molecular crystal treatment at room temperature operation range, the phonon's effect via lattice Hamiltonian $  H_{\mathrm{P}} $ into the electronic Hamiltonian  $ H_{\mathrm{E}} $, through Coulomb integral $U_C$ part and resonance integral $U_J$ part of interaction Hamiltonian $ H_{\mathrm{INT}} $ should be incorporated as investigated in aforementioned work. For the strong interaction coupling regime in the organic molecular crystal operating at an extremely low-temperature range, the phonon drag effect (Polaron Theory) must be incorporated into the electronic Hamiltonian part. Moreover, before applying perturbation theory on the dynamic Hamiltonian part, a canonical transformation is fundamental. As electronic mass is significantly less than phonon drag mass, the direct application of the phonon effect as a perturbation on electronic Hamiltonian is a violation of perturbation theory. Therefore, Dyson's equation's mass operator or self-energy is canonically transformed Polaron mass in a strongly interacting regime to treat the electronic and phonon mass on an equal foundation.

Regarding applying the non-equilibrium Green's function transport framework, it has the advantage of a complete quantum treatment. Furthermore, the spatial resolution of the Fermionic or Bosonic propagator will provide microscopic properties of the device in the spatial dimension. Therefore, this framework is expandable in multi-scale resolution with multi-physics, i.e., heat treatment and many-body, i.e., multi-phonon treatment. However, there is a significant disadvantage in non-equilibrium Green's function coupled system equations are in position, energy, and time-domain and scale in seven dimensionalities for a three-dimensional treatment. Therefore, it will quickly blow up the size of the Hamiltonian in the matrix inversion operation in Recursive Green's function algorithm. Furthermore, in all practical applications, phonon's Green's function is truncated considering phonon in a thermal bath, and electronic self-energy term truncated for second-order perturbation and up to two-phonon interaction by the Migdal's theorem.  

On the other hand, in the semi-classical treatment Boltzmann transport equation has the advantage of being semi-classical. Therefore the left-hand side of the equation is purely governed by Newtonian mechanics, and only the right-hand side of the equation contains interaction mechanics by the quantum treatment. Therefore, it will significantly enhance the time performance for calculation. Furthermore, the equation is in phase space, i.e., six-dimensional position and momentum domain. Nevertheless, the Boltzmann transport equation has the disadvantage as the left-hand side is purely classical, and consequently, the forces and action on the electron-phonon gas are Newtonian. Also, the microscopic properties in the spatial resolution can be investigated only through the stochastic solution, i.e., the Monte-Carlo Algorithm.  

Hereabouts, we also discuss the applicability of the Boltzmann theory in organic molecular crystals. For the organic molecular crystals that have fluctuation around the center of mass in the molecules, and the treatment as electron-phonon gas in the Holstein's molecular-crystal model, here the Boltzmann transport equation is written as the left side of the equation is still purely electronic in nature, i.e., the effect of the external electric field only treated on the electron gas, and phonon gas motion not considered at all in the Newtonian mechanism framework. This procedure inspires by implementing the Boltzmann framework in the solid-state semiconductor field. Therefore, we are still in Born-Oppenheimer approximation, assuming these organic molecular crystals' weak molecular motion. However, in principle, based on the original Boltzmann theory,\cite{boltzmann_theorie_van_1896,boltzmann_theorie_der_1896} we have to write all the species and their Newtonian dynamics motion in the electron-phonon gas system on the left-hand side of equality to balance with the heat-flux exchange part on the right-hand side of the equation. Any attempt to incorporate the phonon gas species drift and diffusion motions through the externally applied generalized field and the non-equilibrium distribution diffusion field will lead to a generalized kinetic equation of motion of electron-phonon gas. Therefore, it will not hold the Boltzmann statistic of species distribution, and a comprehensive distribution function is required to describe such a system. As stated before, our treatment in the organic molecular semiconductor only considers Newtonian dynamics of a purely interacting electron gas in the left-hand side of the Boltzmann transport equation. Furthermore, the right-hand side of the equation contains this electronic gas's interaction with the system's phonon gas through the quantum mechanical prescription.

\bibliographystyle{IEEEtran}
\bibliography{Article}

\begin{thebibliography}{100}
\providecommand{\url}[1]{#1}
\csname url@samestyle\endcsname
\providecommand{\newblock}{\relax}
\providecommand{\bibinfo}[2]{#2}
\providecommand{\BIBentrySTDinterwordspacing}{\spaceskip=0pt\relax}
\providecommand{\BIBentryALTinterwordstretchfactor}{4}
\providecommand{\BIBentryALTinterwordspacing}{\spaceskip=\fontdimen2\font plus
\BIBentryALTinterwordstretchfactor\fontdimen3\font minus
  \fontdimen4\font\relax}
\providecommand{\BIBforeignlanguage}[2]{{%
\expandafter\ifx\csname l@#1\endcsname\relax
\typeout{** WARNING: IEEEtran.bst: No hyphenation pattern has been}%
\typeout{** loaded for the language `#1'. Using the pattern for}%
\typeout{** the default language instead.}%
\else
\language=\csname l@#1\endcsname
\fi
#2}}
\providecommand{\BIBdecl}{\relax}
\BIBdecl

\bibitem{akamatu_new_1947}
\BIBentryALTinterwordspacing
H.~Akamatu and K.~Nagamatsu, ``A new suggestion for a model representing the
  structure of carbon black,'' \emph{Journal of Colloid Science}, vol.~2,
  no.~6, pp. 593--598, Dec. 1947. [Online]. Available:
  \url{http://www.sciencedirect.com/science/article/pii/0095852247900597}
\BIBentrySTDinterwordspacing

\bibitem{akamatu_electrical_1950}
\BIBentryALTinterwordspacing
H.~Akamatu and H.~Inokuchi, ``On the {Electrical} {Conductivity} of
  {Violanthrone}, {Iso}-{Violanthrone}, and {Pyranthrone},'' \emph{The Journal
  of Chemical Physics}, vol.~18, no.~6, pp. 810--811, Jun. 1950. [Online].
  Available: \url{https://aip.scitation.org/doi/10.1063/1.1747780}
\BIBentrySTDinterwordspacing

\bibitem{holstein_studies_1959-1}
\BIBentryALTinterwordspacing
T.~Holstein, ``Studies of polaron motion: {Part} {I}. {The} molecular-crystal
  model,'' \emph{Annals of Physics}, vol.~8, no.~3, pp. 325--342, Nov. 1959.
  [Online]. Available:
  \url{http://www.sciencedirect.com/science/article/pii/0003491659900028}
\BIBentrySTDinterwordspacing

\bibitem{holstein_studies_1959}
\BIBentryALTinterwordspacing
------, ``Studies of polaron motion: {Part} {II}. {The} ``small'' polaron,''
  \emph{Annals of Physics}, vol.~8, no.~3, pp. 343--389, Nov. 1959. [Online].
  Available:
  \url{http://www.sciencedirect.com/science/article/pii/000349165990003X}
\BIBentrySTDinterwordspacing

\bibitem{holstein_theory_1964}
\BIBentryALTinterwordspacing
------, ``Theory of transport phenomena in an electron-phonon gas,''
  \emph{Annals of Physics}, vol.~29, no.~3, pp. 410--535, Oct. 1964. [Online].
  Available:
  \url{http://www.sciencedirect.com/science/article/pii/0003491664900089}
\BIBentrySTDinterwordspacing

\bibitem{yamashita_electronic_1958}
\BIBentryALTinterwordspacing
J.~Yamashita and T.~Kurosawa, ``On electronic current in {NiO},'' \emph{Journal
  of Physics and Chemistry of Solids}, vol.~5, no.~1, pp. 34--43, Jan. 1958.
  [Online]. Available:
  \url{http://www.sciencedirect.com/science/article/pii/002236975890129X}
\BIBentrySTDinterwordspacing

\bibitem{leblanc_band_1961}
\BIBentryALTinterwordspacing
O.~H. LeBlanc, ``Band {Structure} and {Transport} of {Holes} and {Electrons} in
  {Anthracene},'' \emph{The Journal of Chemical Physics}, vol.~35, no.~4, pp.
  1275--1280, Oct. 1961. [Online]. Available:
  \url{https://aip.scitation.org/doi/10.1063/1.1732038}
\BIBentrySTDinterwordspacing

\bibitem{friedman_hall_1963}
\BIBentryALTinterwordspacing
L.~Friedman, ``Hall {Effect} in the {Polaron}-{Band} {Regime},'' \emph{Physical
  Review}, vol. 131, no.~6, pp. 2445--2456, Sep. 1963. [Online]. Available:
  \url{https://link.aps.org/doi/10.1103/PhysRev.131.2445}
\BIBentrySTDinterwordspacing

\bibitem{friedman_transport_1964}
\BIBentryALTinterwordspacing
------, ``Transport {Properties} of {Organic} {Semiconductors},''
  \emph{Physical Review}, vol. 133, no.~6a, pp. A1668--a1679, Mar. 1964.
  [Online]. Available: \url{https://link.aps.org/doi/10.1103/PhysRev.133.A1668}
\BIBentrySTDinterwordspacing

\bibitem{friedman_density_1964}
\BIBentryALTinterwordspacing
------, ``Density {Matrix} {Formulation} of {Small}-{Polaron} {Motion},''
  \emph{Physical Review}, vol. 135, no.~1a, pp. A233--a246, Jul. 1964.
  [Online]. Available: \url{https://link.aps.org/doi/10.1103/PhysRev.135.A233}
\BIBentrySTDinterwordspacing

\bibitem{friedman_electron-phonon_1965}
\BIBentryALTinterwordspacing
------, ``Electron-{Phonon} {Interaction} in {Organic} {Molecular}
  {Crystals},'' \emph{Physical Review}, vol. 140, no.~5a, pp. A1649--a1667,
  Nov. 1965. [Online]. Available:
  \url{https://link.aps.org/doi/10.1103/PhysRev.140.A1649}
\BIBentrySTDinterwordspacing

\bibitem{holstein_hall_1968}
\BIBentryALTinterwordspacing
T.~Holstein and L.~Friedman, ``Hall {Mobility} of the {Small} {Polaron}.
  {II},'' \emph{Physical Review}, vol. 165, no.~3, pp. 1019--1031, Jan. 1968.
  [Online]. Available: \url{https://link.aps.org/doi/10.1103/PhysRev.165.1019}
\BIBentrySTDinterwordspacing

\bibitem{friedman_hall_1971}
\BIBentryALTinterwordspacing
L.~Friedman, ``Hall conductivity of amorphous semiconductors in the random
  phase model,'' \emph{Journal of Non-Crystalline Solids}, vol.~6, no.~4, pp.
  329--341, Nov. 1971. [Online]. Available:
  \url{http://www.sciencedirect.com/science/article/pii/002230937190024X}
\BIBentrySTDinterwordspacing

\bibitem{kubo_statistical-mechanical_1957}
\BIBentryALTinterwordspacing
R.~Kubo, ``Statistical-{Mechanical} {Theory} of {Irreversible} {Processes}.
  {I}. {General} {Theory} and {Simple} {Applications} to {Magnetic} and
  {Conduction} {Problems},'' \emph{Journal of the Physical Society of Japan},
  vol.~12, no.~6, pp. 570--586, Jun. 1957. [Online]. Available:
  \url{https://journals.jps.jp/doi/10.1143/JPSJ.12.570}
\BIBentrySTDinterwordspacing

\bibitem{wannier_structure_1937}
\BIBentryALTinterwordspacing
G.~H. Wannier, ``The {Structure} of {Electronic} {Excitation} {Levels} in
  {Insulating} {Crystals},'' \emph{Physical Review}, vol.~52, no.~3, pp.
  191--197, Aug. 1937. [Online]. Available:
  \url{https://link.aps.org/doi/10.1103/PhysRev.52.191}
\BIBentrySTDinterwordspacing

\bibitem{gosar_linear-response_1966}
\BIBentryALTinterwordspacing
P.~Gosar and S.-i. Choi, ``Linear-{Response} {Theory} of the {Electron}
  {Mobility} in {Molecular} {Crystals},'' \emph{Physical Review}, vol. 150,
  no.~2, pp. 529--538, Oct. 1966. [Online]. Available:
  \url{https://link.aps.org/doi/10.1103/PhysRev.150.529}
\BIBentrySTDinterwordspacing

\bibitem{schnakenberg_derivation_1965}
\BIBentryALTinterwordspacing
J.~Schnakenberg, ``Derivation of the hopping- and band-conductivity of the
  small polaron,'' \emph{Physics Letters}, vol.~14, no.~4, pp. 266--268, Feb.
  1965. [Online]. Available:
  \url{http://www.sciencedirect.com/science/article/pii/0031916365901915}
\BIBentrySTDinterwordspacing

\bibitem{schnakenberg_quasiteilchen-spektrum_1966}
\BIBentryALTinterwordspacing
------, ``Quasiteilchen-{Spektrum} und elektrische {Leitf\"{a}higkeit} des
  kleinen {Polarons},'' \emph{Zeitschrift f\"{u}r Physik}, vol. 190, no.~2, pp.
  209--225, Jun. 1966. [Online]. Available:
  \url{https://doi.org/10.1007/BF01327144}
\BIBentrySTDinterwordspacing

\bibitem{schnakenberg_polaronic_1968}
\BIBentryALTinterwordspacing
------, ``Polaronic {Impurity} {Hopping} {Conduction},'' \emph{physica status
  solidi (b)}, vol.~28, no.~2, pp. 623--633, 1968. [Online]. Available:
  \url{https://onlinelibrary.wiley.com/doi/abs/10.1002/pssb.19680280220}
\BIBentrySTDinterwordspacing

\bibitem{schnakenberg_electron-phonon_1969}
\BIBentryALTinterwordspacing
------, ``Electron-phonon interaction and {Boltzmann} equation in narrow-band
  semiconductors,'' in \emph{Springer {Tracts} in {Modern} {Physics}:
  {Ergebnisse} der exakten {Naturwissenschaften} {Volume} 51}, ser. Springer
  {Tracts} in {Modern} {Physics}.\hskip 1em plus 0.5em minus 0.4em\relax
  Berlin, Heidelberg: Springer, 1969, pp. 74--120. [Online]. Available:
  \url{https://doi.org/10.1007/BFb0107301}
\BIBentrySTDinterwordspacing

\bibitem{silbey_exchange_1965}
\BIBentryALTinterwordspacing
R.~Silbey, J.~Jortner, S.~A. Rice, and M.~T. Vala, ``Exchange {Effects} on the
  {Electron} and {Hole} {Mobility} in {Crystalline} {Anthracene} and
  {Naphthalene},'' \emph{The Journal of Chemical Physics}, vol.~42, no.~2, pp.
  733--737, Jan. 1965. [Online]. Available:
  \url{https://aip.scitation.org/doi/10.1063/1.1695999}
\BIBentrySTDinterwordspacing

\bibitem{silbey_general_1980}
\BIBentryALTinterwordspacing
R.~Silbey and R.~W. Munn, ``General theory of electronic transport in molecular
  crystals. {I}. {Local} linear electron–phonon coupling,'' \emph{The Journal
  of Chemical Physics}, vol.~72, no.~4, pp. 2763--2773, Feb. 1980. [Online].
  Available: \url{https://aip.scitation.org/doi/10.1063/1.439425}
\BIBentrySTDinterwordspacing

\bibitem{munn_theory_1985}
\BIBentryALTinterwordspacing
R.~W. Munn and R.~Silbey, ``Theory of electronic transport in molecular
  crystals. {II}. {Zeroth} order states incorporating nonlocal linear
  electron–phonon coupling,'' \emph{The Journal of Chemical Physics},
  vol.~83, no.~4, pp. 1843--1853, Aug. 1985. [Online]. Available:
  \url{https://aip.scitation.org/doi/10.1063/1.449372}
\BIBentrySTDinterwordspacing

\bibitem{munn_theory_1985-1}
\BIBentryALTinterwordspacing
------, ``Theory of electronic transport in molecular crystals. {III}.
  {Diffusion} coefficient incorporating nonlocal linear electron–phonon
  coupling,'' \emph{The Journal of Chemical Physics}, vol.~83, no.~4, pp.
  1854--1864, Aug. 1985. [Online]. Available:
  \url{https://aip.scitation.org/doi/10.1063/1.449373}
\BIBentrySTDinterwordspacing

\bibitem{coropceanu_charge_2007}
\BIBentryALTinterwordspacing
V.~Coropceanu, J.~Cornil, D.~A. da~Silva~Filho, Y.~Olivier, R.~Silbey, and
  J.-L. Br\'{e}das, ``Charge {Transport} in {Organic} {Semiconductors},''
  \emph{Chemical Reviews}, vol. 107, no.~4, pp. 926--952, Apr. 2007. [Online].
  Available: \url{https://doi.org/10.1021/cr050140x}
\BIBentrySTDinterwordspacing

\bibitem{cheng_unified_2008}
\BIBentryALTinterwordspacing
Y.-C. Cheng and R.~J. Silbey, ``A unified theory for charge-carrier transport
  in organic crystals,'' \emph{The Journal of Chemical Physics}, vol. 128,
  no.~11, p. 114713, Mar. 2008. [Online]. Available:
  \url{https://aip.scitation.org/doi/10.1063/1.2894840}
\BIBentrySTDinterwordspacing

\bibitem{kenkre_unified_1989}
\BIBentryALTinterwordspacing
V.~M. Kenkre, J.~D. Andersen, D.~H. Dunlap, and C.~B. Duke, ``Unified theory of
  the mobilities of photoinjected electrons in naphthalene,'' \emph{Physical
  Review Letters}, vol.~62, no.~10, pp. 1165--1168, Mar. 1989. [Online].
  Available: \url{https://link.aps.org/doi/10.1103/PhysRevLett.62.1165}
\BIBentrySTDinterwordspacing

\bibitem{kenkre_charge_1992}
\BIBentryALTinterwordspacing
V.~M. Kenkre and D.~H. Dunlap, ``Charge transport in molecular solids: dynamic
  and static disorder,'' \emph{Philosophical Magazine B}, vol.~65, no.~4, pp.
  831--841, Apr. 1992. [Online]. Available:
  \url{https://doi.org/10.1080/13642819208204923}
\BIBentrySTDinterwordspacing

\bibitem{kenkre_finite-bandwidth_2002}
\BIBentryALTinterwordspacing
V.~M. Kenkre, ``Finite-bandwidth calculations for charge carrier mobility in
  organic crystals,'' \emph{Physics Letters A}, vol. 305, no.~6, pp. 443--447,
  Dec. 2002. [Online]. Available:
  \url{http://www.sciencedirect.com/science/article/pii/S0375960102015189}
\BIBentrySTDinterwordspacing

\bibitem{giuggioli_mobility_2003}
\BIBentryALTinterwordspacing
L.~Giuggioli, J.~D. Andersen, and V.~M. Kenkre, ``Mobility theory of
  intermediate-bandwidth carriers in organic crystals: {Scattering} by acoustic
  and optical phonons,'' \emph{Physical Review B}, vol.~67, no.~4, p. 045110,
  Jan. 2003. [Online]. Available:
  \url{https://link.aps.org/doi/10.1103/PhysRevB.67.045110}
\BIBentrySTDinterwordspacing

\bibitem{shen_charge_2003}
\BIBentryALTinterwordspacing
Y.~Shen, K.~Diest, M.~H. Wong, B.~R. Hsieh, D.~H. Dunlap, and G.~G. Malliaras,
  ``Charge transport in doped organic semiconductors,'' \emph{Physical Review
  B}, vol.~68, no.~8, p. 081204, Aug. 2003. [Online]. Available:
  \url{https://link.aps.org/doi/10.1103/PhysRevB.68.081204}
\BIBentrySTDinterwordspacing

\bibitem{tovstenko_excitonphonon_2002}
\BIBentryALTinterwordspacing
V.~I. Tovstenko and I.~V. Sekirin, ``Exciton–phonon interaction in molecular
  crystals with noncorrelated couplings between excited states and lattice
  vibrations,'' \emph{The Journal of Chemical Physics}, vol. 117, no.~20, pp.
  9434--9444, Nov. 2002. [Online]. Available:
  \url{https://aip.scitation.org/doi/10.1063/1.1515322}
\BIBentrySTDinterwordspacing

\bibitem{mulliken_formulas_1949}
\BIBentryALTinterwordspacing
R.~S. Mulliken, C.~A. Rieke, D.~Orloff, and H.~Orloff, ``Formulas and
  {Numerical} {Tables} for {Overlap} {Integrals},'' \emph{The Journal of
  Chemical Physics}, vol.~17, no.~12, pp. 1248--1267, Dec. 1949. [Online].
  Available: \url{https://aip.scitation.org/doi/10.1063/1.1747150}
\BIBentrySTDinterwordspacing

\bibitem{frohlich_electrons_1954}
\BIBentryALTinterwordspacing
H.~Fr\"{o}hlich, ``Electrons in lattice fields,'' \emph{Advances in Physics},
  vol.~3, no.~11, pp. 325--361, Jul. 1954. [Online]. Available:
  \url{https://doi.org/10.1080/00018735400101213}
\BIBentrySTDinterwordspacing

\bibitem{frohlich_xx_1950}
\BIBentryALTinterwordspacing
H.~Fr\"{o}hlich, H.~Pelzer, and S.~Zienau, ``{XX}. {Properties} of slow
  electrons in polar materials,'' \emph{The London, Edinburgh, and Dublin
  Philosophical Magazine and Journal of Science}, vol.~41, no. 314, pp.
  221--242, Mar. 1950. [Online]. Available:
  \url{https://doi.org/10.1080/14786445008521794}
\BIBentrySTDinterwordspacing

\bibitem{landau_effective_1948}
L.~Landau and S.~Pekar, ``Effective mass of a polaron,'' \emph{Zh. Eksp. Teor.
  Fiz}, vol.~18, no.~5, pp. 419--423, 1948.

\bibitem{landau_collected_1965}
\BIBentryALTinterwordspacing
L.~D. Landau, \emph{Collected papers of {LD} {Landau}}.\hskip 1em plus 0.5em
  minus 0.4em\relax Pergamon, 1965. [Online]. Available:
  \url{https://www.sciencedirect.com/science/article/pii/B9780080105864500158}
\BIBentrySTDinterwordspacing

\bibitem{lee_motion_1953}
\BIBentryALTinterwordspacing
T.~D. Lee, F.~E. Low, and D.~Pines, ``The {Motion} of {Slow} {Electrons} in a
  {Polar} {Crystal},'' \emph{Physical Review}, vol.~90, no.~2, pp. 297--302,
  Apr. 1953. [Online]. Available:
  \url{https://link.aps.org/doi/10.1103/PhysRev.90.297}
\BIBentrySTDinterwordspacing

\bibitem{osaka_polaron_1959}
\BIBentryALTinterwordspacing
Y.~\={O}saka, ``Polaron {State} at a {Finite} {Temperature},'' \emph{Progress
  of Theoretical Physics}, vol.~22, no.~3, pp. 437--446, Sep. 1959. [Online].
  Available: \url{https://doi.org/10.1143/PTP.22.437}
\BIBentrySTDinterwordspacing

\bibitem{pekar_theory_1969}
S.~Pekar, ``Theory of {Polarons} in {Many}-valley {Crystals} {I}. {Weak}
  {Interaction} {Between} {Electron} and {Lattice} {Polarization} {Field},''
  \emph{Soviet Physics Jetp}, vol.~28, no.~5, 1969.

\bibitem{feynman_slow_1955}
\BIBentryALTinterwordspacing
R.~P. Feynman, ``Slow {Electrons} in a {Polar} {Crystal},'' \emph{Physical
  Review}, vol.~97, no.~3, pp. 660--665, Feb. 1955. [Online]. Available:
  \url{https://link.aps.org/doi/10.1103/PhysRev.97.660}
\BIBentrySTDinterwordspacing

\bibitem{feynman_mobility_1962}
\BIBentryALTinterwordspacing
R.~P. Feynman, R.~W. Hellwarth, C.~K. Iddings, and P.~M. Platzman, ``Mobility
  of {Slow} {Electrons} in a {Polar} {Crystal},'' \emph{Physical Review}, vol.
  127, no.~4, pp. 1004--1017, Aug. 1962. [Online]. Available:
  \url{https://link.aps.org/doi/10.1103/PhysRev.127.1004}
\BIBentrySTDinterwordspacing

\bibitem{luttinger_effect_1951}
\BIBentryALTinterwordspacing
J.~M. Luttinger, ``The {Effect} of a {Magnetic} {Field} on {Electrons} in a
  {Periodic} {Potential},'' \emph{Physical Review}, vol.~84, no.~4, pp.
  814--817, Nov. 1951. [Online]. Available:
  \url{https://link.aps.org/doi/10.1103/PhysRev.84.814}
\BIBentrySTDinterwordspacing

\bibitem{langreth_perturbation_1964}
\BIBentryALTinterwordspacing
D.~C. Langreth and L.~P. Kadanoff, ``Perturbation {Theoretic} {Calculation} of
  {Polaron} {Mobility},'' \emph{Physical Review}, vol. 133, no.~4a, pp.
  A1070--a1075, Feb. 1964. [Online]. Available:
  \url{https://link.aps.org/doi/10.1103/PhysRev.133.A1070}
\BIBentrySTDinterwordspacing

\bibitem{kadanoff_boltzmann_1963}
\BIBentryALTinterwordspacing
L.~P. Kadanoff, ``Boltzmann {Equation} for {Polarons},'' \emph{Physical
  Review}, vol. 130, no.~4, pp. 1364--1369, May 1963. [Online]. Available:
  \url{https://link.aps.org/doi/10.1103/PhysRev.130.1364}
\BIBentrySTDinterwordspacing

\bibitem{ambrosch-draxl_role_2009}
\BIBentryALTinterwordspacing
C.~Ambrosch-Draxl, D.~Nabok, P.~Puschnig, and C.~Meisenbichler, ``The role of
  polymorphism in organic thin films: oligoacenes investigated from first
  principles,'' \emph{New Journal of Physics}, vol.~11, no.~12, p. 125010, Dec.
  2009. [Online]. Available:
  \url{https://doi.org/10.1088\%2F1367-2630\%2F11\%2F12\%2F125010}
\BIBentrySTDinterwordspacing

\bibitem{yanagisawa_homo_2013}
\BIBentryALTinterwordspacing
S.~Yanagisawa, Y.~Morikawa, and A.~Schindlmayr, ``{HOMO} band dispersion of
  crystalline rubrene: {Effects} of self-energy corrections within the {GW}
  approximation,'' \emph{Physical Review B}, vol.~88, no.~11, p. 115438, Sep.
  2013. [Online]. Available:
  \url{https://link.aps.org/doi/10.1103/PhysRevB.88.115438}
\BIBentrySTDinterwordspacing

\bibitem{li_light_2007}
\BIBentryALTinterwordspacing
Z.~Q. Li, V.~Podzorov, N.~Sai, M.~C. Martin, M.~E. Gershenson, M.~Di~Ventra,
  and D.~N. Basov, ``Light {Quasiparticles} {Dominate} {Electronic} {Transport}
  in {Molecular} {Crystal} {Field}-{Effect} {Transistors},'' \emph{Physical
  Review Letters}, vol.~99, no.~1, p. 016403, Jul. 2007. [Online]. Available:
  \url{https://link.aps.org/doi/10.1103/PhysRevLett.99.016403}
\BIBentrySTDinterwordspacing

\bibitem{xi_first-principles_2012}
\BIBentryALTinterwordspacing
J.~Xi, M.~Long, L.~Tang, D.~Wang, and Z.~Shuai, ``First-principles prediction
  of charge mobility in carbon and organic nanomaterials,'' \emph{Nanoscale},
  vol.~4, no.~15, pp. 4348--4369, Jul. 2012. [Online]. Available:
  \url{https://pubs.rsc.org/en/content/articlelanding/2012/nr/c2nr30585b}
\BIBentrySTDinterwordspacing

\bibitem{oberhofer_charge_2017}
\BIBentryALTinterwordspacing
H.~Oberhofer, K.~Reuter, and J.~Blumberger, ``Charge {Transport} in {Molecular}
  {Materials}: {An} {Assessment} of {Computational} {Methods},'' \emph{Chemical
  Reviews}, vol. 117, no.~15, pp. 10\,319--10\,357, Aug. 2017. [Online].
  Available: \url{https://doi.org/10.1021/acs.chemrev.7b00086}
\BIBentrySTDinterwordspacing

\bibitem{abdulla_a_2015}
\BIBentryALTinterwordspacing
M.~Abdulla, K.~Refson, R.~H. Friend, and P.~D. Haynes, ``A first-principles
  study of the vibrational properties of crystalline tetracene under
  pressure,'' \emph{Journal of Physics: Condensed Matter}, vol.~27, no.~37, p.
  375402, sep 2015. [Online]. Available:
  \url{https://doi.org/10.1088/0953-8984/27/37/375402}
\BIBentrySTDinterwordspacing

\bibitem{dorner_the_1982}
\BIBentryALTinterwordspacing
B.~Dorner, E.~L. Bokhenkov, S.~L. Chaplot, J.~Kalus, I.~Natkaniec, G.~S.
  Pawley, U.~Schmelzer, and E.~F. Sheka, ``The 12 external and the 4 lowest
  internal phonon dispersion branches in d10-anthracene at 12k,'' \emph{Journal
  of Physics C: Solid State Physics}, vol.~15, no.~11, pp. 2353--2365, apr
  1982. [Online]. Available: \url{https://doi.org/10.1088/0022-3719/15/11/016}
\BIBentrySTDinterwordspacing

\bibitem{silinsh_molecular_1995}
\BIBentryALTinterwordspacing
E.~Silinsh, A.~Klimkāns, S.~Larsson, and V.~Čápek, ``Molecular polaron
  states in polyacene crystals. formation and transfer processes,''
  \emph{Chemical Physics}, vol. 198, no.~3, pp. 311--331, 1995. [Online].
  Available:
  \url{https://www.sciencedirect.com/science/article/pii/030101049500151D}
\BIBentrySTDinterwordspacing

\bibitem{nematiaram_modeling_2020}
\BIBentryALTinterwordspacing
T.~Nematiaram and A.~Troisi, ``Modeling charge transport in high-mobility
  molecular semiconductors: Balancing electronic structure and quantum dynamics
  methods with the help of experiments,'' \emph{The Journal of Chemical
  Physics}, vol. 152, no.~19, p. 190902, 2020. [Online]. Available:
  \url{https://doi.org/10.1063/5.0008357}
\BIBentrySTDinterwordspacing

\bibitem{feynman_space-time_1948}
\BIBentryALTinterwordspacing
R.~P. Feynman, ``Space-time approach to non-relativistic quantum mechanics,''
  \emph{Rev. Mod. Phys.}, vol.~20, pp. 367--387, Apr 1948. [Online]. Available:
  \url{https://link.aps.org/doi/10.1103/RevModPhys.20.367}
\BIBentrySTDinterwordspacing

\bibitem{schwinger_on_1951}
\BIBentryALTinterwordspacing
J.~Schwinger, ``On gauge invariance and vacuum polarization,'' \emph{Phys.
  Rev.}, vol.~82, pp. 664--679, Jun 1951. [Online]. Available:
  \url{https://link.aps.org/doi/10.1103/PhysRev.82.664}
\BIBentrySTDinterwordspacing

\bibitem{konstantinov_diagram_1961}
O.~Konstantinov and V.~Perel, ``A diagram technique for evaluating transport
  quantities,'' \emph{Journal of Experimental and Theoretical Physics},
  vol.~12, no.~1, pp. 142--149, 1961.

\bibitem{mattuck_guide_1976}
R.~D. Mattuck, \emph{A guide to {Feynman} diagrams in the many-body problem},
  2nd~ed., ser. Advanced book program.\hskip 1em plus 0.5em minus 0.4em\relax
  New York, NY: McGraw-Hill, 1976.

\bibitem{fanchenko_generalized_1983}
\BIBentryALTinterwordspacing
S.~S. Fanchenko, ``Generalized diagram technique of nonequilibrium processes,''
  \emph{Theoretical and Mathematical Physics}, vol.~55, no.~1, pp. 406--409,
  Apr. 1983. [Online]. Available:
  \url{https://link.springer.com/article/10.1007/BF01019028}
\BIBentrySTDinterwordspacing

\bibitem{kohn_analytic_1959}
\BIBentryALTinterwordspacing
W.~Kohn, ``Analytic {Properties} of {Bloch} {Waves} and {Wannier}
  {Functions},'' \emph{Physical Review}, vol. 115, no.~4, pp. 809--821, Aug.
  1959. [Online]. Available:
  \url{https://link.aps.org/doi/10.1103/PhysRev.115.809}
\BIBentrySTDinterwordspacing

\bibitem{giustino_electron_2007}
\BIBentryALTinterwordspacing
F.~Giustino, M.~L. Cohen, and S.~G. Louie, ``Electron-phonon interaction using
  wannier functions,'' \emph{Phys. Rev. B}, vol.~76, p. 165108, Oct 2007.
  [Online]. Available:
  \url{https://link.aps.org/doi/10.1103/PhysRevB.76.165108}
\BIBentrySTDinterwordspacing

\bibitem{marzari_maximally_2012}
\BIBentryALTinterwordspacing
N.~Marzari, A.~A. Mostofi, J.~R. Yates, I.~Souza, and D.~Vanderbilt,
  ``Maximally localized wannier functions: Theory and applications,''
  \emph{Rev. Mod. Phys.}, vol.~84, pp. 1419--1475, Oct 2012. [Online].
  Available: \url{https://link.aps.org/doi/10.1103/RevModPhys.84.1419}
\BIBentrySTDinterwordspacing

\bibitem{stefanucci_vanleeuwen_2013}
G.~Stefanucci and R.~van Leeuwen, \emph{Nonequilibrium Many-Body Theory of
  Quantum Systems: A Modern Introduction}.\hskip 1em plus 0.5em minus
  0.4em\relax Cambridge University Press, 2013.

\bibitem{shavitt_bartlett_2009}
I.~Shavitt and R.~J. Bartlett, \emph{Many-Body Methods in Chemistry and
  Physics: MBPT and Coupled-Cluster Theory}, ser. Cambridge Molecular
  Science.\hskip 1em plus 0.5em minus 0.4em\relax Cambridge University Press,
  2009.

\bibitem{yamashita_heitler-london_1960}
\BIBentryALTinterwordspacing
J.~Yamashita and T.~Kurosawa, ``Heitler-{London} {Approach} {To} {Electrical}
  {Conductivity} and {Application} to d-{Electron} {Conductions},''
  \emph{Journal of the Physical Society of Japan}, vol.~15, no.~5, pp.
  802--821, May 1960. [Online]. Available:
  \url{https://journals.jps.jp/doi/10.1143/JPSJ.15.802}
\BIBentrySTDinterwordspacing

\bibitem{kurosawa_heitler-london_1960}
\BIBentryALTinterwordspacing
T.~Kurosawa, ``Heitler-{London} {Approach} to {Electrical} {Conductivity},
  {II}. {A} {Proof} of the {Hopping} {Motion},'' \emph{Journal of the Physical
  Society of Japan}, vol.~15, no.~7, pp. 1211--1216, Jul. 1960. [Online].
  Available: \url{https://journals.jps.jp/doi/10.1143/JPSJ.15.1211}
\BIBentrySTDinterwordspacing

\bibitem{bateman_tables_1954}
\BIBentryALTinterwordspacing
H.~Bateman, \emph{Tables of {Integral} {Transforms} [{Volumes} {I} \& {II}]},
  A.~Erd\'{e}lyi, Ed.\hskip 1em plus 0.5em minus 0.4em\relax New York:
  McGraw-Hill Book Company, 1954, vol. I \& Ii. [Online]. Available:
  \url{https://resolver.caltech.edu/CaltechAUTHORS:20140123-101456353}
\BIBentrySTDinterwordspacing

\bibitem{feynman_feynman_1965}
\BIBentryALTinterwordspacing
R.~P. Feynman, R.~B. Leighton, and M.~Sands, ``The {Feynman} {Lectures} on
  {Physics}; {Vol}. {I},'' \emph{American Journal of Physics}, vol.~33, no.~9,
  pp. 750--752, Sep. 1965, publisher: American Association of Physics Teachers.
  [Online]. Available: \url{https://aapt.scitation.org/doi/10.1119/1.1972241}
\BIBentrySTDinterwordspacing

\bibitem{gradshteyn_table_1980}
\BIBentryALTinterwordspacing
I.~Ryzhik, ``Table of {Integrals}, {Series}, and {Products},'' in \emph{Table
  of {Integrals}, {Series}, and {Products}}, I.~S. Gradshteyn and I.~M. Ryzhik,
  Eds.\hskip 1em plus 0.5em minus 0.4em\relax Academic Press, 1980, p. vii.
  [Online]. Available:
  \url{https://www.sciencedirect.com/science/article/pii/B9780122947605500040}
\BIBentrySTDinterwordspacing

\bibitem{galitski_exploring_nodate}
\BIBentryALTinterwordspacing
V.~Galitski, B.~Karnakov, and V.~Kogan, \emph{Exploring Quantum
  Mechanics}.\hskip 1em plus 0.5em minus 0.4em\relax Oxford University Press,
  2013, publication Title: Exploring Quantum Mechanics. [Online]. Available:
  \url{https://oxford.universitypressscholarship.com/view/10.1093/acprof:oso/9780199232710.001.0001/acprof-9780199232710}
\BIBentrySTDinterwordspacing

\bibitem{watanabe_definition_2015}
\BIBentryALTinterwordspacing
K.~Watanabe, ``Definition of {Integral} {Transforms} and {Distributions},'' in
  \emph{Integral {Transform} {Techniques} for {Green}'s {Function}}, ser.
  Lecture {Notes} in {Applied} and {Computational} {Mechanics}, K.~Watanabe,
  Ed.\hskip 1em plus 0.5em minus 0.4em\relax Cham: Springer International
  Publishing, 2015, pp. 1--32. [Online]. Available:
  \url{https://doi.org/10.1007/978-3-319-17455-6\%5F1}
\BIBentrySTDinterwordspacing

\bibitem{grosche_handbook_1998}
\BIBentryALTinterwordspacing
C.~Grosche and F.~Steiner, \emph{Handbook of {Feynman} {Path} {Integrals}},
  ser. Springer {Tracts} in {Modern} {Physics}.\hskip 1em plus 0.5em minus
  0.4em\relax Berlin Heidelberg: Springer-Verlag, 1998. [Online]. Available:
  \url{https://www.springer.com/gp/book/9783662147610}
\BIBentrySTDinterwordspacing

\bibitem{berestetskii_preface_1982}
\BIBentryALTinterwordspacing
V.~B. Berestetski, E.~M. Lifshitz, and L.~P. Pitaevski, \emph{Quantum
  {Electrodynamics}}, V.~B. Berestetski, E.~M. Lifshitz, and L.~P. Pitaevski,
  Eds.\hskip 1em plus 0.5em minus 0.4em\relax Oxford: Butterworth-Heinemann,
  1982. [Online]. Available:
  \url{https://www.sciencedirect.com/science/article/pii/B9780080503462500040}
\BIBentrySTDinterwordspacing

\bibitem{bonch-bruevich_green_1962}
V.~Bonch-Bruevich, S.~Tyablikov, N.~Bogolyubov, and D.~Haar, \emph{The Green
  Function Method in Statistical Mechanics}.\hskip 1em plus 0.5em minus
  0.4em\relax {North}-{Holland}, {Amsterdam}; {Interscience} ({Wiley}), {New}
  {York}, 1962.

\bibitem{holmes_on_1999}
\BIBentryALTinterwordspacing
D.~Holmes, S.~Kumaraswamy, A.~J. Matzger, and K.~P.~C. Vollhardt, ``On the
  nature of nonplanarity in the [n]phenylenes,'' \emph{Chemistry – A European
  Journal}, vol.~5, no.~11, pp. 3399--3412, 1999. [Online]. Available:
  \url{https://chemistry-europe.onlinelibrary.wiley.com/doi/abs/10.1002/\%28SICI\%291521-3765\%2819991105\%295\%3A11\%3C3399\%3A\%3AAID-CHEM3399\%3E3.0.CO\%3B2-V}
\BIBentrySTDinterwordspacing

\bibitem{jurchescu_low_2006}
\BIBentryALTinterwordspacing
O.~D. Jurchescu, A.~Meetsma, and T.~T.~M. Palstra, ``Low-temperature structure
  of rubrene single crystals grown by vapor transport,'' \emph{Acta
  Crystallographica Section B}, vol.~62, no.~2, pp. 330--334, Apr 2006.
  [Online]. Available: \url{https://doi.org/10.1107/S0108768106003053}
\BIBentrySTDinterwordspacing

\bibitem{tsubaki_quantum_2020}
\BIBentryALTinterwordspacing
M.~Tsubaki and T.~Mizoguchi, ``Quantum {Deep} {Field} {Data}-{Driven} {Wave}
  {Function} {Electron} {Density} {Generation} and {Atomization} {Energy}
  {Prediction} and {Extrapolation} with {Machine} {Learning},'' \emph{Physical
  Review Letters}, vol. 125, no.~20, p. 206401, Nov. 2020, publisher: American
  Physical Society. [Online]. Available:
  \url{https://link.aps.org/doi/10.1103/PhysRevLett.125.206401}
\BIBentrySTDinterwordspacing

\bibitem{burkle_deep-learning_2021}
\BIBentryALTinterwordspacing
M.~Bürkle, U.~Perera, F.~Gimbert, H.~Nakamura, M.~Kawata, and Y.~Asai,
  ``Deep-{Learning} {Approach} to {First}-{Principles} {Transport}
  {Simulations},'' \emph{Physical Review Letters}, vol. 126, no.~17, p. 177701,
  Apr. 2021, publisher: American Physical Society. [Online]. Available:
  \url{https://link.aps.org/doi/10.1103/PhysRevLett.126.177701}
\BIBentrySTDinterwordspacing

\bibitem{mahan_many-particle_1990}
\BIBentryALTinterwordspacing
G.~D. Mahan, \emph{Many-{Particle} {Physics}}, 2nd~ed., ser. Physics of
  {Solids} and {Liquids}.\hskip 1em plus 0.5em minus 0.4em\relax Springer US,
  1990. [Online]. Available:
  \url{https://www.springer.com/gp/book/9780306434235}
\BIBentrySTDinterwordspacing

\bibitem{cohen_louie_2016}
M.~L. Cohen and S.~G. Louie, \emph{Fundamentals of Condensed Matter
  Physics}.\hskip 1em plus 0.5em minus 0.4em\relax Cambridge University Press,
  2016.

\bibitem{lang_kinetic_1963}
I.~Lang and Y.~A. Firsov, ``Kinetic theory of semiconductors with low
  mobility,'' \emph{Sov. Phys. JETP}, vol.~16, no.~5, p. 1301, 1963.

\bibitem{lang_kinetic_1962}
------, ``Kinetic {Theory} of {Semi} conductors with {Small} {Mobility},''
  \emph{Zh. Eksper. Teor. Fiz}, vol.~43, pp. 1843--1850, 1962.

\bibitem{lang_calculation_1968}
G.~Lang and Y.~A. Firsov, ``Calculation of the activation probability for a
  jump of a small-radius polaron,'' \emph{Sov. Phys.--JETP}, vol.~27, p. 443,
  1968.

\bibitem{bryksin_influence_1968}
V.~Bryksin and Y.~Firsov, ``{Influence} {Of} {Small} {Polarons} {On} {Lattice}
  {Vibrations},'' \emph{Soviet Physics Solid State, Ussr}, vol.~10, no.~5, p.
  1083, 1968.

\bibitem{schwinger_greens-I_1951}
\BIBentryALTinterwordspacing
J.~Schwinger, ``On the {Green}'s functions of quantized fields. {I},''
  \emph{Proceedings of the National Academy of Sciences}, vol.~37, no.~7, pp.
  452--455, 1951. [Online]. Available:
  \url{https://www.pnas.org/content/37/7/452}
\BIBentrySTDinterwordspacing

\bibitem{schwinger_greens-II_1951}
\BIBentryALTinterwordspacing
------, ``On the {Green}'s functions of quantized fields. {II},''
  \emph{Proceedings of the National Academy of Sciences}, vol.~37, no.~7, pp.
  455--459, 1951, publisher: National Academy of Sciences. [Online]. Available:
  \url{https://www.pnas.org/content/37/7/455}
\BIBentrySTDinterwordspacing

\bibitem{martin_theory_1959}
P.~C. Martin, ``Theory of {Many}-{Particle} {Systems}. {I},'' \emph{Physical
  Review}, vol. 115, no.~6, pp. 1342--1373, 1959.

\bibitem{baym_conservation_1961}
\BIBentryALTinterwordspacing
G.~Baym and L.~P. Kadanoff, ``Conservation {Laws} and {Correlation}
  {Functions},'' \emph{Physical Review}, vol. 124, no.~2, pp. 287--299, Oct.
  1961. [Online]. Available:
  \url{https://link.aps.org/doi/10.1103/PhysRev.124.287}
\BIBentrySTDinterwordspacing

\bibitem{kadanoff_theory_1961}
L.~P. Kadanoff, ``Theory of {Many}-{Particle} {Systems}. {II}.
  {Superconductivity},'' \emph{Physical Review}, vol. 124, no.~3, pp. 670--697,
  1961.

\bibitem{keldysh_diagram_1964}
L.~V. Keldysh, ``Diagram technique for nonequilibrium processes,'' \emph{Zh.
  Eksp. Teor. Fiz.}, vol.~47, pp. 1515--1527, 1964.

\bibitem{wick_the_1950}
\BIBentryALTinterwordspacing
G.~C. Wick, ``The evaluation of the collision matrix,'' \emph{Phys. Rev.},
  vol.~80, pp. 268--272, Oct 1950. [Online]. Available:
  \url{https://link.aps.org/doi/10.1103/PhysRev.80.268}
\BIBentrySTDinterwordspacing

\bibitem{binder_nonequilibrium_1995}
\BIBentryALTinterwordspacing
R.~Binder and S.~W. Koch, ``Nonequilibrium semiconductor dynamics,''
  \emph{Progress in Quantum Electronics}, vol.~19, no.~4, pp. 307--462, Jan.
  1995. [Online]. Available:
  \url{http://www.sciencedirect.com/science/article/pii/007967279500001S}
\BIBentrySTDinterwordspacing

\bibitem{kadanoff_quantum_1962}
L.~P. Kadanoff and G.~Baym, \emph{Quantum {Statistical} {Mechanics}: {Green}'s
  {Function} {Methods} in {Equilibrium} and {Nonequilibrium} {Problems}}.\hskip
  1em plus 0.5em minus 0.4em\relax Benjamin-Cummings Publishing Company, 1962.

\bibitem{dyson_the_1949}
\BIBentryALTinterwordspacing
F.~J. Dyson, ``The s matrix in quantum electrodynamics,'' \emph{Phys. Rev.},
  vol.~75, pp. 1736--1755, Jun 1949. [Online]. Available:
  \url{https://link.aps.org/doi/10.1103/PhysRev.75.1736}
\BIBentrySTDinterwordspacing

\bibitem{hedin_new_1965}
\BIBentryALTinterwordspacing
L.~Hedin, ``New method for calculating the one-particle green's function with
  application to the electron-gas problem,'' \emph{Phys. Rev.}, vol. 139, pp.
  A796--a823, Aug 1965. [Online]. Available:
  \url{https://link.aps.org/doi/10.1103/PhysRev.139.A796}
\BIBentrySTDinterwordspacing

\bibitem{hybertsen_electron_1986}
\BIBentryALTinterwordspacing
M.~S. Hybertsen and S.~G. Louie, ``Electron correlation in semiconductors and
  insulators: Band gaps and quasiparticle energies,'' \emph{Phys. Rev. B},
  vol.~34, pp. 5390--5413, Oct 1986. [Online]. Available:
  \url{https://link.aps.org/doi/10.1103/PhysRevB.34.5390}
\BIBentrySTDinterwordspacing

\bibitem{migdal_interaction_1958}
A.~Migdal, ``Interaction between electrons and lattice vibrations in a normal
  metal,'' \emph{Sov. Phys. JETP}, vol.~7, no.~6, pp. 996--1001, 1958.

\bibitem{nery_quasiparticles_2018}
\BIBentryALTinterwordspacing
J.~P. Nery, P.~B. Allen, G.~Antonius, L.~Reining, A.~Miglio, and X.~Gonze,
  ``Quasiparticles and phonon satellites in spectral functions of
  semiconductors and insulators: Cumulants applied to the full first-principles
  theory and the fr\"ohlich polaron,'' \emph{Phys. Rev. B}, vol.~97, p. 115145,
  Mar 2018. [Online]. Available:
  \url{https://link.aps.org/doi/10.1103/PhysRevB.97.115145}
\BIBentrySTDinterwordspacing

\bibitem{augustus1858xx}
\BIBentryALTinterwordspacing
A.~Matthiessen and C.~Wheatstone, ``Xx. on the electric conducting power of the
  metals,'' \emph{Philosophical Transactions of the Royal Society of London},
  vol. 148, pp. 383--387, 1858. [Online]. Available:
  \url{https://royalsocietypublishing.org/doi/abs/10.1098/rstl.1858.0020}
\BIBentrySTDinterwordspacing

\bibitem{jacoboni_monte_1983}
\BIBentryALTinterwordspacing
C.~Jacoboni and L.~Reggiani, ``The monte carlo method for the solution of
  charge transport in semiconductors with applications to covalent materials,''
  \emph{Rev. Mod. Phys.}, vol.~55, pp. 645--705, Jul 1983. [Online]. Available:
  \url{https://link.aps.org/doi/10.1103/RevModPhys.55.645}
\BIBentrySTDinterwordspacing

\bibitem{datta_quantum_2005}
S.~Datta, \emph{{Quantum Transport: Atom to Transistor}}.\hskip 1em plus 0.5em
  minus 0.4em\relax Cambridge, England, UK: Cambridge University Press, Jun
  2005.

\bibitem{lundstrom_fundamentals_2009}
M.~Lundstrom, \emph{{Fundamentals of Carrier Transport}}.\hskip 1em plus 0.5em
  minus 0.4em\relax Cambridge, England, UK: Cambridge University Press, Oct
  2000.

\bibitem{kadanoff_quantum_2018}
L.~P. Kadanoff, \emph{Quantum statistical mechanics}.\hskip 1em plus 0.5em
  minus 0.4em\relax CRC Press, 2018.

\bibitem{bogolyubov_kinetic_1978}
\BIBentryALTinterwordspacing
N.~N. Bogolyubov, ``Kinetic equations for the electron-phonon systems,'' Joint
  Inst. for Nuclear Research, Tech. Rep. Jinr-e–17-11822, 1978. [Online].
  Available:
  \url{http://inis.iaea.org/Search/search.aspx?orig\%5Fq=RN:10427633}
\BIBentrySTDinterwordspacing

\bibitem{bogolyubov_generalized_1979}
\BIBentryALTinterwordspacing
------, ``Generalized kinetic equation for a dynamic system interacting with a
  phonon field,'' \emph{Theoretical and Mathematical Physics}, vol.~43, no.~1,
  pp. 283--292, 1979. [Online]. Available:
  \url{https://doi.org/10.1007/BF01018458}
\BIBentrySTDinterwordspacing

\bibitem{baumann_quantum_1962}
\BIBentryALTinterwordspacing
K.~Baumann and J.~Ranninger, ``Quantum theory of transport coefficients. {I},''
  \emph{Annals of Physics}, vol.~20, no.~1, pp. 157--170, Oct. 1962. [Online].
  Available:
  \url{http://www.sciencedirect.com/science/article/pii/0003491662901215}
\BIBentrySTDinterwordspacing

\bibitem{baumann_quantum_1963}
\BIBentryALTinterwordspacing
K.~Baumann, ``Quantum theory of transport coefficients. {II},'' \emph{Annals of
  Physics}, vol.~23, no.~2, pp. 221--232, Aug. 1963. [Online]. Available:
  \url{https://www.sciencedirect.com/science/article/pii/0003491663901933}
\BIBentrySTDinterwordspacing

\bibitem{wagner_expansions_1991}
\BIBentryALTinterwordspacing
M.~Wagner, ``Expansions of nonequilibrium {Green}'s functions,'' \emph{Physical
  Review B}, vol.~44, no.~12, pp. 6104--6117, Sep. 1991. [Online]. Available:
  \url{https://link.aps.org/doi/10.1103/PhysRevB.44.6104}
\BIBentrySTDinterwordspacing

\bibitem{fetter_quantum_1971}
A.~L. Fetter and J.~D. Walecka, \emph{Quantum {Theory} of {Many}-particle
  {Systems}}.\hskip 1em plus 0.5em minus 0.4em\relax McGraw-Hill, 1971.

\bibitem{datta_electronic_1997}
S.~Datta, \emph{Electronic transport in mesoscopic systems}.\hskip 1em plus
  0.5em minus 0.4em\relax Cambridge university press, 1997.

\bibitem{lake_single_1997}
\BIBentryALTinterwordspacing
R.~Lake, G.~Klimeck, R.~C. Bowen, and D.~Jovanovic, ``Single and multiband
  modeling of quantum electron transport through layered semiconductor
  devices,'' \emph{Journal of Applied Physics}, vol.~81, no.~12, pp.
  7845--7869, Jun. 1997. [Online]. Available:
  \url{https://aip.scitation.org/doi/10.1063/1.365394}
\BIBentrySTDinterwordspacing

\bibitem{wacker_semiconductor_2002}
\BIBentryALTinterwordspacing
A.~Wacker, ``Semiconductor superlattices: a model system for nonlinear
  transport,'' \emph{Physics Reports}, vol. 357, no.~1, pp. 1--111, Jan. 2002.
  [Online]. Available:
  \url{http://www.sciencedirect.com/science/article/pii/S0370157301000291}
\BIBentrySTDinterwordspacing

\bibitem{danielewicz_quantum-I_1984}
\BIBentryALTinterwordspacing
P.~Danielewicz, ``Quantum theory of nonequilibrium processes, i,'' \emph{Annals
  of Physics}, vol. 152, no.~2, pp. 239--304, 1984. [Online]. Available:
  \url{https://www.sciencedirect.com/science/article/pii/0003491684900927}
\BIBentrySTDinterwordspacing

\bibitem{danielewicz_quantum-II_1984}
\BIBentryALTinterwordspacing
------, ``Quantum theory of nonequilibrium processes ii. application to nuclear
  collisions,'' \emph{Annals of Physics}, vol. 152, no.~2, pp. 305--326, 1984.
  [Online]. Available:
  \url{https://www.sciencedirect.com/science/article/pii/0003491684900939}
\BIBentrySTDinterwordspacing

\bibitem{gebauer_current_2004}
\BIBentryALTinterwordspacing
R.~Gebauer and R.~Car, ``Current in open quantum systems,'' \emph{Phys. Rev.
  Lett.}, vol.~93, p. 160404, Oct 2004. [Online]. Available:
  \url{https://link.aps.org/doi/10.1103/PhysRevLett.93.160404}
\BIBentrySTDinterwordspacing

\bibitem{burke_density_2005}
\BIBentryALTinterwordspacing
K.~Burke, R.~Car, and R.~Gebauer, ``Functional theory of the electrical
  conductivity of molecular devices,'' \emph{Phys. Rev. Lett.}, vol.~94, p.
  146803, Apr 2005. [Online]. Available:
  \url{https://link.aps.org/doi/10.1103/PhysRevLett.94.146803}
\BIBentrySTDinterwordspacing

\bibitem{frederiksen_inelastic_2007}
\BIBentryALTinterwordspacing
T.~Frederiksen, M.~Paulsson, M.~Brandbyge, and A.-P. Jauho, ``Inelastic
  transport theory from first principles: Methodology and application to
  nanoscale devices,'' \emph{Phys. Rev. B}, vol.~75, p. 205413, May 2007.
  [Online]. Available:
  \url{https://link.aps.org/doi/10.1103/PhysRevB.75.205413}
\BIBentrySTDinterwordspacing

\bibitem{landauer_spatial_1957}
R.~Landauer, ``Spatial variation of currents and fields due to localized
  scatterers in metallic conduction,'' \emph{IBM Journal of Research and
  Development}, vol.~1, no.~3, pp. 223--231, 1957.

\bibitem{beenakker_quantum_1991}
\BIBentryALTinterwordspacing
C.~W.~J. Beenakker and H.~van Houten, ``Quantum {Transport} in {Semiconductor}
  {Nanostructures},'' in \emph{Solid {State} {Physics}}, ser. Semiconductor
  {Heterostructures} and {Nanostructures}, H.~Ehrenreich and D.~Turnbull,
  Eds.\hskip 1em plus 0.5em minus 0.4em\relax Academic Press, Jan. 1991,
  vol.~44, pp. 1--228. [Online]. Available:
  \url{http://www.sciencedirect.com/science/article/pii/S0081194708600910}
\BIBentrySTDinterwordspacing

\bibitem{weinmann_quantum_1994}
D.~Weinmann, \emph{Quantum transport in nanostructures}.\hskip 1em plus 0.5em
  minus 0.4em\relax Wirtschaftsverl. NW, Verlag f\"{u}r Neue Wiss., 1994.

\bibitem{datta_nanoscale_2000}
\BIBentryALTinterwordspacing
S.~Datta, ``Nanoscale device modeling: the {Green}'s function method,''
  \emph{Superlattices and Microstructures}, vol.~28, no.~4, pp. 253--278, Oct.
  2000. [Online]. Available:
  \url{http://www.sciencedirect.com/science/article/pii/S0749603600909200}
\BIBentrySTDinterwordspacing

\bibitem{anantram_modeling_2008}
M.~Anantram, M.~S. Lundstrom, and D.~E. Nikonov, ``Modeling of nanoscale
  devices,'' \emph{Proceedings of the IEEE}, vol.~96, no.~9, pp. 1511--1550,
  2008.

\bibitem{kubis_quantum_2009}
T.~Kubis, \emph{Quantum {Transport} in {Semiconductor} {Nanostructures}}, ser.
  Selected topics of semiconductor physics and technology.\hskip 1em plus 0.5em
  minus 0.4em\relax Walter Schottky Institut, Technische Universit\"{a}t
  M\"{u}nchen, 2009.

\bibitem{luisier_quantum_2006}
\BIBentryALTinterwordspacing
W.~Fichtner, \emph{Quantum {Transport} for {Nanostructures}}.\hskip 1em plus
  0.5em minus 0.4em\relax Integrated Systems Laboratory, ETH Z\"{u}rich, Sep.
  2006. [Online]. Available: \url{https://nanohub.org/resources/1792}
\BIBentrySTDinterwordspacing

\bibitem{hirsbrunner_review_2019}
\BIBentryALTinterwordspacing
M.~R. Hirsbrunner, T.~M. Philip, B.~Basa, Y.~Kim, M.~J. Park, and M.~J.
  Gilbert, ``A review of modeling interacting transient phenomena with
  non-equilibrium {Green} functions,'' \emph{Reports on Progress in Physics},
  vol.~82, no.~4, p. 046001, Mar. 2019. [Online]. Available:
  \url{https://doi.org/10.1088\%2F1361-6633\%2Faafe5f}
\BIBentrySTDinterwordspacing

\bibitem{craig_perturbation_1968}
\BIBentryALTinterwordspacing
R.~A. Craig, ``Perturbation {Expansion} for {Real}-{Time} {Green}'s
  {Functions},'' \emph{Journal of Mathematical Physics}, vol.~9, no.~4, pp.
  605--611, Apr. 1968. [Online]. Available:
  \url{https://aip.scitation.org/doi/abs/10.1063/1.1664616}
\BIBentrySTDinterwordspacing

\bibitem{danielewicz_quantum_1984}
\BIBentryALTinterwordspacing
P.~Danielewicz, ``Quantum theory of nonequilibrium processes, {I},''
  \emph{Annals of Physics}, vol. 152, no.~2, pp. 239--304, Feb. 1984. [Online].
  Available:
  \url{http://www.sciencedirect.com/science/article/pii/0003491684900927}
\BIBentrySTDinterwordspacing

\bibitem{haug_quantum_2008}
\BIBentryALTinterwordspacing
H.~Haug and A.-P. Jauho, \emph{Quantum {Kinetics} in {Transport} and {Optics}
  of {Semiconductors}}, 2nd~ed., ser. Springer {Series} in {Solid}-{State}
  {Sciences}.\hskip 1em plus 0.5em minus 0.4em\relax Berlin Heidelberg:
  Springer-Verlag, 2008. [Online]. Available:
  \url{https://www.springer.com/gp/book/9783540735618}
\BIBentrySTDinterwordspacing

\bibitem{haydock_recursive_1980}
\BIBentryALTinterwordspacing
R.~Haydock, ``The recursive solution of the {Schr\"{o}dinger} equation,''
  \emph{Computer Physics Communications}, vol.~20, no.~1, pp. 11--16, Sep.
  1980. [Online]. Available:
  \url{http://www.sciencedirect.com/science/article/pii/0010465580901010}
\BIBentrySTDinterwordspacing

\bibitem{teichert_improved_2017}
\BIBentryALTinterwordspacing
F.~Teichert, A.~Zienert, J.~Schuster, and M.~Schreiber, ``Improved recursive
  {Green}'s function formalism for quasi one-dimensional systems with realistic
  defects,'' \emph{Journal of Computational Physics}, vol. 334, pp. 607--619,
  Apr. 2017. [Online]. Available:
  \url{http://www.sciencedirect.com/science/article/pii/S0021999117300347}
\BIBentrySTDinterwordspacing

\bibitem{thouless_conductivity_1981}
\BIBentryALTinterwordspacing
D.~J. Thouless and S.~Kirkpatrick, ``Conductivity of the disordered linear
  chain,'' \emph{Journal of Physics C: Solid State Physics}, vol.~14, no.~3,
  pp. 235--245, Jan. 1981. [Online]. Available:
  \url{https://doi.org/10.1088\%2F0022-3719\%2F14\%2F3\%2F007}
\BIBentrySTDinterwordspacing

\bibitem{mackinnon_calculation_1985}
\BIBentryALTinterwordspacing
A.~MacKinnon, ``The calculation of transport properties and density of states
  of disordered solids,'' \emph{Zeitschrift f\"{u}r Physik B Condensed Matter},
  vol.~59, no.~4, pp. 385--390, Dec. 1985. [Online]. Available:
  \url{https://doi.org/10.1007/BF01328846}
\BIBentrySTDinterwordspacing

\bibitem{lee_nonequilibrium_2002}
\BIBentryALTinterwordspacing
S.-C. Lee and A.~Wacker, ``Nonequilibrium {Green}'s function theory for
  transport and gain properties of quantum cascade structures,'' \emph{Physical
  Review B}, vol.~66, no.~24, p. 245314, Dec. 2002. [Online]. Available:
  \url{https://link.aps.org/doi/10.1103/PhysRevB.66.245314}
\BIBentrySTDinterwordspacing

\bibitem{rahman_theory_2003}
A.~Rahman, J.~Guo, S.~Datta, and M.~S. Lundstrom, ``Theory of ballistic
  nanotransistors,'' \emph{Electron Devices, IEEE Transactions on}, vol.~50,
  no.~9, pp. 1853--1864, 2003.

\bibitem{cauley_distributed_2011}
S.~Cauley, M.~Luisier, V.~Balakrishnan, G.~Klimeck, and C.~K. Koh,
  ``Distributed non-equilibrium {Green}'s function algorithms for the
  simulation of nanoelectronic devices with scattering,'' \emph{Journal of
  Applied Physics}, vol. 110, no.~4, p. 043713, 2011.

\bibitem{caroli_direct_1971}
\BIBentryALTinterwordspacing
C.~Caroli, R.~Combescot, P.~Nozieres, and D.~Saint-James, ``Direct calculation
  of the tunneling current,'' \emph{Journal of Physics C: Solid State Physics},
  vol.~4, no.~8, pp. 916--929, Jun. 1971. [Online]. Available:
  \url{https://doi.org/10.1088\%2F0022-3719\%2F4\%2F8\%2F018}
\BIBentrySTDinterwordspacing

\bibitem{meir_landauer_1992}
\BIBentryALTinterwordspacing
Y.~Meir and N.~S. Wingreen, ``Landauer formula for the current through an
  interacting electron region,'' \emph{Physical Review Letters}, vol.~68,
  no.~16, pp. 2512--2515, Apr. 1992. [Online]. Available:
  \url{https://link.aps.org/doi/10.1103/PhysRevLett.68.2512}
\BIBentrySTDinterwordspacing

\bibitem{boltzmann_theorie_van_1896}
\BIBentryALTinterwordspacing
L.~Boltzmann, ``Theorie van der waals; gase mit zusammengesetzten molekülen;
  gasdissociation; schlussbemerkungen,'' in \emph{Theorie van der Waals; Gase
  mit Zusammengesetzten Molekülen; Gasdissociation; Schlussbemerkungen}, ser.
  Vorlesungen über gastheorie.\hskip 1em plus 0.5em minus 0.4em\relax Leipzig:
  verlag von Johann Ambrosius Barth, 1896, pp. 1--265. [Online]. Available:
  \url{https://gutenberg.beic.it/webclient/DeliveryManager?pid=10946008}
\BIBentrySTDinterwordspacing

\bibitem{boltzmann_theorie_der_1896}
\BIBentryALTinterwordspacing
------, ``Theorie der gase mit einatomigen molekülen, deren dimensionen gegen
  die mittlere weglänge verschwinden,'' in \emph{Theorie der Gase mit
  einatomigen Molekülen, deren Dimensionen gegen die mittlere weglänge
  Verschwinden}, ser. Vorlesungen über gastheorie.\hskip 1em plus 0.5em minus
  0.4em\relax Leipzig: verlag von Johann Ambrosius Barth, 1896, pp. 1--204.
  [Online]. Available:
  \url{https://gutenberg.beic.it/webclient/DeliveryManager?pid=10990650}
\BIBentrySTDinterwordspacing

\bibitem{ponce_epw_2016}
\BIBentryALTinterwordspacing
S.~Ponc\'{e}, E.~R. Margine, C.~Verdi, and F.~Giustino, ``{EPW}:
  {Electron}–phonon coupling, transport and superconducting properties using
  maximally localized {Wannier} functions,'' \emph{Computer Physics
  Communications}, vol. 209, pp. 116--133, Dec. 2016. [Online]. Available:
  \url{http://www.sciencedirect.com/science/article/pii/S0010465516302260}
\BIBentrySTDinterwordspacing

\bibitem{jin_perturbo_2021}
\BIBentryALTinterwordspacing
J.-J. Zhou, J.~Park, I.-T. Lu, I.~Maliyov, X.~Tong, and M.~Bernardi,
  ``Perturbo: A software package for ab initio electron–phonon interactions,
  charge transport and ultrafast dynamics,'' \emph{Computer Physics
  Communications}, vol. 264, p. 107970, 2021. [Online]. Available:
  \url{https://www.sciencedirect.com/science/article/pii/S0010465521000837}
\BIBentrySTDinterwordspacing

\end{thebibliography}
\end{document}